\documentclass[11pt]{article}
\pdfoutput=1
\usepackage{jcapmod}
\usepackage{graphicx}
\usepackage{amsmath}
\usepackage{amsfonts}
\usepackage{amssymb}
\usepackage{color}
\usepackage{epsf}
\usepackage{array}
\usepackage{longtable}
\newcolumntype{L}[1]{>{\raggedright\let\newline\\\arraybackslash\hspace{0pt}}m{#1}}
\newcolumntype{C}[1]{>{\centering\let\newline\\\arraybackslash\hspace{0pt}}m{#1}}
\newcolumntype{R}[1]{>{\raggedleft\let\newline\\\arraybackslash\hspace{0pt}}m{#1}}
\usepackage{float}
\usepackage{pifont}
\usepackage{dcolumn}
\usepackage{bm}
\usepackage{epsfig}
\usepackage{amssymb,latexsym,mathrsfs}
\usepackage{hyperref}
\usepackage{etoolbox}
\newcolumntype{C}[1]{>{\centering\let\newline\\\arraybackslash\hspace{0pt}}m{#1}}

\def\beq{\begin{equation}}
\def\eeq{\end{equation}}
\def\be{\begin{equation}}
\def\ee{\end{equation}}
\def\bea{\begin{eqnarray}}
\def\eea{\end{eqnarray}}

\newcommand{\smallfoot}[1]{\let\thefootnote\relax\footnotetext{\scriptsize{#1}}}

\definecolor{darkgreen}{rgb}{0,0.5,0}
\definecolor{darkblue}{rgb}{0,0,0.75}
\definecolor{royalblue}{RGB}{0,0,128}
\definecolor{darkred}{RGB}{210,0,0}

\newcommand\degree       {{\ifmmode^\circ\else$^\circ$\fi}}  
\newcommand\arcm         {{\ifmmode {'\ }\else$'     $\fi} } 
\newcommand\arcs         {{\ifmmode{''\ }\else$''    $\fi} } 

\newcommand\cge          {{$_ >\atop{^\sim}$}}
\newcommand\cle          {{$_ <\atop{^\sim}$}}
\newcommand\degsq        {{$deg^{2}$}}
 
\newcommand\eg           {{\it e.g.}, }
\newcommand\etal         {{et\thinspace al.}}     
\newcommand\Ha           {{$H\alpha$} }
\newcommand\Hb           {{$H\beta$} }

\newcommand\ie           {{\it i.e.}, }

\newcommand\Lstar        {{$L^{*}$} }

\newcommand\MAB          {{$M_{AB}$} }

\newcommand\mum          {{$\mu$m}}
\newcommand\Mo           {{\ $M_{\odot}$}}

\newcommand\Mstar        {{$M^{*}$} }

\newcommand\Phistar      {{$\Phi^{*}$} }

\def\apjl{\rmfamily{ApJ~}}        
\def\apjs{\rmfamily{ApJS~}}       

\def\aap{\rmfamily{A\&A~}}        

\def\ao{\rmfamily{Appl.~Opt.~}}   

\begin{document}
\setcounter{page}{1} \baselineskip=14.5pt \thispagestyle{empty}

\bigskip\


\title{Science Impacts of the SPHEREx All-Sky Optical to Near-Infrared
  Spectral Survey: \\
\vspace{1cm}
\emph{Report of a Community Workshop Examining
  Extragalactic, Galactic, Stellar and Planetary Science}
}


\abstract{
SPHEREx, the Spectro-Photometer for the History of the
  Universe, Epoch of Reionization, and Ices Explorer, is a proposed
  SMEX mission selected for Phase A study pointing to a downselect in
  early CY2017, leading to launch in CY2021. SPHEREx will
  carry out the first all-sky spectral survey at wavelengths between
  0.75 and 4.18 $\mu$m [with spectral resolution R$\sim$41.4] and 4.18 and 5.00 $\mu$m [with
  R$\sim$135].  At  the end of its two year mission,  SPHEREx will provide 0.75-to-5.00
  $\mu$m spectra of every 6.2$\times$6.2 arcsec pixel on the sky.  SPHEREx will
  easily obtain spectra with S/N $>$ 50 per frequency element of all sources in the 2MASS
  catalog and spectra with S/N $>$5 per frequency element of the faintest sources detected by
  WISE in its short wavelength channels at 3.4 and 4.8 $\mu$m.   More
  details concerning SPHEREx are available at
  \href{[Link]}{http://spherex.caltech.edu}.  The SPHEREx team has
  proposed three specific science investigations to be carried out
  with this unique data set: cosmic inflation,  interstellar and circumstellar ices, and the extra-galactic
  background light.

It is readily apparent, however, that many other questions in both
astrophysics and planetary sciences could be addressed with the
SPHEREx data. The SPHEREx team convened a community workshop in
February 2016, with the intent of enlisting the aid of a larger group
of scientists in defining these questions, and this paper summarizes
the results of that workshop
\href{[Link]}{http://spherex.caltech.edu/Workshop.html}. A rich and
varied menu of investigations was laid out, including studies of the composition of main belt and
Trojan/Greek asteroids; mapping the zodiacal light with higher spatial
and spectral resolution than has been done previously; identifying and
studying very low-metallicity stars; improving stellar parameters in
order to better characterize transiting exoplanets; studying aliphatic
and aromatic carbon-bearing molecules in the interstellar medium;
mapping star formation rates in nearby galaxies; determining the
redshift of clusters of galaxies; identifying high redshift quasars
over the full sky; and providing a NIR spectrum for all eROSITA X-ray
sources.  All of these investigations, and others not listed here, can
be carried out with the all-sky spectra to be produced by SPHEREx;
none of them imposes any additional observational requirements or
operating modes.  In addition, the workshop defined enhanced data
products and user tools which would facilitate some of these
scientific studies.  Finally, the workshop noted the high degrees of
synergy between SPHEREx and a number of other current or forthcoming
programs, including JWST, WFIRST, Euclid, GAIA, K2/Kepler, TESS,
eROSITA and LSST. 
}

\vspace{0.6cm}

\author{Olivier  Dor\'e$^{1,21}$, Michael W. Werner$^{1,21}$, Matt
  Ashby$^{26}$, Pancha Banerjee$^{23}$,  Nick  Battaglia$^{16}$, James Bauer$^1$,
  Robert A. Benjamin$^{34}$, Lindsey E. Bleem$^{36,37}$, Jamie Bock$^{21}$,
  Adwin Boogert$^2$, Philip Bull$^{1,21}$, Peter Capak$^8$, Tzu-Ching
  Chang$^3$, Jean Chiar$^{38}$, Seth H. Cohen$^{30}$,  Asantha Cooray$^{28}$, Brendan Crill$^{1,21}$, Michael
  Cushing$^4$, Roland de Putter$^{21}$, Simon P. Driver$^{32}$, Tim
  Eifler$^{21}$, Chang Feng$^{28}$, Simone
  Ferraro$^{19}$,  Douglas Finkbeiner$^{31}$, B. Scott Gaudi$^{18}$,
  Tom  Greene$^5$,  Lynne Hillenbrand$^{21}$, Peter A.~H\"
  oflich$^{6}$, Eric Hsiao$^6$, Kevin Huffenberger$^6$, Rolf
  A. Jansen$^{30}$, Woong-Seob Jeong$^7$, Bhavin Joshi$^{30}$, Duho
  Kim$^{30}$, Minjin Kim$^7$,  J.Davy Kirkpatrick$^8$, Phil
  Korngut$^{21}$, Elisabeth Krause$^9$, Mariska Kriek$^{25}$, Boris
  Leistedt$^{10}$, Aigen Li$^{40}$, Carey M. Lisse$^{11}$, Phil Mauskopf$^{30}$, Matt
  Mechtley$^{30}$, Gary Melnick$^{26}$, Joseph Mohr$^{35,41}$, Jeremiah
  Murphy$^6$, Abraham Neben$^{13}$, David	Neufeld$^{24}$, Hien 
  Nguyen$^{1,21}$, Elena Pierpaoli$^{23}$, Jeonghyun Pyo$^{7}$, Jason
  Rhodes$^{1,21}$, Karin Sandstrom$^{15}$, Emmanuel Schaan$^{16}$, 
  Kevin C. Schlaufman$^{22}$, John  Silverman$^{12}$, Kate Su$^{33}$,
  Keivan Stassun$^{17}$, Daniel Stevens$^{18}$, Michael
  A. Strauss$^{16}$, Xander Tielens$^{39}$, Chao-Wei Tsai$^{20}$, Volker Tolls$^{26}$,
  Stephen  Unwin$^{1}$, Marco Viero$^9$, Rogier A. Windhorst$^{30}$, Michael
  Zemcov$^{27,1}$\\ 
\vspace{0.75cm}
\emph{Editors and corresponding authors: Olivier Dor\'e$^{1,21}$,
  Michael Werner$^{1,21}$ (\texttt{olivier.dore@caltech.edu, michael.w.werner@jpl.nasa.gov})}}

\affiliation{$^1$Jet Propulsion Laboratory, 4800 Oak Grove Drive, Pasadena, CA 91109, USA\\
$^2$Universities Space Research Association, Stratospheric
Observatory for Infrared Astronomy, NASA Ames Research Center, MS 232-11, Moffett
Field, CA 94035, USA\\
$^3$ASIAA, AS/NTU, 1 Roosevelt Rd Sec 4, Taipei, 10617, Taiwan\\
$^4$Department of Physics and Astronomy, University of Toledo, 2801
W. Bancroft St., Toledo, OH 43606, USA\\
$^5$NASA Ames Research Center, MS 245-6, Moffett Field, CA 94035, USA\\
$^6$Department of Physics, Florida State University, PO Box 3064350, Tallahassee, Florida 32306-4350, USA\\
$^7$Korea Astronomy and Space Science Institute, Daejeon, 34055, Korea\\
$^8$IPAC, Caltech, 770 S. Wilson Ave, Pasadena, CA 91125, USA\\
$^9$Kavli Institute for Particle Astrophysics \& Cosmology, P. O. Box 2450, Stanford University, Stanford, CA 94305, USA\\
$^{10}$Center for Cosmology and Particle Physics, Department of Physics, New York University, New York, NY 10003, USA\\
$^{11}$JHU-APL, SES/SRE, Bldg 200/E206, 11100 Johns Hopkins Road,
Laurel, MD 20723, USA \\
$^{12}$Kavli Institute for the Physics and Mathematics of the Universe
(Kavli IPMU),The University of Tokyo, 5-1-5 Kashiwanoha
Kashiwa, 277-8583, Japan\\
$^{13}$Massachusetts Institute of Technology, Department of Physics, 70 Vassar Street, Bldg. 37-656, Cambridge, MA 02139, USA\\
$^{14}$Dept. of Physics and Astronomy, Johns Hopkins University, 3400 N. Charles St., Baltimore, MD 21218, USA\\
$^{15}$University of California, 9500 Gilman Drive, San Diego, La Jolla, CA 92093, USA\\
$^{16}$Department of Astrophysical Sciences, Peyton Hall, Princeton University, Princeton, NJ 08544, USA\\
$^{17}$Physics \& Astronomy Department, Vanderbilt University, Nashville, TN 37235, USA\\
$^{18}$Department of Astronomy, The Ohio State University, 140 W. 18th
Avenue, Columbus, OH 43210 USA\\
$^{19}$Berkeley Center for Cosmological Physics and Department of
Astronomy, University of California, Berkeley, CA 94720, USA\\
$^{20}$Department of Physics and Astronomy, UCLA, Los Angeles, CA 90095-1547, USA\\
$^{21}$California Institute of Technology, 1200 E. California Blvd, Pasadena, CA 91125, USA\\
$^{22}$Observatories of the Carnegie Institution for Science, 813
Santa Barbara St., Pasadena, CA 91101\\
$^{23}$Department of Physics and Astronomy, University of Southern California, 3620 McClintock Avenue, Los Angeles, CA 90089, USA\\
$^{24}$Dept. of Physics and Astronomy, Johns Hopkins University, 3400 N. Charles St., Baltimore, MD 21218, USA\\
$^{25}$Department of Astronomy, University of California, Berkeley, CA
94720, USA\\
$^{26}$ Harvard Smithsonian CfA, 60 Garden St., Cambridge, MA 02138, USA\\
$^{27}$ Rochester Institute of Technology, College of Science, 74 Lomb
Memorial Drive, Rochester, NY 14623, USA\\
$^{28}$University of California Irvine, 4186 Frederick Reines Hall,
Irvine, CA 92697, USA\\
$^{30}$Arizona State University, Department of Physics, P.O. Box
871504, Tempe, AZ 85287, USA\\
$^{31}$ Harvard-Smithsonian, Center for Astrophysics, 60 Garden
St. Cambridge, MA 02138, USA\\
$^{32}$ ICRAR M468, The University of Western Australia, 35 Stirling
Hwy, Crawley, Western Australia, 6009, Australia\\
$^{33}$ Steward Observatory, University of Arizona, Tucson, AZ 85721,
USA\\
$^{34}$ University of Wisconsin - Whitewater, Whitewater, WI 53190,
USA\\
$^{35}$ Faculty of Physics, Ludwig-Maximilians-Universitaet,
Scheinerstrasse 1, 81679 Munich, Germany\\
$^{36}$ Argonne National Laboratory, High-Energy Physics Division,
9700 S. Cass Avenue, Argonne, IL 60439, USA\\
$^{37}$ Kavli Institute for Cosmological Physics, University of
Chicago, 5640 South Ellis Avenue, Chicago, IL 60637, USA\\
$^{38}$ SETI Institute, Carl Sagan Center, 189 Bernardo Avenue, Mountain View,
CA 94043, USA\\
$^{39}$ Leiden Observatory, Leiden University, PO Box 9513, NL-2300RA
Leiden, The Netherlands\\
$^{40}$ Department of Physics and Astronomy, University of Missouri, Columbia,
MO 65211, USA\\
$^{41}$ Max Planck Institute for Extraterrestrial Physics, Giessenbachstrasse 1, 85478 Garching, Germany
}

\maketitle

\section{SPHEREx and the Decade of the Surveys}
\label{sec:intro}

\begin{figure}[!th]
\centering
\includegraphics[width=0.7\textwidth]{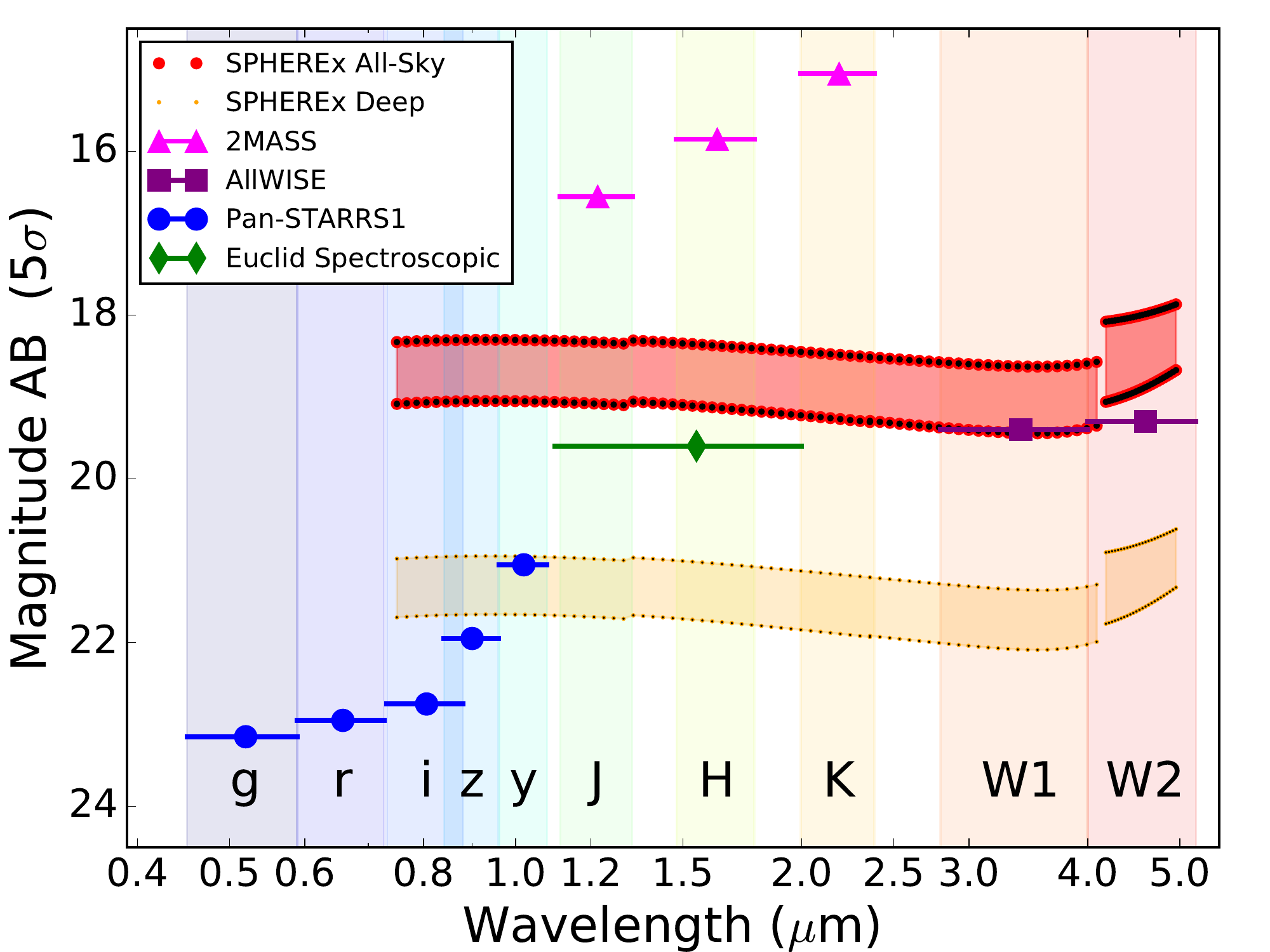}
\caption{Sensitivity of SPHEREx and current surveys (all at 5$\sigma$). The SPHEREX
sensitivity is quoted for  each $\lambda/\Delta\lambda$ = 41.4 spectral channel for 0.75 $<$
z $<$ 4.18 $\mu$m and in each $\lambda/\Delta\lambda$ = 135 spectral
channel for 4.18 $< z <$ 5.0 $\mu$m. The bottom red
  curve corresponds to the current best estimate sensitivity over the
  whole sky while the top curve corresponds to the instrument
  sensitivity based on specifications that each sub-system can meet
  with contingency over the  whole sky. The orange dots correspond to
  the analogous sensitivity curves over the deep regions. The statistical
  sensitivity does not include the effects of astrophysical source
  confusion, which is significant at the deep survey depth.} 
\label{fig:lam_mag}
\end{figure}

SPHEREx, the Spectro-Photometer for the History of the Universe, Epoch
of Reionization, and Ices Explorer is a proposed SMEX mission selected
for Phase A study pointing to a downselect in early CY2017, leading to
launch in CY2020.  The Principal Investigator is Professor 
Jamie Bock of Caltech.  SPHEREx will carry out the first
all-sky spectral survey at wavelengths between 0.75 and 4.18 $\mu$m
[with spectral resolution R$\sim$41.4] and 4.18 and 5.00 $\mu$m [with R$\sim$135].  At the end of
its two year mission, SPHEREx will obtain 0.75-to-5$\mu$m spectra of
every 6.2$\times$6.2 arcsec pixel on the sky, with a 5-sigma
sensitivity AB$\simeq$18-19 per spectral/spatial resolution element
(see Fig.~\ref{fig:lam_mag}). SPHEREx thus will easily obtain spectra
with S/N $>$ 50 per frequency element of all sources in the 2MASS catalog and spectra with S/N $>$ 5 of the faintest
sources detected by WISE in its short wavelength channels at 3.4 and
4.8 $\mu$m. Such a rich archival spectral database will support
numerous scientific investigations (e.g., Table \ref{tab:legacy}) of
great interest to the wider scientific community. We outline below a few examples.

The SPHEREx mission and its core science objectives are described in
detail on this website
\href{http://spherex.caltech.edu}{http://spherex.caltech.edu} and in \cite{Dore:2014cca}. The
  proposing team defined three high priority science investigations on
  which to focus its efforts, which in turn relate to the three major
  astrophysics themes of the NASA Science Plan: 
\begin{enumerate}
\item Test models of inflation by mapping the 3D distribution of
galaxies to measure primordial non-Gaussianity, to probe the running of the
primordial power spectrum spectral index, and to search for departures from geometric flatness
[NASA Science Plan Theme:  Probe the origin and destiny of the
universe]; and
\item Investigate the connection between ices in interstellar clouds,
  in planet-forming disks, and in our own Solar System [NASA Science
  Plan Theme:  Explore whether planets around other stars could harbor
  life]; and 
\item Determine the origin of the large-scale cosmic infrared
  background fluctuations, and constrain the history of its production
  [NASA Science Plan Theme: Explore the origin and evolution of galaxies].
\end{enumerate}

As important as these three investigations are, they represent only a
fraction of the science which would be enabled by SPHEREx' unique
all-sky spectroscopic data base as illustrated in Table~\ref{tab:legacy} extracted
from \cite{Dore:2014cca}.  From 24 to 26 February 2016, the
SPHEREx team convened a workshop held at the Keck Center on the
Caltech campus.   Experts in a wide range of astrophysical and planetary science disciplines attended to
discuss other scientific areas in which the SPHEREx data would have a
great impact, and to define software tools and ancillary data sets
which would enable the exploitation of the data in order to address
these other questions. Agenda and presentations can be found here
\href{http://spherex.caltech.edu/Workshop.html}{http://spherex.caltech.edu/Workshop.html}. This
white paper, based in large part on contributions from workshop participants, describes how the SPHEREx
data could be used to advance our understanding of key questions in
more than a dozen areas identified during the workshop.  Many
of these ideas will be presented as part of a Science Enhancement
Option (SEO) to be included in the SPHEREx study report which will be evaluated
by NASA in the process of selecting a mission for flight. In addition,
the white paper addresses the synergyism between SPHEREx and other current and future
NASA and ground-based programs.   

During the workshop, Michael Strauss of Princeton referred to the
2020s as the decade of the surveys.  He had in mind such programs
as WFIRST, Euclid, and LSST, each of which will survey large areas of
the sky and produce rich data sets which are synergistic with SPHEREx,
as discussed further below.  

SPHEREx would be the latest in a series of all-sky surveys to be carried out under NASA's Explorer program; previous
surveys include IRAS, COBE, WMAP, GALEX, and WISE.  2MASS is an
additional important survey, not carried out from space but supported
in part by NASA funds.  SPHEREx, based on its orbit and wavelength
coverage, might be thought of as a spectroscopic version of WISE or
2MASS.  Like those missions, SPHEREx will produce a treasure trove of
well-characterized, uniformly calibrated data which will be used by
the scientific community for years to come.  In addition, based on its
exploration of a previously unexplored region of scientific phase
space, SPHEREx, like the predecessor surveys, has the potential to
discover new and exciting phenomena as it extends our understanding of
the astronomical Universe. 

\begin{center}
\footnotesize{
\begin{longtable}{|L{4.0cm}|L{2.0cm}|L{4.0cm}|C{4.0cm}|}
 \hline
 \hline
 Object & \# Sources & Legacy Science & Reference \\
\hline
\hline
Detected galaxies & 1.4 billion & Properties of distant and heavily
obscured galaxies & Simulation based on COSMOS and Pan- STARRS\\
\hline
Galaxies with $\sigma (z)/(1+z)<0.1$ & 301 million & Study large scale
clustering of galaxies & Simulation based on COSMOS and Pan-STARRS\\
\hline
Galaxies with $\sigma (z)/(1+z)<0.03$ & 120 million & Study
(H$\alpha$, H$\beta$, CO, OII, OIII, SII,
H$_2$O) line and PAH emission by galaxy type. Explore galaxy and AGN life
cycle & Simulation based on COSMOS and Pan- STARRS\\
\hline
Galaxies with $\sigma(z)/(1+z)<0.003$ & 9.8 million & Cross check of
Euclid photo-$z$. Measure dynamics of groups and map
filaments. Cosmological galaxy clustering, BAO, RSD. &
Simulation based on COSMOS and Pan- STARRS\\
\hline
QSOs & $>$ 1.5 million & Understand QSO lifecycle, environment, and
taxonomy & \cite{Ross:2012dt} plus simulations\\
\hline
QSOs at $z > 7$ & 1-300 & Determine if early QSOs exist. Follow-up
spectro- scopy probes EOR through Ly$\alpha$ forest  &
\cite{Ross:2012dt} plus simulations\\
\hline
Clusters with $\geq$ 5 members & 25,000 & Redshifts for all eRosita
clusters. Viral masses and merger dynamics & \cite{Geach:2011gu} \\
\hline
Main sequence stars & $>100$ million & Fundamental parameters
                                       (mass, radius and effective
                                       temperature) of a large sample
                                       of stars.  Test the uniformity
                                       of the Galactic stellar mass
                                       function as input to
                                       extragalactic studies & 2MASS 
catalogs\\
\hline
Mass-losing, dust forming stars &	Over 10,000 of all types &
                                                                   Spectra
                                                                   of
                                                                   M
                                                                   supergiants, OH/IR stars, Carbon stars. Stellar atmospheres, dust return rates, and composition of dust &	Astrophysical Quantities, 4th edition [ed. A.Cox] p. 527\\  
\hline
Brown dwarfs & $>$400, incl. $>$40 of types T and Y & Atmospheric
structure and composition; search for hazes. Informs studies of giant
exoplanets & {\texttt dwarfarchives.org} and J.D. Kirkpatrick,
priv. comm.\\
\hline
Stars with hot dust & $>$1000 & Discover rare dust clouds produced by
cataclysmic events like the collision which produced the Earth’s moon
& Kennedy \& Wyatt (2013)\\
\hline
Young Stars with Accretion Disks & Over 20,000 of all types & Probe
                                                              strong
                                                              accretion/outflow
                                                              signatures,
                                                              star/disk
                                                              atmospheres,
                                                              extinction
                                                              & C2D
                                                                and
                                                                related
                                                                2MASS+Spitzer/WISE
                                                                surveys\\

\hline
Diffuse ISM &	Map of the Galactic plane & Study diffuse emission
from interstellar clouds and nebulae; hydro-carbon emission in the 3$\mu$m
region &	GLIMPSE survey (Churchwell et al. 2009) \\
\hline
\hline
\caption{SPHEREx spectral database populations (extract from \cite{Dore:2014cca}).} 
\label{tab:legacy}
\end{longtable}
}
\end{center}

\section{Extra-galactic Science}

\subsection{Mapping Star Formation Rates in Nearby Galaxies}

Star formation is one of the fundamental drivers of galaxy evolution.
Mapping the distribution of the current star formation rate (SFR) in
galaxies is central to many observational studies, however it is
subject to major systematic uncertainties.  Two of the key systematic
uncertainties are 1) the timescale over which a tracer is produced
(i.e. hydrogen recombination lines have timescales $<10$ Myr while
far-UV emission has timescales of $\sim 100$ Myr) and 2) how to
correct for extinction.  The latter issue has been the major impetus
for ``hybrid'' star formation rate tracers which combine H$\alpha$ or
far-UV with mid-IR dust emission, representing the obscured portion of
the SF \citep{2007ApJ...666..870C}.  The ``hybrid'' tracers are
typically calibrated with near-IR hydrogen recombination line emission
such as Paschen-$\alpha$, since it  is mostly insensitive to
extinction and dominated by recent $<10$Myr star
formation. Unfortunately, the inclusion of mid-IR emission into these
tracers opens up a host of additional issues related to how much of
the IR emission is powered by star formation rather than the radiation
field generated by older stars \citep[a.k.a. the ``cirrus''
correction][]{2009ApJ...703.1672K,2011ApJ...730...72R,2011ApJ...735...63L,Calzetti:2012ux,2012AJ....144....3L}.   

SPHEREx observations will allow a major step forward in studying star
formation in nearby galaxies by making 6 arcsec maps of the
Paschen-$\alpha$ and Brackett-$\alpha$ lines over the entire sky.  The sensitivity of the
line observations is well matched to cover the main star forming area
of nearby galaxies, detecting the near-IR recombination lines in most
regions where molecular gas dominates the ISM gas content.  In
addition, the stellar continuum observed by SPHEREx will provide a
gold standard stellar mass map, avoiding issues with contamination by
hot dust and PAH emission at 3.3 $\mu$m
\citep{2012ApJ...744...17M,2015ApJS..219....5Q}.  Combining this 
information, SPHEREx will provide resolved maps of the specific star
formation rate in the full local galaxy population, giving the
definitive account of where $z=0$ galaxies are forming stars.  SPHEREx
observations of near-IR hydrogen recombination lines will also be the
premier dataset for calibrating other SFR tracers for use in samples
of more distant galaxies. 

\subsection{ Star-Forming and Early-Type Galaxies with SPHEREx at
  z\cle 0.5}
\label{sec:gal0.5}

\begin{figure}
\center
\includegraphics[width=0.45\textwidth,angle=-0]{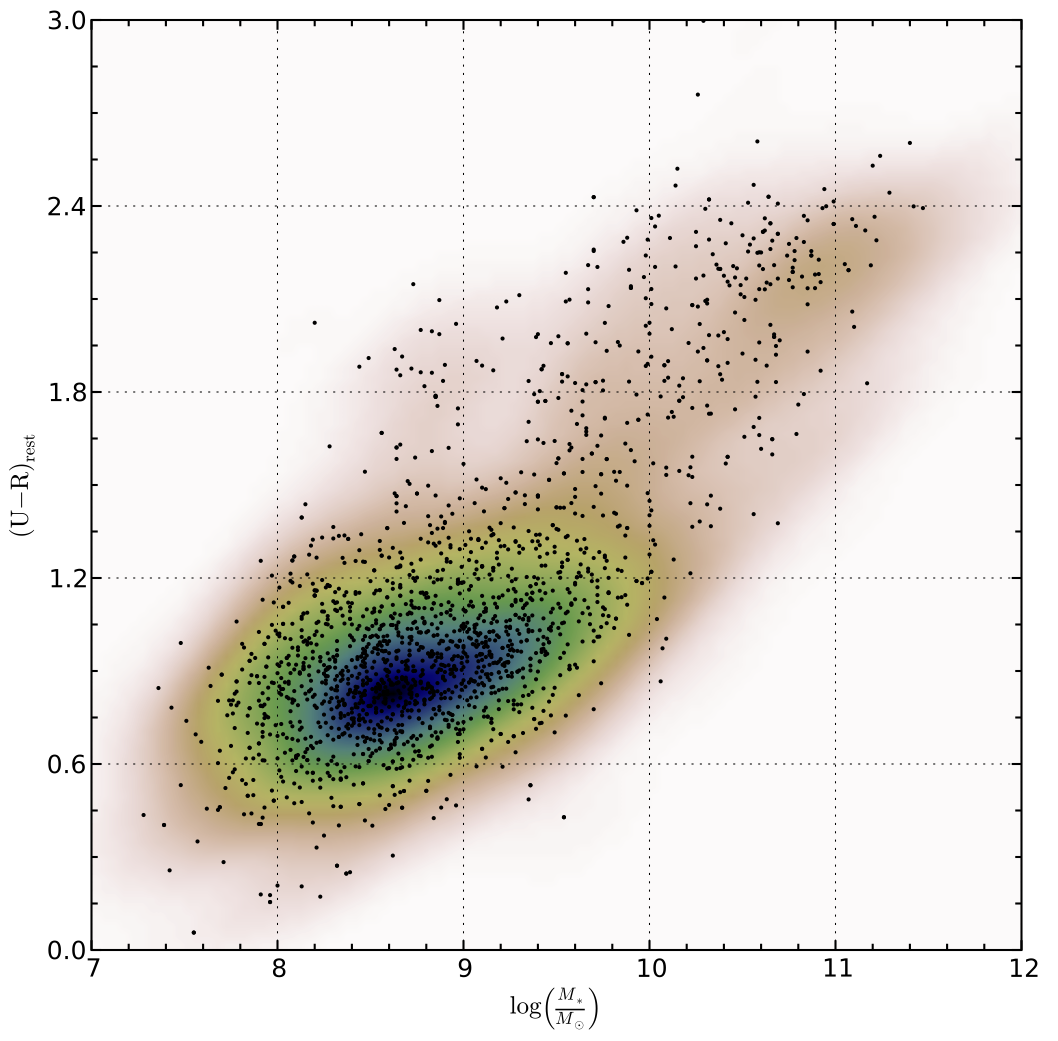}
\includegraphics[width=0.45\textwidth,height=0.45\textwidth,angle=-0]{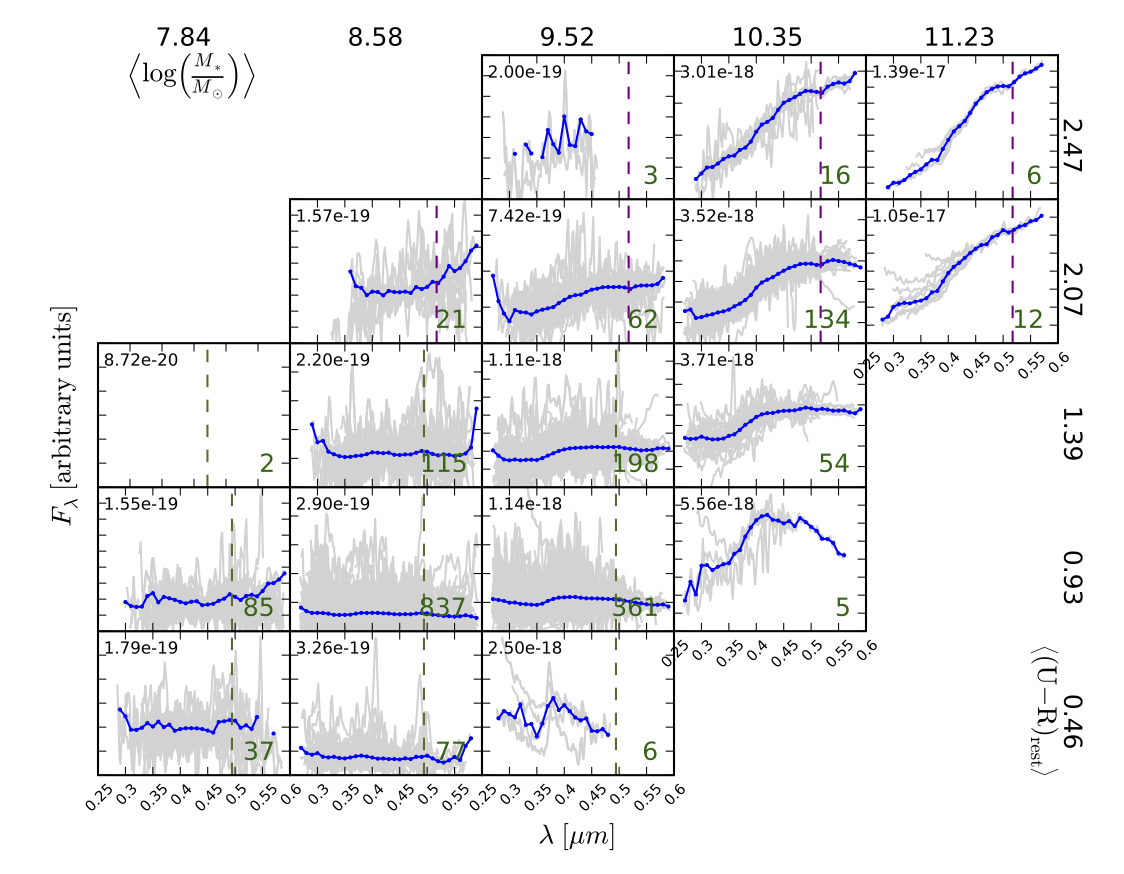}
\caption{\emph{Left:} Rest-frame (U--B)$_{rest}$ color vs. stellar mass
  (\Mstar/\Mo) for all galaxies to AB\cle 25 mag from the HST ACS
  ``PEARS'' G800L grism survey \cite{Malhotra:2005}, as analyzed in a 5$\times$5 
color-mass grid by (Joshi et al. 2016, in preparation) which samples the ``blue cloud'', ``green
valley'' and the narrower ``red cloud''.  \emph{Right:} Clipped-average rest-frame spectra for each of the 
(U--B)$_{rest}$ color and stellar mass bins [log(M$_{stellar}$)] in
the left panel. Green numbers indicate the number of similar rest-frame spectra averaged in 
each bin, the vertical green dashed line the location of the combined \Hb 4861
+ OIII 5007\AA\ lines, and the vertical purple line the location of the Mg 5175
\AA\ absorption feature. SPHEREx spectra averaged over $\sim$10$^6$ 
galaxies per color-mass bin will outline how galaxy mass assembly, gas
accumulation, metal build-up, and extinction build-up have taken place in the
last 5 billion years.} 
\label{fig:sfg}
\end{figure}

Fig.~\ref{fig:sfg} left shows the rest-frame (U--B) color vs. stellar mass for all galaxies
to AB\cle 25 mag from the HST ACS ``PEARS'' grism survey \cite{Malhotra:2005}. The galaxy stellar mass estimates and
spectroscopic redshifts (0.6\cle z\cle 1.2) come from 3D-HST \cite{Skelton:2014}, and were analyzed in a
5$\times$5 color-mass grid as described in Joshi \etal 2016 (in preparation). The large
``blue cloud'' and the much smaller ``red cloud'', as well as the ``green valley''
in between are visible, and each are sampled by several grid
boxes. Fig.~\ref{fig:sfg} right shows the clipped-average rest-frame spectra for each of the (U--B)$_{rest}$
color and stellar mass bins in Fig.~\ref{fig:sfg} left. Some of the low-mass star-forming
(blue-cloud) galaxies show a Balmer 3648 \AA\ break and a weak \Hb 4861 + O III
5007 line in their averaged grism spectra, while the more massive, early-type
(red-cloud) galaxies show a clear 4000\AA\ break and Mg 5175\AA\ in absorption (for
details, see  Joshi \etal 2016, in preparation). HST PEARS has done this spectral averaging for
$\sim$1000 galaxies to AB\cle 25 mag at 0.6\cle z\cle 1.2 in 40 arcmin$^2$. 
{\it SPHEREx will do this spectral averaging for \cge 10$^8$ galaxies across
the sky to AB\cle 19 mag to z\cle 0.5} (see Fig. 12 of \cite{Windhorst:2011}
and Fig.~\ref{fig:sfg} here). The spectral stacks from the SPHEREx 0.75--5.0 \mum\ bands
will cover restframe $\sim$0.45-3.0 \mum\ (see Fig.~\ref{fig:seds} and
\ref{fig:groups}). That is, SPHEREx will study the average low-resolution spectra of millions of galaxies per bin in a
very fine color-mass grid, tracing average spectral features like Mg 5175, Na
5895, the \Ha 6563 line (and for a higher-redshift subset also the \Hb 4861 + O
III 5007 line), the Calcium triplet (8500, 8544, 8665), and the near-IR break
at 2.33 \mum, all as a function of galaxy mass, color (\ie current SFR or SED
age), metallicity, and dust extinction ($A_V$). Monte Carlo tests with accurate
spectroscopic galaxy templates from SDSS, GAMA and WIGGLEz plus SED
models \cite{Bruzual:2003tq} will be used to help calibrate the
observed average absorption features in the SPHEREx data.  With very
careful attention to systematics (see the object confusion discussion in
Sec.~\ref{sec:cross-cal}), the resulting average SPHEREx spectra, averaged over $\sim$10$^6$ galaxies per
color-mass bin will outline how galaxy mass assembly, gas accumulation (from
emission lines, including those from SDSS, GAMA and WIGGLEz), metal build-up,
and extinction build-up have taken place in the last 5 billion years.

\subsection{The Galaxy Population at $z\sim 1$ with SPHEREx}
\label{sec:gal1.0}

Our understanding of the galaxy population at z$\sim$1 has grown
tremendously in the past decade. Photometric surveys have constrained
the overall mass growth, while spectroscopic surveys provided insights
into the physical processes governing galaxy growth. However, both
approaches have their limitations; while photometric surveys lack the
spectroscopic information needed to derive redshifts and physical
properties, spectroscopic surveys result in small and biased galaxy
samples. To compromise between these two techniques, slitless
low-resolution spectroscopic and photometric surveys at near-IR (e.g.,
rest-frame optical) wavelengths have become increasingly more
popular (see e.g., PRIMUS \cite{Cool:2013}, COSMOS \cite{Ilbert:2009},
NMBS \cite{vanDokkum:2009}, SHARDS \cite{Cava:2015}). SPHEREx will build upon
the success of such programs, and will provide a unique view to the galaxy population at $z\sim 1$.

While previous low-resolution surveys obtained deep spectra on
relatively small fields, the power of SPHEREx is in its large area. Studies of individual galaxies will be limited to the
brightest systems. Based on previous surveys, we estimate that in the
SPHEREx-DEEP survey, spectral features will be detected for
$\sim$10,000 individual galaxies at $0.7<z<1.45$ per patch. These
observations allow us to better constrain the massive end of the galaxy distribution at these redshifts.

However, much of the power of SPHEREx in studying $z\sim 1$ galaxies
comes from stacking techniques. Deep optical photometry from LSST and
other optical surveys will allow us to stack galaxies in bins of
stellar mass or color-color space and accordingly study their infrared
counterparts. The downside of such methods is that intrinsically
different spectral types will be combined, and thus information will
get lost. 

To retain the diversity of the galaxy population at $z\sim 1$ and
study galaxy properties during all possible evolutionary phases, more
clever matching techniques should be applied. For example, by matching
galaxies using their combined LSST-SPHEREx rest-frame spectral energy
distributions (SEDs), we can group them by their full spectral type
(see \cite{Kriek:2011}). For each spectral type, we can stack the
SEDs. This technique will enable the measurement of spectroscopic
features which can not be measured for individual galaxies, such as
rest-frame optical emission lines, metal absorption lines, or
molecular bands. 

This technique opens up a whole range of science applications. Most
importantly, it will give a census of the star formation rate and ages
for the complete galaxy population at z$\sim$1. Furthermore, the wide
wavelength coverage provides additional unique science cases. For
example, for post-starburst galaxies it will allow us to constrain the
thermally-pulsing asymptotic giant branch phase, and for star-forming
galaxies we can access the rest-frame near-infrared emission
lines. Finally, the width of the spectral features will provide a
direct constraint on the accuracy of the used spectroscopic redshifts
that are used. 

\subsection{Environmental Dependence of Galaxy Evolution}

SPHEREx will independently identify tens of thousands of mostly
low-redshift galaxy clusters spanning a large range of masses and
about 75$\%$ of the sky (not at low galactic latitudes). These identifications will facilitate the study
of galaxy properties such as stellar mass and galaxy spectral type
(also measurable by SPHEREx) in the densest environments.

SPHEREx will also be able to reconstruct the galaxy density field even into
low-density environments. High precision ($\sigma_z /(1+z) < 0.005$)
spectroscopic redshifts can be used to measure projected densities with
sufficient accuracy to rival spectroscopic surveys \cite{Cooper05}.
SPHEREx will measure redshifts at this precision or better
for over 9 million galaxies. This represents a doubling of the SDSS
spectroscopic sample over the SDSS footprint. Combining the two data
sets will allow for high quality measurements of the density field
\cite{Kovac10,Erwin11,Chen15}.
This will allow the study of the joint galaxy mass and star formation rate
functions as a function of environment for environments that span the
range from clusters to filaments to very low density regions. These
are key measurements for disentangling mass and environmental star
formation quenching, and for making precise measurements of galaxy
conformity, the relationship between the star formation occurring in
central galaxies within a dark matter halo and their satellite
galaxies \cite{Kauffmann13}.

In addition, the wavelength range covered by SPHEREx will include the
Paschen-$\alpha$ line at 1.8 micron. Given a 5$\sigma$ detection limit
of 18.5 AB mag, SPHEREx will measure Paschen-$\alpha$ lines
corresponding to star formation rates of about 10 M$_{sun}$/year at $z
= 0.1$. This will allow the identification of the most extreme line
emitters across the entire sky. Such galaxies are sometimes thought to be
forming their first generation of stars, and are useful analogs for
high redshift galaxies. 


\subsection{AGN Science}

According to the unified theory of active galactic nuclei (AGN), the 
central part of AGN is composed of an accretion disk
around a supermassive black hole (SMBH), a broad line region (BLR),
and a dusty torus (Fig.~\ref{fig:agn}).  The accretion disk, the BLR, and the torus are the
energy sources powering a UV/optical continuum (the big blue bump),
broad emission lines (e.g., hydrogen recombination  lines), and an IR
continuum, respectively. To investigate the detailed structure of the
central structure of AGNs it is crucial to better understand the
various aspects of AGNs. Thanks to its wide wavelength coverage,
SPHEREx will uniquely lead to a better understanding of the AGN
structure and also of the coevolution of BHs and its host
galaxies. SPHEREx will also provide a powerful data set for
optical identification and follow-up of AGNs selected from other surveys.  

\subsubsection{IR and Optical Reverberation Mapping of Bright AGNs}

\begin{figure}[!ht]
\center
\includegraphics[width=0.6\textwidth,angle=90]{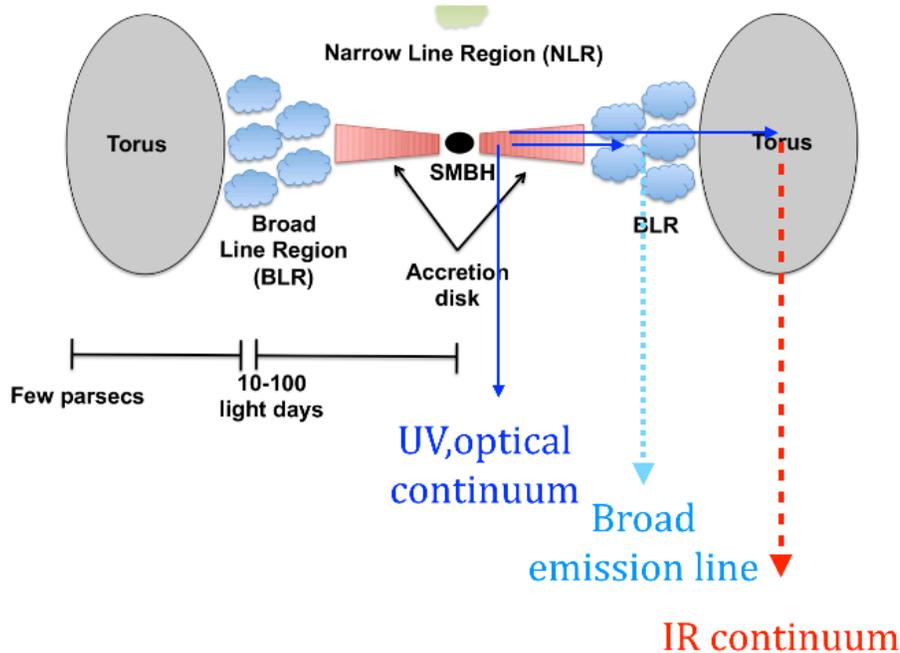}
\caption{Schematic diagram of the central structure of an AGN.  Using the time
lags among the light in different wavelengths, we will be able  to
estimate the physical size of each component (Image credit : \href{http://www.isdc.unige.ch/~ricci/Website/Active_Galactic_Nuclei.html}{Claudio
Ricci)}}
\label{fig:agn}
\end{figure}

Using time lags in variability between the UV/optical continuum, broad emission lines,
and NIR/MIR continuum, one is able to estimate the sizes of broad line
regions and tori in AGN, the technique called the ``reverberation mapping method''
(Fig.~\ref{fig:agn}). Reverberation mapping has been applied for a
limited number ($\sim 100$) of nearby (less luminous) type 1 AGNs,
since it requires a significant amount of telescope time to obtain  
multi-epoch data. In addition, investigating the time lag between the 
UV/optical continuum and NIR continuum is even more difficult to do with 
ground-based telescopes because of the high sky background and low
atmospheric transparency. Having access to multiple passes over the
deep survey regions in the mid-IR, allows us to detect AGN varying in
the rest-frame optical/UV at redshift higher than 6. SPHEREx will certainly
increase the statistics of the few objects currently having
reberberation mapping data between NIR and optical.

The spectral information provided by SPHEREx from optical 
to NIR, will give us not only the fluxes of the Balmer and Paschen  hydrogen
recombination lines but also the brightness of the IR continuum
simultaneously. Moreover, variation of the shape 
of  the IR continuum will give us crucial information about the clumpiness of the 
dusty torus. In SPHEREx' ecliptic polar regions, multi-epoch data, for 2 years will 
be available, which will enable us to calibrate the time lags in conjunction 
with complementary ground-based UV/optical data. The density of the 
type 1 AGNs from SDSS shows that there will be approximately 200 QSOs 
($i<$ 18 mag) in the polar regions, for which we can obtain high enough S/N 
data for the reverberation mapping of both broad emission lines and IR 
continuum.  We can increase S/N of the data by applying spectral binning so that, we 
can make use of fainter QSOs ($i\sim19$ mag) at least for the IR reverberation 
mapping. 

Additional multi-epoch UV/optical imaging data from ground-based 
telescopes is crucial for the time lag measurements. This is possible
using several small class (0.5-1.6m) telescopes mostly operated by the
Korean community but also the Las Cumbres facility. There is
possibility that the time lags can be larger than the operation time 
of SPHEREx (2 years) especially for the most luminous and distant QSOs. In order to 
account for this extreme cases, we plan to collect ground-based data at 
least 1-2 years prior to the launch of SPHEREx. The spectral resolution 
of SPHEREx is not high enough to measure the width of emission lines. 
Thus, we also plan to obtain the spectra of target AGNs using several mid-class
ground-based telescopes (e.g., MMT). 

\subsubsection{Systematic Study of Optically Selected Type 2 QSOs} 

AGNs are commonly classified in two types (type 1 and 2), based on the 
presence of UV/optical featureless continuum and broad emission lines. The
conventional model for AGN unification suggests that the type of 
AGNs is be simply determined by the line-of-sight relative to the dusty torus. 
However, recent observational studies show that the fraction of type 2 to type 1 
strongly depends on AGN luminosity, suggesting that the conventional AGN 
unification might be over-simplified and that the torus structure depends on the AGN 
brightness (e.g., receding torus model). However, these studies are mainly 
based on IR/X-ray selection methods because it is extremely hard to 
search for type 2 AGNs using optical data due to the faintness of the optical
counterpart. For example, SDSS discovered more than 400,000 type 1 
QSOs, but only a few thousand type 2 QSOs. Thanks to the infrared spectral information, 
SPHEREx will be an ideal tool for a unbiased search for type 2 QSOs. The 
whole sky survey data will enable us to discover at least $\sim3,000$ 
distant ($z>0.8$) type 2 QSOs, assuming a conservative 10$\%$ type 2 fraction at
high luminousity.
  
\subsubsection{Star Formation Activity in Bright AGNs}

Supermassive black holes are ubiquitous in the
center of massive galaxies, and that their mass is strongly 
correlated with the mass of the host galaxy
\citep[e.g.][]{2013ARAA..51..511K}. This indicates that  
SMBH and galaxies are closely linked in their evolution, although the detailed 
mechanism of the coevolution is still under debate. One way to address this 
issue is to investigate how BH growth and galaxy growth evolve with
cosmic time. As such, it is worthwhile 
measuring both the BH accretion rate and the star formation rate in
AGNs, where the BH  is actively growing in mass. This is challenging
because most SFR indicators are also very prominent in AGNs. There are a few 
exceptions, such as [O II], PAH emission, and FIR continuum. Based on the existing 
SDSS QSO catalogue, we expect to detect the flux of PAH $3.3\mu m$ emissions 
for $\sim 1000$ QSOs. In addition, the WISE AGN selection can
significantly increase the sample size by at least  by a factor of 10,
which will enable us to pursue the statistical study of the correlation
between SFR and BH growth rate in nearby AGNs.  

\subsubsection{Synergies with eROSITA}

eROSITA  (extended ROentgen Survey with an  Imaging Telescope Array)
will be the primary  instrument on the Russian
Spektrum-Roentgen-Gamma  (SRG) mission. 
eROSITA will provide an all-sky Xray survey, which will go $\sim$ 30 times deeper
than ROSAT \citep{1999AA...349..389V,2016AA...588A.103B} in the soft band (0.5-2 keV)  and
the first ever all-sky survey in the hard band (2-10 keV). It is
expected to detect $10^5$ clusters of galaxies up to redshift  $\sim$
1, about 3 million AGN to z$\sim$6, and about 500,000 stars.

The design of eROSITA is driven by effiiency in area coverage, which
comes at the cost of resolution, with a half energy width (HEW) of
16'' for pointed observations and about 26'' in survey mode. Hence
complete and deep multi-wavelength coverage of the sky is essential to
identify the correct counterparts to the point-like X-ray sources like
AGN and stars detected with eROSITA. SPHEREx will be  crucial for the following reasons. 

\begin{itemize}
\item 
SPHEREx will cover the entire sky, and provide full-sky object catalogs with higher 
angular resolution than eROSITA and homogeneous selection.
Such homogeneity is fundamental for selecting the counterpart to the X-ray sources
in a probabilistic way, using for example Maximum Likelihood
Ratio \citep{2007ApJS..172..353B} or Bayesian approaches \cite[e.g.][]{2008ApJ...679..301B},
in which the likelihood of an optically selected object being the correct counterpart 
to an eROSITA detection is tested against the likelihood of it being
an unassociated source.

\item SPHEREx will cover the longer optical wavelengths, NIR and
MIR. At these wavelengths, AGNs will be easily identified: either due to their intrinsic
SED \citep[e.g.][]{2012ApJ...753...30S}, or due to the redshifting of 
the strong UV emission of high-redshift AGNs into the SPHEREx bands.
Additionally, the number of field sources decreases towards at longer wavelengths, 
which improves the success rate of counterpart identification methods. 

\item eROSITA and SPHEREx will both be deeper at the ecliptic poles. Here
SPHEREx will reach a depth of $\sim$21 mag AB which will allow the
identification of the faintest eROSITA sources. 
Currently, there are no suitable multi-wavelength data sets at the
poles. In particular, at the SEP there are still a few thousand square degrees of areas
where only WISE/GALEX is available for AGN identification; these are
reliable but are too shallow for eROSITA.

\item SPHEREx will continuously revisit the sky with a six month cadence (at
a given position and given wavelength element; and 30
times more frequently at the ecliptic poles). This multi-epoch data set will contribute to
understanding the physics regulating the activity of black hole, and provides
valuable additional information on X-ray sources with ambiguous optical associations: 
between 2 sources with the same magnitude/colors,
the one varying in time is more likely the  AGN.

\item SPHEREx will allow reliable spectroscopic redshift determination for X-ray sources.
It is well established that intermediate and narrow band surveys
are ideal for photometric redshifts \citep{2009ApJ...692L...5B} and
many surveys are build with this in mind at optical wavelengths
(COMBO-17 \cite{Wolf:2004}, COSMOS-21 \cite{Taniguchi:2005},  SHARDS 
\cite{Cava:2015}). This method relies on the capability for unambiguous identification 
emission and absorption lines. In particular for AGN, the emission lines 
from the host and from the vicinity of the central BH, combined with
proper SEDs, will allow the determination of accurate spectroscopic redshifts
\citep{2009ApJ...690.1250S,2010ApJS..189..270C,2011ApJ...742...61S,2012ApJS..198....1F,2014ApJ...796...60H}. It will bring much improvement on, for
example, the X-ray AGN Luminosity Function, which is still very uncertain at z$>$3
\citep{2016ApJ...817...34M,2011ApJ...741...91C,2015MNRAS.453.1946G}.  
\end{itemize}

\begin{figure}[!t]
\center
\includegraphics[width=1.00\textwidth,angle=-0]{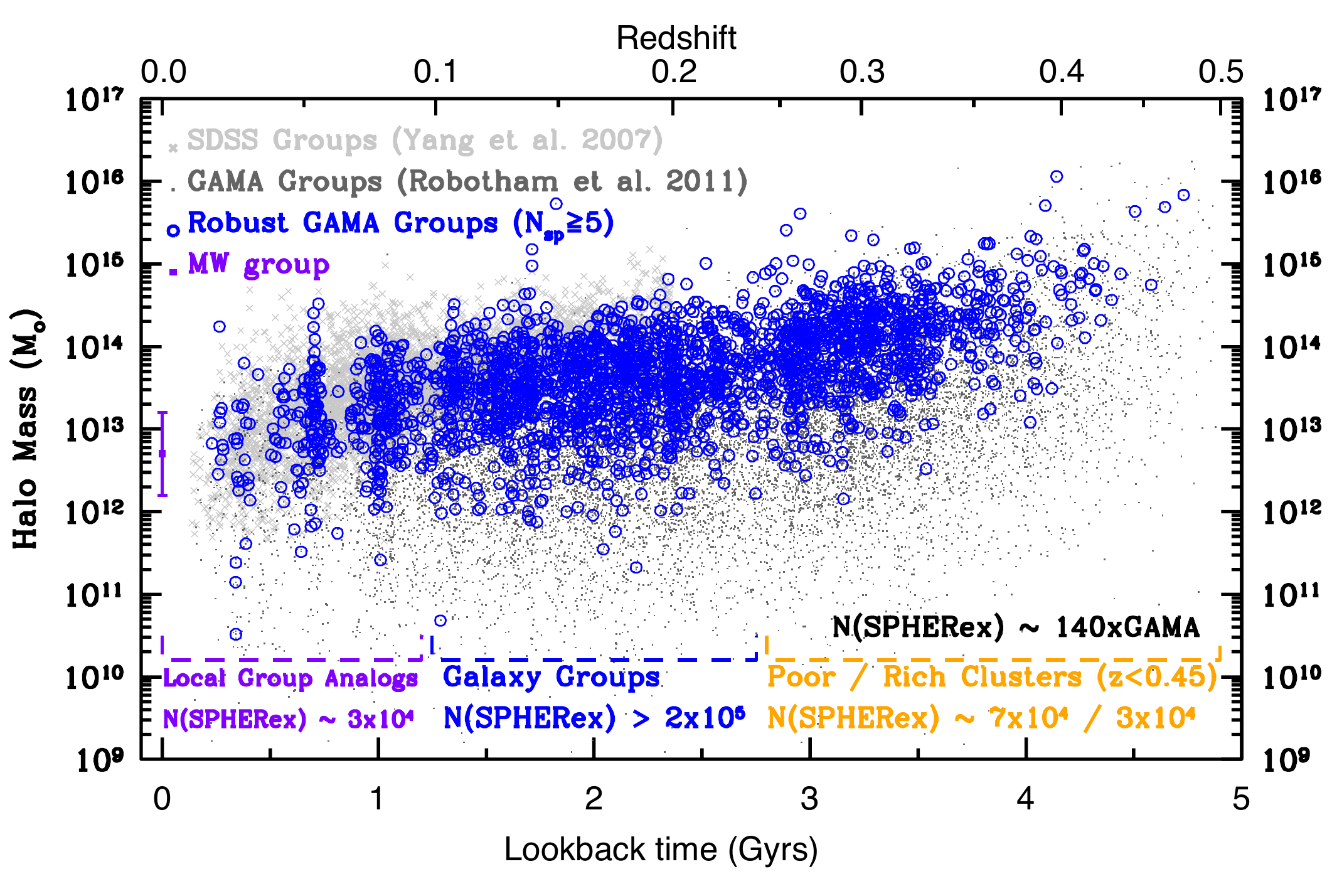}
\caption{Comparison of spectroscopically confirmed galaxy groups at z\cle
0.2 from SDSS \cite{Yang:2007} (light grey crosses) to galaxy groups at z\cle
0.45 from the GAMA survey \cite{Robotham:2011}; black dots and blue open
circles). The purple data point is our Local Group. SPHEREx will cover more
than 100$\times$ the GAMA area to similar depths at 0.7--5.0 \mum. Hence,
SPHEREx will be able to estimate the stellar masses of about
3$\times$10$^4$ Local Group Analogs (LGAs) to z $\leq$ 0.1, about 2$\times$10$^5$
galaxy groups, $\sim$7$\times$10$^4$ poor clusters and 3$\times$10$^4$ rich
clusters out to z$\leq$ 0.5.}
\label{fig:groups}
\end{figure}

\subsection{Galaxy Groups with SPHEREx at z\cle 0.5}
\label{sec:groups}

Fig.~\ref{fig:groups} shows a comparison of spectroscopically confirmed galaxy groups at 
z\cle 0.2 from SDSS \cite{Yang:2007} to galaxy groups at z\cle 0.45
from the Galaxy And Mass Assembly (GAMA) survey
\cite{Robotham:2011}. Compared to Sloan, GAMA finds galaxy groups 
much further out, sampling a depth and distance that is more
comparable to the SPHEREx all-sky depth of AB\cle 19 mag. GAMA covered $\sim$300 \degsq, and
identified more than 24,000 groups from a total of 360,000 GAMA galaxy 
redshifts \cite{Driver:2016}, including 2400 reliable groups with at least
N=5 reliable spectroscopic redshifts to AB\cle 19.8 mag, resulting in unbiased
{\it dynamical} group and cluster mass-estimates \cite{Robotham:2011}.
WIGGLEz has provided about 100,000 redshifts to z\cle 1
\cite{Drinkwater:2010}. SPHEREx will cover more than 100$\times$ the
GAMA area to similar depths at 0.75--5.0 \mum. Hence, SPHEREx will be
able to estimate the {\it stellar} masses of about 3$\times$10$^4$
Local Group Analogs (LGAs) to z\cle 0.1, and about 2$\times$10$^5$ galaxy groups, $\sim$7$\times$10$^4$ poor clusters
and 3$\times$10$^4$ rich clusters out to z\cle 0.5. For all 5080 SDSS groups at
z\cle 0.2 and all 2400 GAMA groups to z\cle 0.45 --- each with accurate
dynamical masses from high resolution spectroscopy --- SPHEREx will estimate
reliable total stellar luminosities and total stellar masses, constraining the
mass-to-light (M/L) ratios and dark matter properties for galaxy groups and
clusters as a function of their environment. That is, SPHEREx will
complete the stellar mass census for the best spectroscopically studied regions
in the sky where the dark-matter halo mass is known dynamically (\eg\ from 
SDSS, GAMA and WIGGLEz).

SPHEREx will measure groups like our own 
Milky Way + M31 group out to the Coma distance (z$\simeq$0.023), detecting its
LMC-like dwarf galaxies (\ie\ \MAB$\simeq$--15 mag) about $\sim$6 mag below
$M^\star$. Fig.~\ref{fig:groups} shows that in total SPHEREx will detect about
3$\times$10$^4$ slightly more 
massive Local Group Analogs (LGA) to twice the Coma distance (z\cle 0.05), where
dwarf galaxies surrounding Milky Way type \Lstar galaxies can be detected to
\MAB$\simeq$--16.5 mag. The spectroscopic redshift of individual
companion dwarf galaxies to AB\cle 19 mag will still be accurate enough to
statistically constrain the number of dwarf galaxies around each
\Lstar galaxy in the  universe at z\cle 0.05, about 4--6 mag below
\Mstar for 3$\times$10$^4$ nearly Local Group Analogs. This will
immediately constrain $\Lambda$CDM models, where the number of dwarf
galaxies is predicted to be an order of magnitude higher than
observed. Hence, SPHEREx has the potential to directly address a key
remaining uncertainty that the $\Lambda$CDM model has left unresolved in the local universe.  

\subsection{SPHEREx Galaxy Cluster Selection and Redshift Measurements
  for Clusters Selected in the X-ray, Sunyaev-Zel'dovich Effect and
  Optical} 
\label{sec:cluster}

SPHEREx will see numerous clusters of galaxies in its all-sky
survey. We first look at the clusters SPHEREx will detect on its
own. We calculate the number of clusters SPHEREx will independently detect 
over the all-sky survey by extrapolating from two cluster  catalogs  built on  SDSS DR8: 
The RedMapper catalog from \cite{Rykoff:2013ovv}, and the AMF catalog
from Banerjee et al (in prep).  These catalogs, restricted to the
10400 square  degrees DR8 area,  contain  26000  galaxy   clusters
(redMaPPer) and about 43,000 (AMF). 

We apply the SPHEREx flux limit of $z_{AB}\lesssim18.5$ mag to the
potential galaxy clusters. For  the reMaPPer catalog,  we identify
9800  clusters with 5 or more cluster members brighter than this flux
limit and with probability, $p_{member}$, 
of belonging to the cluster greater than 70\%.  Extrapolating this to
the 75\%\ of the  sky  gives an estimate of over 29,000 galaxy clusters
that will be independently detected with SPHEREx and have
high-precision redshift measurements. The median redshift of these
clusters is $z_{med}\simeq0.2$.

The AMF catalog, when restricting cluster membership  to the 10
galaxies that mostly contribute to the overall likelihood, finds 18,700
clusters. Out of the clusters that are found in AMF and not in
RedMaPPer, about half are also found in the WHL \citep{Wen:2012tm}
catalog. The AMF finding would  imply that  SPHEREx  will detect over
75\%of the sky  about 56,000 galaxy clusters and groups. Therefore we overall expect 
SPHEREx  to detect at least 30,000 galaxy clusters and groups, and perhaps substantially more.

But while SPHEREx can be used to select tens of thousands of clusters,
another exciting use of its data will be to 
estimate and/or improve redshift measurements for clusters discovered
in other surveys.  In particular, upcoming millimeter-wave cluster
surveys (such as SPT-3G and AdvACTpol) and the eROSITA X-ray mission
will discover 10,000-100,000 systems \cite{benson14,
  henderson15,Merloni12} and, as the intracluster medium-based
observables in these surveys contain limited-to-no redshift
information, auxiliary data is required to obtain robust redshifts.  

The sheer number of clusters necessitates that the bulk of these
redshifts be obtained from other wide area surveys (rather than
pointed observations), and data from current and next generation
optical imaging surveys such as the Dark Energy Survey, Pan-STARRS, and LSST
will naturally fill this role \cite{desref,kaiser:02,SciBook}.
In such surveys, clusters are identified as significant over-densities
in photometric redshift space \cite[e.g.,][]{Wen:2012tm} or as
concentrations of red-sequence galaxies \cite[e.g.,][]{gladders05,
  koester07a, bleem15, rykoff16}, and modern optical cluster catalogs
  have excellent redshift performance  ($\sigma_z/(1+z) \sim$ 0.01
  -0.02).  

\begin{figure}[!th]
\center
\includegraphics[width=0.7\textwidth]{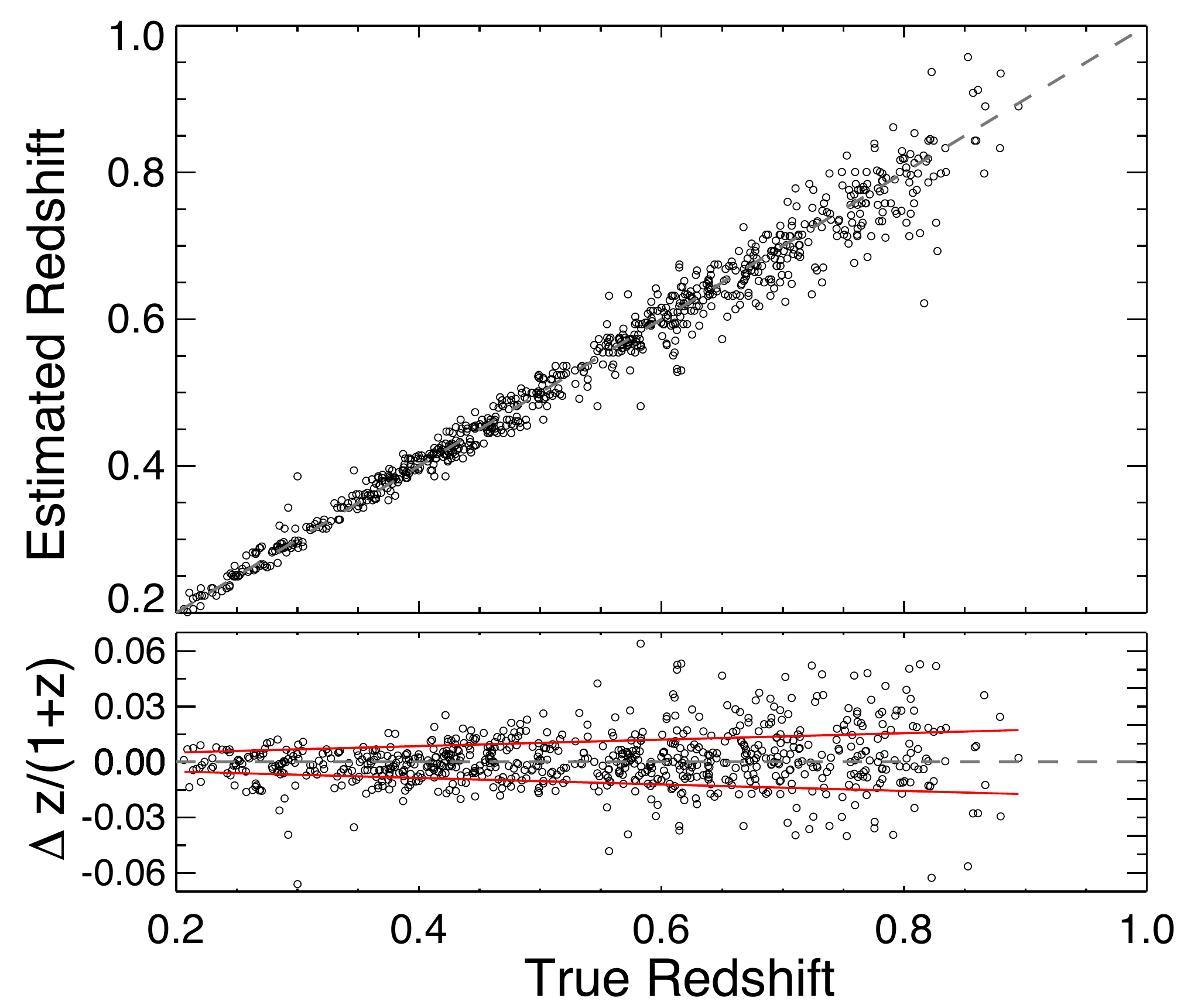}
\caption{Simulation of cluster redshift recovery using SPHEREx data. 
\emph{Top:} Recovered versus input cluster redshift for simulated SPHEREx
observations of $\sim$800 clusters selected with the redMaPPer cluster
identification algorithm in Dark Energy Survey data ($M_{500c} \ge
10^{14} M_{\odot}$, see \cite{rykoff16}). \emph{Bottom:} Similar to above,
now plotting the difference between input and recovered redshifts
scaled by 1/(1+$z$). Redshifts were estimated using SPHEREx spectra of
galaxies identified in DES data as having high probability of cluster
membership (interlopers can be present). The precision of the
simulated SPHEREx redshift estimates exceeds that of current
generation optical surveys at $z\lesssim0.6$ and remains excellent
($\sigma_z/(1+z) < 0.03$) to $z\sim0.9$, the highest redshift of
clusters in this sample.}\label{fig:cluster}
\end{figure}

In Fig.~\ref{fig:cluster} we demonstrate how SPHEREx can further
improve cluster redshift estimates in the presence of such imaging
data. This simulated SPHEREx cluster data was created using
photometric source, photo-$z$, and the redMaPPer cluster catalogs from
the recent Science Verification release of the Dark Energy survey
\cite{rykoff16,bonnett15} and full details will be provided in Bleem
(in prep).  These simulations suggest that the precision of cluster
redshifts from SPHEREx should equal or exceed that of current
generation optical surveys to $z\lesssim 0.6$ and remain
$\sigma_z/(1+z) < 0.03$ to $z\sim0.9$. As the fits shown here were
conducted only using data  $1 \mu$m $< \lambda < 4.1 \mu$m (though
restricted to galaxies whose optical colors indicate high probability
of cluster membership), these constraints can be further combined with
the optical measurements to improve the precision. But these results
indicate that by itself, SPHEREx could become a unique redshift
machine for surveys such as SPT-3G, AdvACTpol and
eROSITA. Perhaps more importantly for future cosmological constraints, 
however, SPHEREx data provides a useful cross-check of the redshifts---especially for
clusters at redshifts near where the 4000\AA\ break transitions between
optical filters---as the smooth shifting of the 1.6 $\mu$m feature 
through the SPHEREx bands should not suffer similar
discontinuities. For future surveys, ensuring low bias in redshifts
(rather than improving the already excellent precision) will be
important \cite{huterer04, lima07}.   

Furthermore, using the cluster redshifts obtained by SPHEREx in
combination with the next generation of CMB 
experiments allows for the tomographic reconstruction of the tSZ
signal from these clusters \cite{Shao2009}. A 3-dimensional tSZ map would
enable us to study the evolution of the thermal properties of the
intracluster medium (ICM), hence furthering our understanding of the
thermodynamic processes involved in galaxy formation. Additionally, a
3D map such as this can be used to constrain cosmological parameters
directly \cite{HBSBPS2013}, or in combinations with other statistical
measurements of the tSZ like the power spectrum e.g.,
\cite{Sievers2013,George2015,PlancktSZPS2015}, bispectrum or skewness
\cite{Wilson2012,Crawford2014}, and the temperature histogram
\cite{Hill2014} all which are limited by the astrophysical
uncertainties in the ICM modeling \cite{BBPSS2010,HP2013,McCarthy2014}.  

\subsection{Synergy with future CMB experiments}

\subsubsection{kSZ signal for next CMB experiments} 

The amount of ionized gas observed in galaxies and clusters falls
short of the cosmological abundance \cite{2004ApJ...616..643F},
especially for group-sized halos or smaller.  The resulting
uncertainty in the baryon profile of $M^*$ halos will be a limiting
systematic effect for future weak lensing experiments, such as Euclid,
LSST and WFIRST \cite{2015ApJ...806..186O, 2015MNRAS.454.2451E}. A
large fraction of the gas is thought to reside in the outskirts of the
halo, in a form that is too diffuse and too cold to be effectively
imaged with X-ray or thermal Sunyaev-Zel'dovich (tSZ)
observations. SPHEREx in conjunction with the next generation of CMB
experiments will greatly increase our ability to measure this gas
component via the kinematic Sunyaev-Zel'dovich effect.

The kinematic Sunyaev-Zel'dovich effect, which is the Doppler boosting
of CMB photons scattering off electrons with a non-zero peculiar
velocity, is a unique probe of low density and low temperature
regions, its amplitude being directly proportional to the electron
number density, and independent of the electron temperature. It is
thus an unbiased probe of the total electron abundance associated with
the halo, as well as of the gas profile.  

In temperature units, the shift $\Delta T^{\rm kSZ}(\bm{\hat{n}})$
produced by the kSZ effect is sourced by the free electron
\textit{momentum field}  $n_e \bm{v}_e$, and is given by \cite{sun72,
  1986ApJ...306L..51O} 
\be
\frac{\Delta T^{\rm kSZ}(\bm{\hat{n}})}{T_{\rm CMB}}  = -  \sigma_T
\int \frac{d \chi}{1+z} e^{-\tau(\chi)} n_e(\chi\hat{\bm{n}},\chi) \
\frac{\bm{v}_e}{c} \cdot \bm{\hat{n}}, 
\label{eq:kSZdef}
\ee
where $\sigma_T$ is the Thomson scattering cross section, $\chi(z)$ is
the comoving distance to redshift $z$, $\tau$ is the optical depth to
Thomson scattering, $n_e$ and $\bm{v}_e$ are the \textit{free}
electron physical number density and peculiar velocity, and
$\bm{\hat{n}}$ is the line-of-sight direction. 

Since the kSZ signal at a galaxy location is proportional to the
galaxy peculiar velocity, which is equally likely to be positive or
negative, a simple cross correlation (or stacking) between the tracer
position and a CMB temperature map will vanish.  To remedy this, a
number of estimators have been proposed in the literature
\cite{2009arXiv0903.2845H, 2011MNRAS.413..628S, 2014MNRAS.443.2311L,
  1999ApJ...515L...1F, 2008PhRvD..77h3004B}, typically requiring
either spectroscopic redshifts or very accurate photometric redshifts
(say $\sigma(z) \lesssim 0.01$). 

The signal to noise ratio $S/N$ of the kSZ signal is very sensitive to both the actual
electron profile, which is uncertain for low mass galaxy groups, and
to the noise in the CMB temperature maps. For current and upcoming
ground-based CMB experiments, the atmospheric noise gives an important
contribution on top of the usual white noise. Therefore, the most
realistic and conservative forecast for SPHEREx is obtained by
rescaling existing measurements. We use the recent ACTPol measurement
on CMASS galaxies \cite{2015arXiv151006442S}  and use the following
scaling: 
\begin{equation}
S/N \propto \sqrt{\rm{Volume}} \ \sqrt{\frac{b^2 P(k_*)}{b^2 P(k_*) + 1/\bar{n}}} \propto \sqrt{N b^2 P(k_*)}
\label{eq:rescale}
\end{equation}
Here $\bar{n}$ is the mean number density, $N$ is the total number of
objects and $b$ is the mean bias of the sample. Here $k_* \sim
\ell_{\rm kSZ} / \chi(\bar{z}) \sim 3.5\ h/$Mpc, where we have assumed
that most of the kSZ signal comes from $\ell_{\rm kSZ} \sim 3000$ or
larger. Note that the second equality in Eq. \ref{eq:rescale} holds in
the shot-noise dominated regime, which is the case for both CMASS and
SPHEREx galaxies at the scales of interest\footnote{Since SPHEREx is
  in the shot-noise dominated regime on this scale, the result is
  independent of the value of $k_*$}. 

With the velocity reconstruction method, we require very good redshift
determination, and therefore we include only the two highest redshift
accuracy samples from \cite{2014arXiv1412.4872D}. For this population,
we find $b \approx 1.1$ and $\bar{z} \approx 0.3$, with a total number
of galaxies on half of the sky $N \approx 24.5$ million. With the
simple rescaling described above, we forecast $S/N \sim 55$ on half of
the sky, assuming a CMB map noise of 14 $\mu$K-arcmin. Note that the
$S/N$ will also depend on the particular mass and redshift
distribution of galaxies, which is not considered here. Future CMB
experiments are expected to have smaller map noise and should allow an
even higher $S/N$ detection, even though the amplitude of the noise
from fluctuations in the atmosphere is still uncertain. For other
recent detections using ACT, Planck and SPT data, see
\cite{Handetal2012, 2015arXiv150403339P, 2016arXiv160303904S}. 

Because of the stringent requirement on redshift errors, in our
forecast so far we have only used a small fraction of objects in the
SPHEREx catalog. A different technique can be used in absence of
redshift information \cite{2004ApJ...606...46D}. This method only
requires a statistical redshift distribution and will allow the use of the
full SPHEREx catalog. It has been recently used to detect the kSZ
signal from WISE selected galaxies in combination with Planck CMB data
\cite{2016arXiv160301608H}. The basic idea consists of cross
correlating tracers with the square of an appropriately filtered CMB
map. The squaring operation removes the information about the
direction of the peculiar velocity and therefore allows a detection in
cross-correlation. Recent work \cite{Ferraro:2016} suggests that a very large
signal to noise ($S/N > 100$) can be achieved by combining SPHEREx
with future CMB experiments.  This method requires a very good frequency cleaning of the foregrounds
in the CMB temperature map, so the limiting factor will likely be the
ability to do component separation in CMB maps on small scales.  

As we have seen, the combination of current and future CMB experiments
with large scale structure surveys like SPHEREx will yield very high
$S/N$ detections of the kSZ effect. This in turn will allow precision
measurement of the baryon abundance (in the ionized state), of the
baryon profile out to several virial radii and of the scale dependence
of the velocity correlation function, potentially a powerful probe of
scale-dependent modified gravity. 

The ionization fraction and baryon profile is expected to depend both
on mass and redshift of the host halo and could depend on other galaxy
properties such as star formation rate, color, presence of an active
AGN etc. The unique spectral coverage of SPHEREx over the full sky
allows for a better characterization of several galaxy properties
compared to a photometric survey, and will allow us to select and compare
different populations, shedding light on the effect of feedback and
star formation on the gas. When combined with tSZ measurements, the
temperature of the IGM as well as the amount of energy injection will
be constrained. 

\subsubsection{Galaxy -- CMB lensing cross-correlations} 

In this subsection we consider the cross-correlation of positions of
SPHEREx galaxies, with CMB lensing convergence maps, which probes the
galaxy--matter cross-power spectrum at the redshift of the galaxy
sample.

\begin{figure}[!th]
\begin{center}
\includegraphics[width=0.45\textwidth]{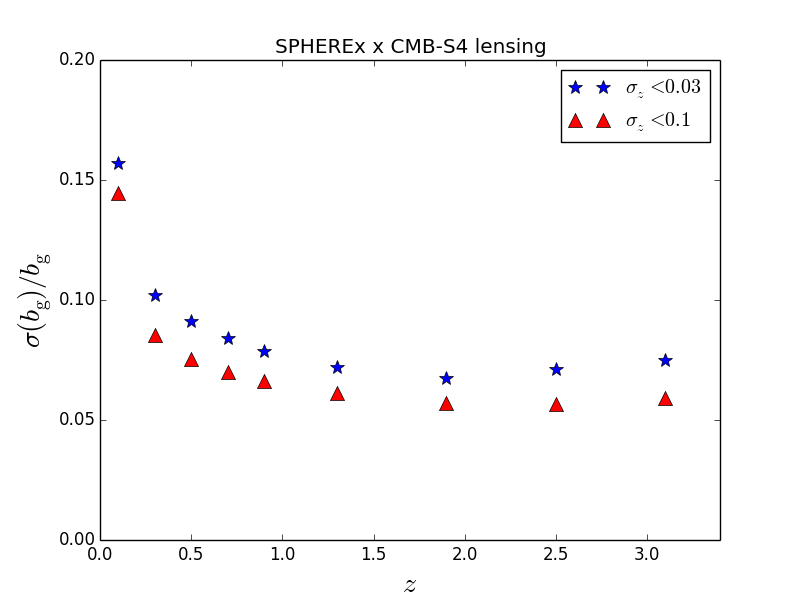}
\includegraphics[width=0.45\textwidth]{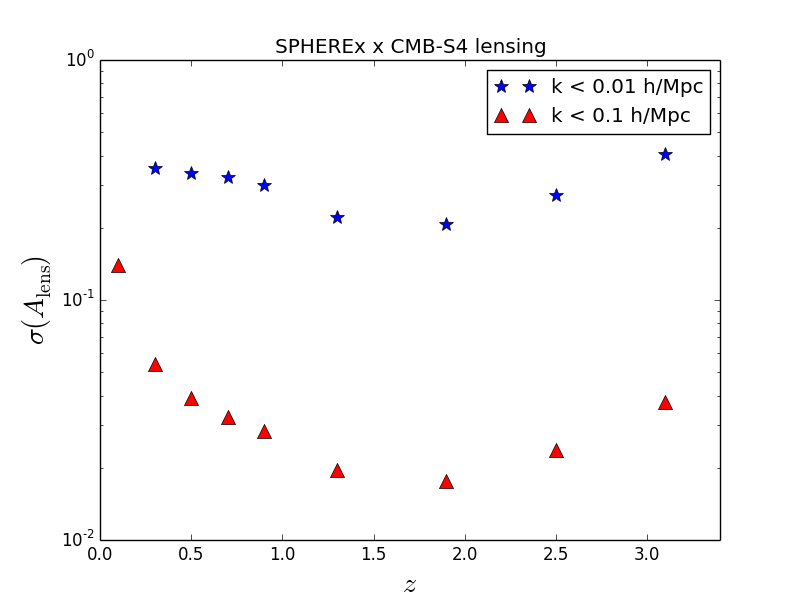}
\end{center}
\caption{\emph{Left:} Fractional constraints on galaxy bias in the
  nominal SPHEREx cosmology analysis redshift bins, derived by
  cross-correlating the SPHEREx galaxy catalog with the CMB lensing
  alone. The blue stars show forecasts for the   $\sigma_z <
  0.03(1+z)$ galaxy sample, which is representative of   galaxies in
  the bispectrum analysis, and the red triangles show   forecasts for
  the $\sigma_z < 0.1(1+z)$ galaxy sample, which is   representative
  of galaxies in the power spectrum analysis   \emph{Right:}
  Constraints on the $A_\mathrm{lens}$ parameter as a   function of
  redshift and wavenumber. The constraints shown with blue   stars
  include only large scales, approximately $k < k_\mathrm{eq}$,
  while the red triangles also include modes down to the quasi-linear regime. }  
\label{fig:Alens}
\end{figure}

The amplitude of the galaxy--CMB lensing signal is proportional to
galaxy bias, and assuming a cosmological model, it provides an
independent validation of the bias parameters derived in the
SPHEREx-internal cosmology analysis. The left panel of
Fig.~\ref{fig:Alens} shows forecasts of the fractional constraints on
galaxy bias of the SPHEREx cosmology analysis galaxy samples from
galaxy--CMB lensing alone, assuming $f_{\mathrm{sky}} =0.5$ overlap
between SPHEREx and a CMB-S4-like survey. These constraints are
marginalized over $w$CMD cosmology parameters, and the redshift
uncertainty of the SPHEREx galaxies. 

Alternatively, if galaxy bias is known (e.g., determined from the
galaxy power spectrum analysis), the galaxy--CMB lensing signal can
used to constrain an amplitude parameter $A_\mathrm{lens}$ \citep{2015arXiv150201591P}. In general
relativity, we expect $A_\mathrm{lens} = 1$, and a measurement
deviating from this value would imply modifications of the relation
between the metric potentials $\Phi$ and $\Psi$, termed gravitational
slip \citep{Caldwell:2007cw,Bertschinger:2006aw,PhysRevD.64.083004,PhysRevD.65.023003}(or unaccounted systematic errors, such as a mis-calibration of
the galaxy redshift distribution). The right panel of
Fig.~\ref{fig:Alens} forecasts the uncertainty of $A_\mathrm{lens}$
measurements as a function of redshift and physical scale, assuming that
galaxy bias is determined in the galaxy clustering analysis, and
marginalizing over the residual uncertainty in galaxy bias and $w$CDM
parameters. On these very large scales, the clustering measurement of
SPHEREx galaxies is cosmic variance limited and the constraints on $A_{\mathrm{lens}}$
approach its (half-sky) cosmic variance limit.  

\subsection{Constraining Structure Growth with SPHEREx}

The growth rate of structure is an important window on the behavior of
gravity on large scales. A wide variety of alternative theories of
gravity (e.g. f(R), symmetron, and massive gravity) have been
developed over the last decade or so, predominantly motivated by the
need to explain cosmic acceleration/dark energy \cite{2012PhR...513....1C}. In many cases, these
theories can be tuned to give a cosmic expansion history that is
almost identical to that of a cosmological constant-dominated universe
in General Relativity. The growth rate, on the other hand, is often
modified away from its GR behavior in an unambiguous manner, making it
a highly valuable diagnostic for exotic gravitational physics \cite{2008PhRvD..78b4015B}. 

The premier probe of cosmic growth is the redshift-space distortion
(RSD) technique. This uses the fact that the apparent clustering
pattern of galaxies is distorted due to their infall onto matter
over-densities, by an amount that is proportional to the linear growth
rate, $f(z)$ \cite{2011RSPTA.369.5058P}. The RSD technique is typically applied only to
spectroscopic galaxy surveys, as a loss of information in the radial
direction (e.g. due to photometric errors) can easily wash out the
signal. The two SPHEREx samples with the highest photometric precision
will have sufficient redshift accuracy to make RSD measurements
practical, however. 

\begin{figure}[!t]
\centering
\includegraphics[width=0.78\textwidth]{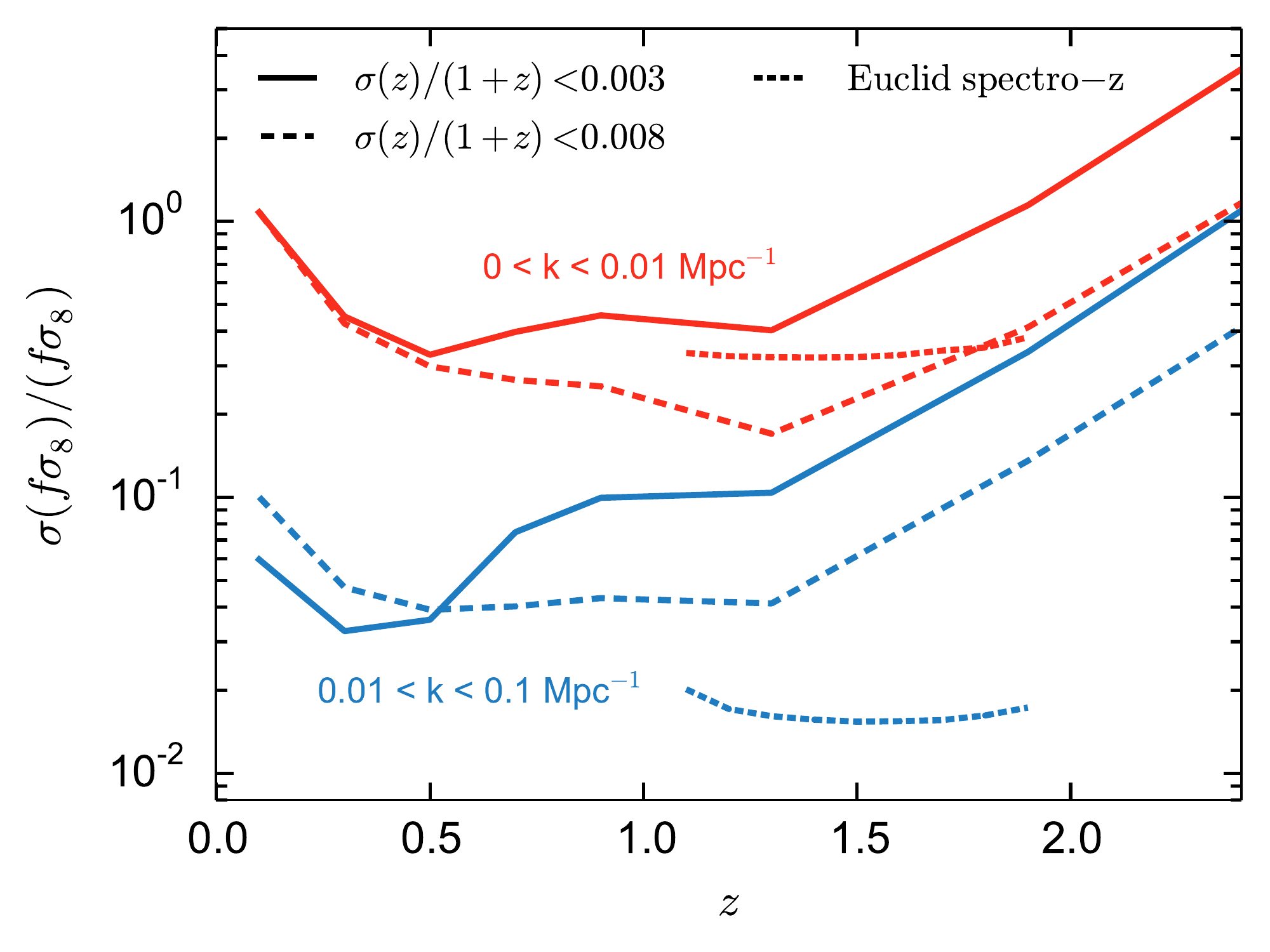}
\caption{Fractional errors on the growth rate as a function of
  redshift, binned by wavenumber. The first bin (upper curves) are
  large scales, approximately $k < k_{\rm eq}$. The second bin (lower
  curves) are mostly linear scales, down to the quasi-linear
  regime. The solid and long-dash curves correspond to SPHEREx while
  the short-dash curves correspond to Euclid. The combination of
  SPHEREx and Euclid allows a complete measurement of the error growth rate
over the redshift range when dark energy dominates.} 
\label{fig:fs8sd}
\end{figure}

Fig.~\ref{fig:fs8sd} presents Fisher forecasts for the fractional
errors on the growth rate (from SPHEREx galaxy clustering), using a
flat sky Fisher code, with no wide-angle or relativistic corrections,
and with evolution neglected within each redshift bin. Photo-z errors
and non-linear damping are included. Various nuisance parameters
(including the bias in each redshift bin) have been
marginalized. Similar predictions for a spectroscopic H-$\alpha$
galaxy survey with the Euclid satellite are included for
comparison. On intermediate scales (blue lines), SPHEREx is expected
to be strongly complementary to Euclid, providing $\sim$few percent
constraints in the $z < 1$ range where the growth rate is changing
most rapidly. 

The full-sky coverage of SPHEREx, coupled with its high source
density, will also allow ``ultra-large'' scales (i.e. approaching the
cosmological horizon size) to be probed down to the cosmic variance
limit for $z \lesssim 1$. Novel relativistic phenomena arise in this
regime, comparable in size to the scale-dependent bias effect caused
by primordial non-Gaussianity
\cite{2009PhRvD..80h3514Y,2014CQGra..31w4005V,2015ApJ...814..145A,2016JCAP...05..009R}. Many
alternative theories of gravity deviate from GR in a scale-dependent
manner, so it is possible that these effects could be drastically
modified in ways that are unobservable on smaller scales
\cite{2013PhRvD..87j4019L,2015ApJ...811..116B,2016arXiv160403487R}. Though
precision is intrinsically limited here due to cosmic variance, SPHEREx will nevertheless offer
the best constraints on growth on ultra-large scales of any individual
survey. 

\subsection{Synergy with Future 21 cm Surveys}

New low frequency interferometers such as the MWA \cite{tingay13},
PAPER \cite{ali2015}, LOFAR \cite{lofar}, and GMRT
\cite{Paciga2011} are nearing sufficient sensitivity to see
redshifted 21\,cm emission from the Epoch of Reionization, and the
next-generation HERA \cite{PoberNextGen} (now under construction)
will deliver $\text{SNR}>10$ constraints on reionization parameters
\citep{neben16,ewallwice16} in the coming years. However, cross
correlations with other probes will be needed to confirm any purported
detection and establish a consistent picture of the astrophysics of
the EOR. Of particular interest is the cross correlation between
redshifted 21cm emission from the neutral IGM (between ionized
bubbles) and redshifted Ly-$\alpha$ emission from the galaxies (inside
the ionized bubbles) that SPHEREx will map.  

Simulations and theory predict a negative correlation on scales of
order the typical bubble size at any given redshift
\citep{silva,fernandez14,mao14}. \cite{silva} self-consistently
simulate 21\,cm IGM emission and galactic Ly-$\alpha$ emission,
predicting that the anti-correlation peaks at 5\,Mpc (2 arcmin) scales
at $z=10$, and at 50\,Mpc (20 arcmin) at $z=7$. \cite{fernandez14}
show that this anti-correlation evolves in strength as reionization
progresses, peaking near the midpoint of reionization. On much smaller
scales, of order the galaxy size, the matter field overdensity that
gave rise to the galaxy in the first place is dense enough to stave
off reionization, thus giving rise to a local peak in 21cm emission. This
latter effect should yield a positive correlation on angular scales
smaller than 15 arcsec.  

Beyond shedding light on the astrophysics of reionization, the
21\,cm--infrared correlation is likely the only way to probe the EOR
component of the near infrared background (NIRB). While early studies
proposed that the angular fluctuations on few arcmin scales were
sourced by these first sources \cite[e.g.,][]{kash1,kash2,kash3},
recent observations and modeling propose alternative explanations
\cite[e.g.,][]{kash4} such as intermediate redshift sources with
significant intrahalo light \cite{cooray12, zemcov14}. The EOR
component is predicted to be a factor of roughly 50 lower in surface
brightness, and can only be recovered through cross correlation with
other EOR probes, such as 21\,cm maps.  

The main obstacle in measuring the 21\,cm--infrared correlation is
foregrounds. Even if low radio frequency foregrounds were perfectly
uncorrelated with near infrared foregrounds, their presence in a cross
correlation estimate contributes a sort of sample variance noise. This
noise can only be mitigated through better foreground
subtraction/masking, or using a larger sky area. A rough estimate of
the SNR of a cross correlation measurement, assuming the noise is
dominated by residual uncorrelated foregrounds, is given by  
\begin{equation}
\text{SNR}\sim3 \left(\frac{P_{\text{IR},\text{cosmo}}/P_{\text{IR},\text{FG}}}{10^{-3.5}}\right)^{1/2}
\left(\frac{P_{\text{21},\text{cosmo}}/P_{\text{21},\text{FG}}}{10^{-2}}\right)^{1/2}\left(\frac{\theta_\text{FOV}}{70^\circ}\right)
\left(\frac{3'}{\theta_\text{PSF}}\right)
\end{equation}
These numbers assume 50\% of the infrared pixels are masked due to
point sources \cite{zemcov14}, and 99.9\% of radio foregrounds are
subtracted ($(10\text{mK})^2/(0.1\%\times100\text{K})^2$). The image
resolution is limited by  that of the radio interferometer, typically
a few arcmin for the MWA and HERA. First generation 21cm experiments
are subtracting $\sim90$\% of foregrounds \cite{beardsleythesis}, and
99.9\% subtraction is not far out of reach for next generation
experiments. It may be possible to take advantage of the frequency
dimension of 21\,cm image cubes to subtract foregrounds even further. 

What relevant capabilities does SPHEREx bring? Of all the current and
future wide-field near infrared sky surveys (DES, Pan-STARRS, TESS,
WFIRST, EUCLID, ...) SPHEREx is the most promising for this cross
correlation measurement. Its 6'' resolution is coarse enough to
permit a survey of the whole sky, yet fine enough to mask out infrared
point source foregrounds while leaving at least 50\% of the sky
unmasked \citep{zemcov14}. Further, the spectral resolving power of
R$\simeq$40 is equivalent to a redshift resolution of $\Delta z=0.2$ for
Ly-$\alpha$, permitting studies of the evolution of the cross
correlation over the course of the EOR. The frequency dependence of
the cross correlation is also a check on whether a purported cross
correlation is due to foregrounds. Lastly, ground-based IR surveys see
a bright and variable airglow of OH lines around 1$\mu$m, and mosaics
from wide field ground-based IR surveys typically combine postage
stamp (10s of arcmin wide) images from many different observing
conditions. This has the effect of entirely washing out diffuse
structures on larger scales. SPHEREx sees only the zodiacal
background, a factor of $\sim5-10$ lower in surface brightness than
airglow, and its large instantaneous field of view permits
observations of $\sim5^\circ$ scale regions of sky under the same
observing conditions.  


\subsection{Supernovae Investigations}

During routine observations, SPHEREx will provide a unique opportunity
to constrain the physics of supernovae. These stellar explosions
contribute to the origin of elements, influence star formation, and
are one of the main laboratories in astrophysics and cosmology. For
both thermonuclear and core-collapse supernovae, SPHEREx is ideally
suited to constrain overall rates, nucleosynthesis processes, and
production of dust.  Because classification of supernovae by late-time
spectra is not time-sensitive, SPHEREx can make these measurements for
free during its normal survey operations.  
 
Thermonuclear supernovae of white dwarf stars, so called SNe Ia, have light curves
powered by radioactive decay.  The vast majority of them are ``standardizable candles,'' probing the
expansion history of the Universe to high precision.  However, discoveries of sub-types
of SNe~Ia \citep[e.g.,~SNe Iax,][]{2003PASP..115..453L} and the subtle
dependence of calibrated supernova luminosities on host galaxy properties
\citep{2010MNRAS.406..782S} have highlighted the diversity of the progenitor channels and explosion mechanisms.
Understanding this mix will be key to improve precision measurements
of the properties of Dark Energy. 

Core-collapse supernovae, on the other hand, mark the end of life of massive stars.  They are very diverse and play important roles in
 star formation.  Although the explosion mechanism is not yet fully understood, the frequency, high
luminosity, and short progenitor lifetime of core-collapse supernovae make them prime candidates for future studies of
the first stars during the dark ages.  Key products of the nuclear burning---visible in the near- and mid-IR---can probe
explosion mechanisms and the onset of dust formation, basic
ingredients in the life cycle of the elements. 

SPHEREx provides a unique spectral window for a large, uniformly-selected sample of local supernovae, and addresses these questions:

\begin{enumerate}
\item What are the local rates of supernovae of different types and sub-types?
The rate measurement is a powerful tool to distinguish between
progenitor channels.  Comparison of rates between sub-types of
supernovae provides clues to the origins of the observed diversity.

\item Are the ``unusual'' reddening laws seen in supernovae produced by
freshly synthesized dust or by heating of dust in the circumstellar
environment? Dust will reveal its formation and destruction history by
the energy distribution in the mid-IR, i.e., a few microns. The existence of dust in the
circumstellar medium is intricately linked to the evolution of the
progenitor system leading up to the explosion.

\item Are CO and SiO able to form in all types of supernovae? 
These molecules play important roles in the cooling of supernovae.
The timing of the onset of cooling provides constraints to the
explosion physics.

\item 
What is the distribution of argon lines in different types of
supernovae?  Argon is a key indicator of the conditions that cause nuclear
burning breakout from quasi-nuclear equilibrium.  Argon lines are
observable only in the mid-IR. Study of supernova argon lines
provides insight on mixing processes during nuclear burning. 

\end{enumerate}

\begin{figure}[!th]
\begin{center}
\includegraphics[width=0.5\textwidth]{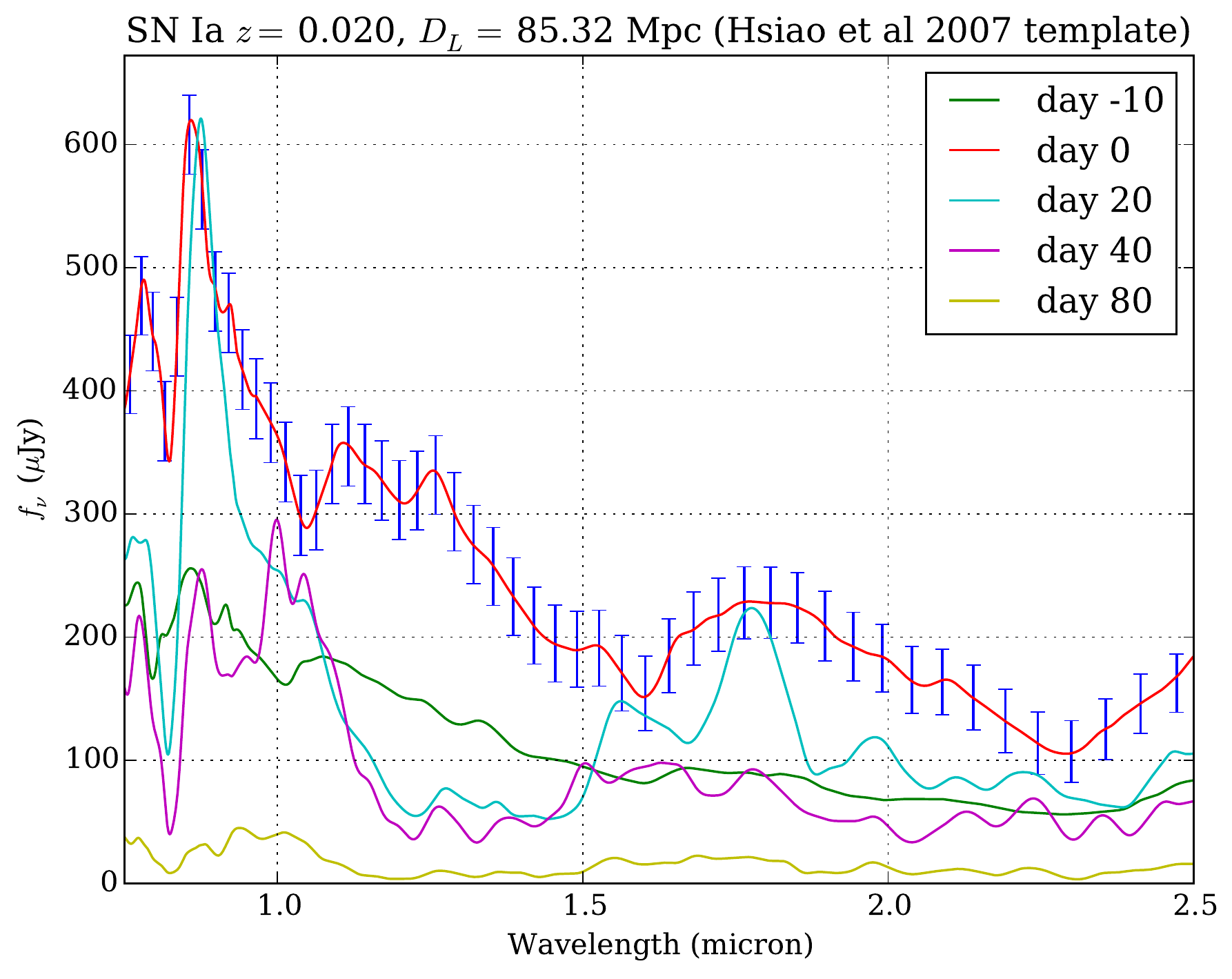}
\end{center}
\caption{{Demonstration of SPHEREx spectral classification for SN rate
    measurements.  This spectral template prediction for a local SN Ia
    shows broad iron features at $\sim$1.6 $\mu$m, whereas a Ib/c
    would show a carbon-oxygen dominated spectrum, and a SN II would
    show strong Paschen lines.  SPHEREx error bars for a
    survey-difference search for transients are compared to the
    evolution from ten days before to eighty days after maximum
    (optical) light. }} 
\label{fig:SNIa_templates}
\end{figure}

SPHEREx will provide near- and mid-IR spectra for supernovae in larger
numbers than will be feasible with Spitzer, JWST, or WFIRST.  From a
local SN Ia rate of $0.55^{+0.50}_{-0.29}$(stat.)$\pm 0.20$(sys.)
Mpc$^{-3}$ year$^{-1}$\citep[][based on just 3 SNe]{2015AA...584A..62C},  we expect $\sim
135$ events on the sky each year brighter than the one plotted in
Fig.~\ref{fig:SNIa_templates}, showing the template of
\cite{2007ApJ...663.1187H} with SPHEREx error bars for a local SN Ia
at $z=0.02$.  The error bars are based on a spectral difference of two
surveys to look for transients.  Such  nearby supernovae are
relatively bright; at B-band maximum light (day 0), the 
SPHEREx signal-to-noise ratio is 124 for a template amplitude fit to
0.75--2.5 $\mu$m.
{Such a supernova may be visible and
classifiable for {perhaps 30 days}}, while nearer ones will
last longer.  Core-collapse supernovae are 3--4 times more common
though typically dimmer.  Thus SPHEREx will see many supernovae per
year.  Most importantly, all SPHEREx supernovae will have
spectroscopic information.  By contrast, for most supernova surveys
the efficiency for selecting events for spectroscopic follow-up
(depending on human decision-making) is low inevitably introducing a large
systematic error into the rate measurements.

\begin{figure}[!th]
\begin{center}
\includegraphics[width=0.5\textwidth]{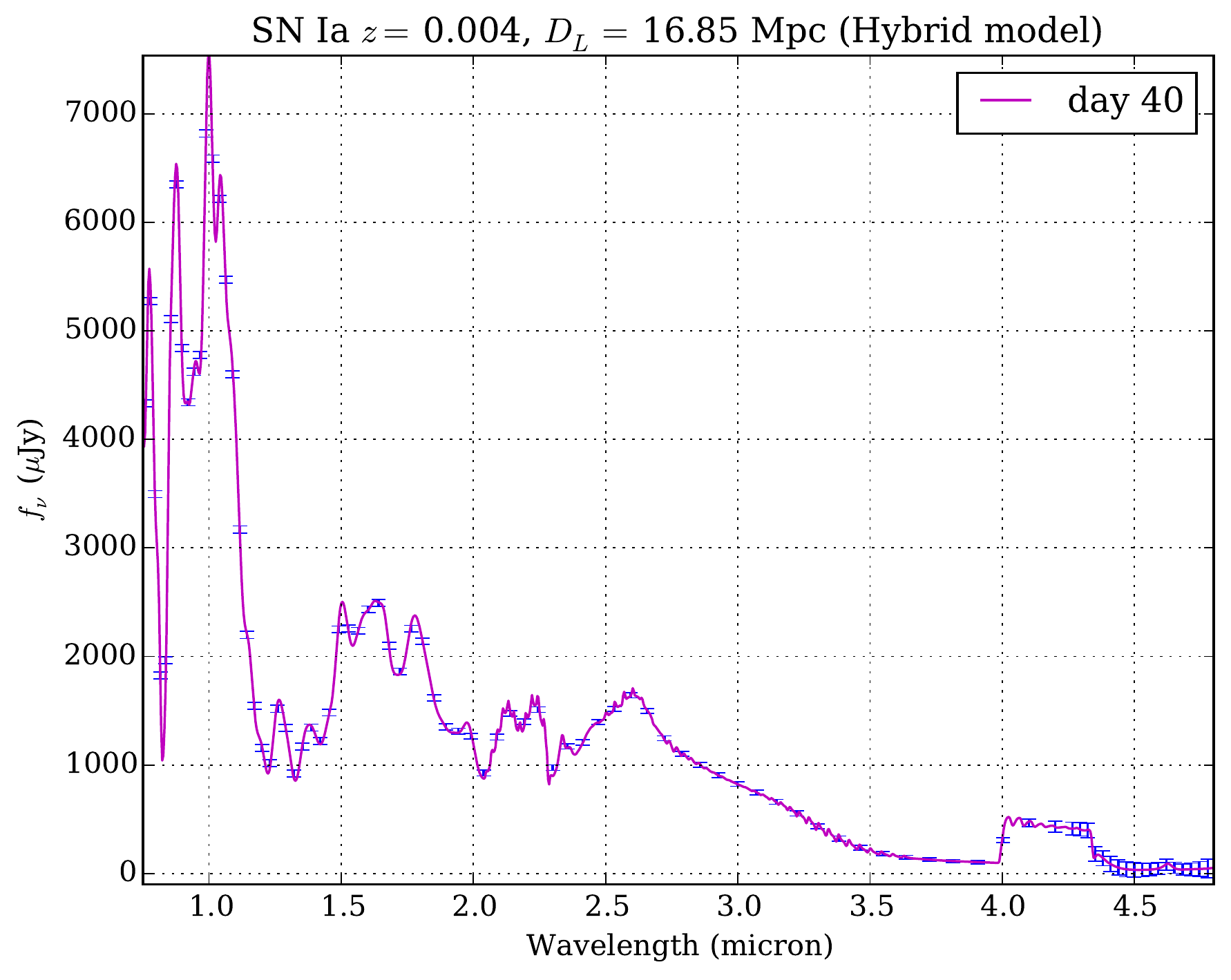}
\end{center}
\caption{Onset of CO formation is more visible from the  fundamental
  band at 4.3 $\mu$m rather than from the first overtone at 2.3
  $\mu$m, which is heavily blended by strong iron group features
  (2.3--2.5 $\mu$m).  Here we give a template spectrum at the onset of
  CO formation modeled at day 40 for a nearby, sub-luminous SN Ia
  compared to SPHEREx errors for a survey difference.} 
\label{fig:SNIa_MIR_model}
\end{figure}

Previous detailed modeling suggests that IR supernovae
spectra evolve slowly with time and homogeneous light curves
\citep{1995ApJ...444..831H,2006ApJ...649..939K,2007ApJ...661..995G,2015ApJ...798...93T}.   
SPHEREx can establish this similarity over a wide range of
progenitors, or prove it wrong and provide a new tool to probe the
variety of SNe Ia, with obvious consequences for the use of mid-IR for supernova
cosmology. Detection of the dust precursor CO  in molecular bands
(Fig.~\ref{fig:SNIa_MIR_model}) would indicate significant amounts of 
carbon, low temperatures, and sufficiently high densities.  For both core-collapse and thermonuclear
supernovae, CO also 
carries information about the local ionizing radiation
environment, because C$+$O$^+\rightarrow$CO$^+$ formation is significantly faster than C$+$O$\rightarrow$CO 
\citep{1995ApJ...444..831H,1990ApSS.171..213S,2000AJ....119.2968G}. Carbon
allows us to detect specific subclasses of SNe Ia/Iax.  Molecular
cooling lowers the temperature in the outer regions, modifying the
transparency of the outer layers and setting the stage for dust
formation. Similarly, for core-collapse supernovae, CO and
dust-formation have already been observed, for example in SN1987A and
SN1998S. Only the much weaker, and heavily-blended, overtone of CO can
be observed from the ground, giving SPHEREx---observing the
fundamental---a useful capability in concert with ground-based
supernova studies (e.g., $\sim$ Carnegie Supernova Project). 
  
SNe Ia show absorption laws unusual for the ISM, in particular for
cases of high reddening/extinction \citep{ stritz99ee99ex02,
  stritzinger11,2015AA...573A...2S}.  SPHEREx can probe for dust in
large numbers of SNe Ia. The circumstellar dust may originate from
red-giant winds of a donor star or the stellar evolution of the stars of either single white dwarf
or double white dwarf progenitor systems.  Argon lines are excellent tools to
separate different regimes of burning, namely to probe the region of
incomplete Si burning. Argon lines will provide new insights and
constraints on the instability and mixing of nuclear flames,
off-center {delayed-detonation transitions}, mergers, and rotation.

Few studies of supernovae have been published for wavelengths beyond
the K band at about 2 micron.  Though upcoming missions (like JWST and WFIRST)
 will open this window, providing exquisite data for a small
number of objects, SPHEREx in the normal course of operations will
yield a large, uniform sample, probing the  intrinsic and
apparent diversity of objects key to high-precision cosmology.




\section{Galactic Science}

We divide the discussion of Galactic Science enabled by SPHEREx into
two sections:  Stars (Sec.~\ref{sec:stellar}) xsand Interstellar and Circumstellar Matter (this section).
SPHEREx will provide low resolution spectra of millions of stars
which, particularly when combined with spectrophotometric and
astrometric data from Gaia, will address problems related to
exoplanets, planetary systems, and Galactic structure as well as
providing improved data on fundamental stellar parameters.  At the
same time, SPHEREx will obtain data on emission and absorption in
interstellar and circumstellar space which can be used to study the
energetics of interstellar shocks and outflows, to study the
distribution of hydrocarbons and to search for deuterated species
amongst them, and to characterize the properties of interstellar
grains in a variety of environments. 

\subsection{Aromatic and aliphatic hydrocarbons}
\label{ssec_hydro}

Polycyclic Aromatic Hydrocarbons (PAHs) play a key role in the heating, chemistry and ionization of the
interstellar gas.  Their emissions dominate the flux observed from the
Galactic plane in the 5.8 and 8~$\mu$m bands of Spitzer/IRAC \citep{2012AA...545A..39R} and the band of WISE. PAH emissions
reveal a population of bubbles associated with young stars in regions
of high-mass star-formation\citep{2006ApJ...649..759C}. 
Whereas the IRAC 8~$\mu$m band is sensitive to the C-C mode at
7.7~$\mu$m, SPHEREx can map the aromatic C-H stretch at 3.3~$\mu $m,
and distinguish it from its aliphatic counterparts in the 3.40 -
3.56~$\mu$m range.  The Galactic plane survey, GLIMPSE \citep{2009PASP..121..213C}, performed with IRAC at Galactic latitudes $\vert b \vert
\le 1^0$, suggests that PAH emissions above the SPHEREx detection
threshold will be present up to $\vert b \vert \sim 5^0$ or greater,
the SPHEREx sensitivity to PAH emission being a factor $\sim 2$ better
than GLIMPSE for extended sources.  This opens the possibility of
studying UV-heated PAHs much higher above the Galactic plane than in
previous surveys, and of determining the ratio of aliphatic to
aromatic C-H sites. 

In addition, the SPHEREx bandpass covers the analogous C-D stretching
modes at 4.40~$\mu$m (aromatic sites), and the 4.56 - 4.85~$\mu$m
range (aliphatic sites).  The latter were detected toward two
photodissociation regions (the Orion Bar and M17) by
\cite{2004ApJ...604..252P}, using ISO, implying that PAHs were
deuterated at the several $\times 10\%$ level.  This led to the suggestion that a significant fraction, or even the majority, of the  Galactic
deuterium might be sequestered on PAHs \citep{2006ASPC..348...58D}, providing an
explanation for the surprisingly large variations (from 5 to 22 ppm)
observed in the atomic D/H ratio \citep[e.g.,][]{2006ASPC..348...47H}. Models
for the  preferential incorporation of D into PAHs provide theoretical
support for the possibility of such a high level of deuteration
\citep{2006ASPC..348...58D,2016AA...586A..65D}. More recently, however, AKARI has
surveyed $\sim 50$ HII regions for the C-D stretching mode of
deuterated PAHs and obtained only six ``unambiguous'' detections,
suggesting that high levels of deuteration are sometimes present but
are not ubiquitous \citep{2016AA...586A..65D}. The large statistical power
provided by the SPHEREx' all-sky survey, targeting both the C-H and C-D
stretching modes,  promises to resolve the question of how
  widespread deuterium  sequestration onto PAHs is. 

\subsection{SPHEREx Observations of Hydrocarbon Grains}

\begin{figure}[h!]     
\center
\includegraphics[width=0.6\textwidth, angle=90]{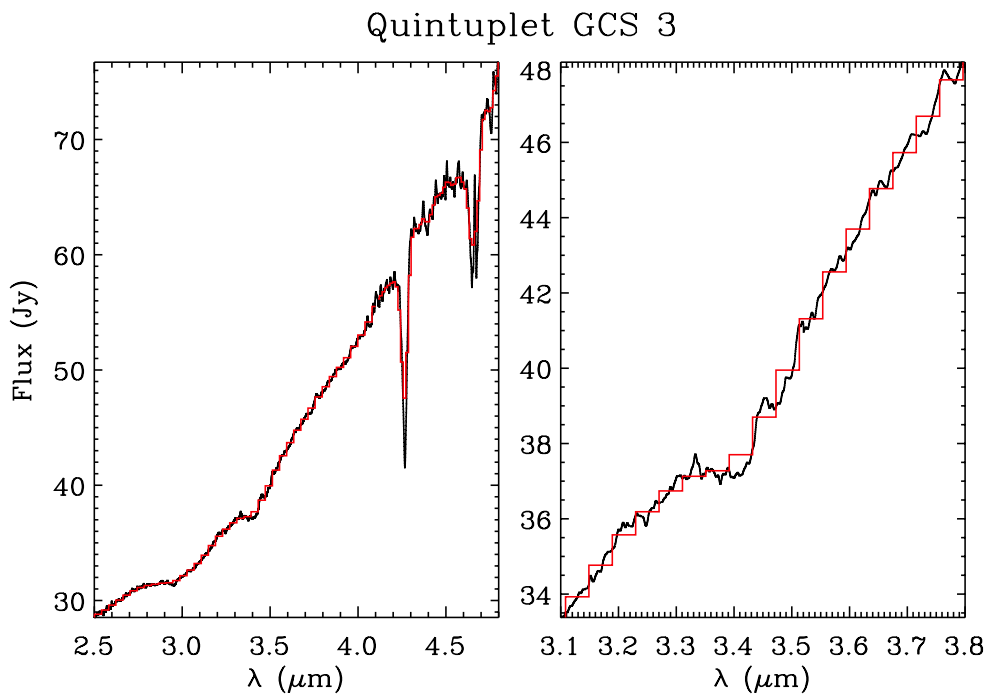}
 \caption{ISO/SWS observation of the Quintuplet cluster source GCS 3
   at a resolving power of 400 (black) and convolved for a portion of
   the wavelength range, to the SPHEREx
   resolving power (red). The feature near 3.4 $\rm \mu m$ is due to
   hydrocarbons along the diffuse ISM sight-line, while features near
   3.0, 4.3, and 4.7 $\rm \mu m$ are due to H$_2$O, CO$_2$, CO, and
   OCN$^-$ ices respectively. The aliphatic group substructure will not be resolved
   by SPHEREx, but the overall feature is easily
   detectable.}\label{f1}
\end{figure}           

The aliphatic hydrocarbon absorption features near 3.4 $\rm \mu m$,
observed in the diffuse ISM, have an abundance of a few to 10\%
relative to carbon in aromatic compounds (PAHs, which consume 10-20\%
of the cosmic carbon).  To date, the aliphatics have been looked for
and thus seen only along selected lines of sight, so the SPHEREx all
sky survey can greatly increase our understanding of their prevalence and
significance.  The origin of the aliphatic hydrocarbons was initially 
thought to be in heavily processed ices, but this is disputed based on
a lack of polarization in the observed features
\citep{1999ApJ...512..224A,2006ApJ...651..268C}.  Instead, aliphatic
hydrocarbon dust might evolve from the 
interplay of H-atom reaction with and UV photolysis of graphite dust
in the ISM.  Production in carbon-rich evolved stars is also a
possibility given the detection toward the young planetary nebula CRL
618 \citep{1998ApJ...507..281C}. However, these studies are based on the
detection of the 3.4 $\rm \mu m$ absorption in just $\sim$10 lines of
sight in the diffuse ISM, and in one evolved stellar envelope. A
SPHEREx all-sky survey would enable a study of the hydrocarbon
abundance across the Galactic Disk.  For example, do aliphatic
hydrocarbons evolve from PAH species driven by strong UV fields in the
ISM? Are they strong near massive stars with strong UV fields? Are
they preferably associated with AGB stars? Is there a gradient in
strength with Galactocentric radius, related to the fraction of carbon
stars among AGB stars and thus the star formation history?

Contamination with PAH emission would complicate the analysis; an
all-sky survey is needed to find sight-lines suitable for aliphatic hydrocarbon
study.  Although at a resolving power of 40, the CH, CH$_2$ and CH$_3$
groups would not be resolved, SPHEREx is well suited to determine the
total volume of aliphatic hydrocarbon dust through the 3.4 $\mu$m absorption
feature (Fig.~\ref{f1}) and this can be compared to the total volume of dust
determined from the continuum extinction.  SPHEREx will enable this
work to be extended to extragalactic nuclei, which often show
prominent 3.4 $\rm \mu$m bands.  Extrapolating from the Galactic
SPHEREx results, we will learn what this feature tells us about the physical and
star formation history in external galaxies.

\subsection{The Life-cycle of Interstellar Dust \& the 3.3 $\mu$m  PAH feature in nearby galaxies}

The combination of wavelength coverage and angular resolution provided
by SPHEREx opens up a wide range of exciting science questions in
nearby galaxies.  SPHEREx sensitivities will allow widespread
detections of the 3.3 $\mu$m PAH feature as well as the near-IR H recombination lines, which provide
extinction-insensitive tracers of massive star formation.  At
6 arcsec resolution, SPHEREx resolves the scale of individual star
forming regions and molecular clouds in all of the galaxies of the
Local Group. Nearby objects like the Magellanic Clouds are highly
resolved by SPHEREx into $\leq 2$ pc resolution elements.  For the
LMC, this will mean $>10^8$ individual resolution elements across the
main body of the galaxy. In galaxies out to 10 Mpc, SPHEREx resolves
galactic disks at hundreds of pc resolution, and can distinguish
arm/interarm regions and separate nuclear regions from the disk.
Beyond widespread PAH and H recombination lines, bright regions of
nearby galaxies will show a wealth of near-IR lines including H$_2$
vibrational lines, [Fe II] 1.257 \& 1.646 $\mu$m, and many others.
These lines will be of great interest for diagnosing physical
conditions in the gas, particularly relating to shocks and feedback
from massive star formation and active galactic nuclei.  Here we focus
on two questions related to the 3.3 $\mu$m PAH feature and maps of
star formation in nearby galaxies. 

PAHs play key roles in the interstellar medium.  In dense clouds they
influence chemical reactions, most importantly the formation of H$_2$
\citep{1998ApJ...499..258B,2008ApJ...680..384W,2009ApJ...704..274L}. Across
many ISM phases, PAHs are the dominant contributor to the
photoelectric heating rate
\citep{1994ApJ...427..822B,1995ApJ...443..152W}.  PAHs also radiate a
significant fraction of the total infrared emission from galaxies
\citep[$\sim10-20$\%][]{2007ApJ...656..770S}, making them a
potentially useful tracer of star formation out to high redshift
\citep{2007ApJ...666..870C}.  The properties of PAHs (size, charge,
abundance) influence how effectively they carry out these roles.  For
example, the photoelectric heating rate depends on the size
distribution of the grains and their charge
\citep{Tielens:2008fx}---e.g. when PAHs are more highly charged, they
are less effective sources of photoelectric heating.  Correlations of
[CII]-to-PAH emission with PAH band ratios observed in nearby galaxies
\citep{2012ApJ...747...81C,2012ApJ...751..144B} suggests that 
the physical properties of the PAHs are indeed playing a role in the heating
and cooling of the ISM.  The key open question we need to answer is
how and why the properties of PAHs vary between galaxies and with
environmental conditions within galaxies. 


The mid-IR PAH emission bands provide a suite of useful diagnostic
features for the population of PAHs.  Much work has been done
attempting to understand how PAH properties change with environment
making use of the mid-IR lines with ISO and Spitzer observations
\citep[e.g.][]{2007ApJ...656..770S,2012ApJ...744...20S}.  However, the
mid-IR bands suffer from some degeneracies between the features tied
to charge and those tied to PAH size.  For example, while the 7.7/11.3
$\mu$m ratio is often used as a tracer of PAH ionization, this ratio
can also vary if the size distribution of PAHs changes \citep[see][for
more detail]{2007ApJ...657..810D}.  The 3.3 $\mu$m feature is a
crucial diagnostic in this regard because its emission is strongly
associated with small, neutral PAHs, making it one of few mostly
unambiguous tracers for the PAH population.  Ratios of 3.3/11.3
$\mu$m features, for instance, are a clean tracer of size,
since both 3.3 and 11.3 are from neutral PAHs.  Observations of the
3.3 $\mu$m feature can therefore serve as an anchor for interpreting
the other band ratios, separating the effects of PAH size and charge
variation.  Unfortunately, the 3.3 $\mu$m feature was not covered by
Spitzer and previous and subsequent surveys (e.g., with ISO, AKARI)
have had limited coverage of the local galaxy population.  SPHEREx
will change this.   

The combination of all-sky SPHEREx mapping of the 3.3 $\mu$m feature
with existing data from WISE and the Spitzer archive will enable a
definitive study of the variability of PAH properties across all local
galaxies.  Since SPHEREx maps the whole sky, every spectral mapping
observation with Spitzer-IRS will gain coverage of the 3.3 $\mu$m
feature---this will give a wide ranging basis set of full PAH coverage 
(from 3.3 through 17.0 $\mu$m).  Moving beyond the targets with
direct spectroscopic PAH coverage, Spitzer mapped a large sample of
nearby galaxies with the IRAC 8 $\mu$m band, which primarily samples
the 7.7 $\mu$m PAH complex.  With spectroscopic-to-photometric PAH
calibrations \citep[see for example][]{2012ApJ...747...81C}, this
provides 3.3/7.7 diagnostics for a huge number of galaxies.  Likewise,
in the largest possible sample, the WISE all-sky survey band 3 covers
a combination of the 7.7 and 11.3 features.  Combining these datasets
will let us map the variation of PAH populations across all nearby
galaxies and tie the resolved changes to local environment. 

One specific question that needs to be addressed about PAHs is what
causes their disappearance at low metallicity
\citep{2005ApJ...628L..29E,Madden:2006fj,Engelbracht:2008cl,2007ApJ...657..810D,2010ApJ...715..701S}.
The deficit of PAHs at low metallicity is a long-standing
puzzle---some combination of PAH production, destruction or ISM
processing changes dramatically at $12+$log(O/H)$\sim8.1$ . Studies
with Spitzer-IRS suggest that low metallicity PAHs have {\em smaller}
average sizes compared to PAHs in higher metallicity galaxies,
inferred from the relative weakness of the 17.0 $\mu$m feature
\citep{2007ApJ...656..770S}.  This observation has important
implications for the PAH lifecycle---larger PAHs are the most
difficult to destroy, so a size distribution shifted towards smaller
sizes most likely indicates a deficit in large PAH production
\citep{2012ApJ...744...20S}.  However, ambiguities in the relative
importance of size and charge in setting the ratio of 17.0 to other
bands remain an obstacle.  SPHEREx will observe 3.3/11.3 ratios both
spectroscopically (in all Spitzer-IRS maps) and photometrically (in
combination with WISE band 3) giving the definitive dataset for
answering why PAHs disappear at low metallicity. 


\subsection{Probing the Gas Phase: Synergy with GLIMPSE and MIPSGAL}

The SPHEREx results will build upon, and be enhanced by, numerous
previous surveys. The GLIMPSE\footnote{http://www.astro.wisc.edu/glimpse/} and MIPSGAL\footnote{http://mipsgal.ipac.caltech.edu/} surveys of the galactic
plane carried out by Spitzer provide an excellent example of this
synergy.  GLIMPSE surveyed the entire galactic plane, covering 360
degrees in longitude and at least $\pm$ 0.5 degree in latitude in the
Spitzer bands at 3.6 and 4.5 $\mu$m [and much of the inner Galaxy at 5.8
and 8 $\mu$m and over a larger latitude range], providing a data base which
might be considered an extension of 2MASS.  The GLIMPSE beam size is
$\simeq$ 2''.  The GLIMPSE data base contains tens of millions of sources, and
the brightness limits of the catalog show that SPHEREx will achieve
more than 5$\sigma$ per spectral resolution element for GLIMPSE 3.6
and 4.5 $\mu$m point sources, even from 4.18 to 5 $\mu$m where SPHEREx's  continuum
sensitivity is reduced due to the higher spectral resolution.
MIPSGAL, synergistic with GLIMPSE, surveyed a goodly portion of
the inner galactic plane at 24 and 70~$\mu$m; although these wavelengths
are far outside of the SPHEREx band, the MIPSGAL discoveries described
below have been found to be bright at J, H, and K and are therefore
candidates for study from SPHEREx. 

A potentially exciting example of the synergy between GLIMPSE and
SPHEREx comes from a class of object discovered by GLIMPSE to show
extended regions of excess 4.5 $\mu$m emission.  Dubbed 
``Extended Green Objects (EGOs)'' by  \cite[hereafter
C08]{2008AJ....136.2391C} -- ``green" because the 4.5~$\mu$m band has traditionally been displayed as the green channel
in false-color images -- many of these objects show a clear
association with massive young stellar objects (MYSOs).  The
association of 4.5~$\mu$m emission with MYSOs suggests the influence of
shock waves.  Shock waves are widespread in the ISM, and are often
associated with supernova remnants and supersonic protostellar
outflows.  They may compress, heat and ionize the interstellar gas
through which they propagate. Slow shocks propagating in molecular
gas radiate strongly in rovibrational and pure rotational transitions
of molecular hydrogen, many of which lie within the SPHEREx wavelength
range: the S(9) pure rotational transition at 4.69~$\mu$m is expected
to be particularly strong. C08 posited  line emissions from
outflow-driven shock waves as the origin of the observed IRAC
4.5~$\mu$m excess, suggesting H$_2$ pure rotational lines (the S(9),
S(10) and S(11) transitions accessible to SPHEREx) as dominant contributors to the
4.5~$\mu$m flux.  Subsequently, \cite{2012ApJS..200....2L} conducted a 
near-IR search for H$_2$ vibrational emissions from EGOs; roughly one-third of the targeted sources showed
detectable H$_2$ vibrational emissions, although in many cases the
H$_2$  line emission showed a morphology differing from that of the
IRAC 4.5~$\mu$m emission. This led \cite{2012ApJS..200....2L} to argue that scattered
continuum emission might contribute significantly to the  4.5~$\mu$m
fluxes measured in many EGOs (although the effects of differential
dust extinction between 2.1 and 4.5~$\mu$m might provide an alternative
explanation for at least part of the observed morphological
differences.) 

A catalog of roughly 300 EGOs detected in the GLIMPSE survey has been
presented by C08. For this sample, the median solid angle subtended
by the sources was 114 sq. arcsec, corresponding to $\sim 3$ SPHEREx
pixels and suggesting that the {\it typical} source will only be
marginally resolved at the angular resolution of SPHEREx, but sources
as large as 1767 sq. arcsec ($\sim 50$ SPHEREx pixels) were
catalogued.  For the larger EGOs in which the diffuse emission can be
resolved spatially from the associated MYSOs, SPHEREx has the
sensitivity to readily detect line-dominated sources in the C08
catalog, and will potentially add a large number of additional sources
at higher latitudes ($\vert b \vert > 1^0$).  Results from SPHEREx
will have the potential of (1) revealing the relative 
contribution of H$_2$ line emission to the IRAC-detected 4.5~$\mu$m
excesses, and thus resolving an outstanding question about their
nature; (2) expanding the sample of MYSOs, and (3) yielding H$_2$
line ratios that could constrain the physical conditions in the
emitting gas.  

In addition to the pure rotational lines of H$_2$ that fall within the
SPHEREx wavelength range, several other spectral signatures of warm,
shocked-heated gas may  be detected.  These include several H$_2$
rovibrational lines, the strongest being the $v=1-0$~S(1) line at
2.12~$\mu$m, and two near-IR [FeII] lines at 1.26~$\mu$m and
1.64~$\mu$m.  Because the latter two lines are optically thin and
originate in the same upper state of Fe$^+$, their intrinsic ratio is
a constant determined by atomic physics: the observed line ratio is
therefore a useful probe of extinction.  Line maps obtained by SPHEREx
will complement smaller ground-based surveys (UWISH2 and UWIFE) of the
2.12~$\mu$m H$_2$ and 1.64~$\mu$m [FeII] lines that have revealed a
large sample of PDRs, jet outflow regions, planetary nebulae, and
supernova remnants \citep{2015MNRAS.454.2586F,2014MNRAS.443.2650L}.

The combined GLIMPSE and MIPSGAL surveys have identified other
potentially interesting classes of objects.  One group is known as
Yellowballs due to their appearance in a false-color presentation of
GLIMPSE+MIPSGAL data; another is the more prosaically named MIPSGAL
bubbles \citep{2010AJ....139.1542M}. The MIPS bubbles have been
studied spectroscopically from the ground and searched for in other
catalogs and found to be a heterogeneous group
\citep[e.g.,][]{2014AJ....148...34F}. They include planetary nebulae,
shells around massive, mass-losing-stars such as Wolf-Rayet stars and
Luminous Blue Variables, and shocked-excited nebulae powered by winds ad outflows
from high mass young stellar objects. The Yellowballs are a more recent discovery; circumstantially
they are identified as a previously unknown stage in massive star
formation \citep{2015ApJ...799..153K}, but extensive ground-based
spectroscopy has yet not been reported.  

The remarkable characteristic of SPHEREx is that it will return
spectra of most or all of these nebulae and of the EGOs, well over one
thousand spectra in all.  In addition to the shock diagnostics
discussed above, the SPHEREx wavelength range encompasses Paschen and Brackett lines
diagnostic of gas ionized by embedded massive stars; helium emission
lines which might arise from hydrogen poor planetary nebulae or
Wolf-Rayet stars; and PAH emission at 3.3 $\mu$m which could arise at the
interface between a nebula and adjacent neutral material.  Thus we can
anticipate a treasure trove of nebular spectra from SPHEREx
observations of these families of compact nebulae, spectra which for
one object, might bolster and refine a identification, but, for
another, demonstrate a highly unusual or perhaps previously unknown
type of nebula in the galaxy. 


\subsection{Mapping Dust Variations}

Many galactic and extragalactic astrophysical objects and processes
are affected by dust. It is thus important to know the composition of
dust, the shape and size distribution of dust grains, and the physical
and chemical processes that form, modify and destroy dust.  Much of
the dust is formed in the ejecta of evolved stars: AGB stars and
supernovae.  The type of dust produced is set by the conditions and
composition of the ejecta. Carbon-rich shells produce graphite, and
aliphatic and aromatic soot-like materials, while silicate dust is
formed in oxygen-rich shells. After ejection into the diffuse ISM, a
hydrogen-rich surface layer may be formed, balanced by hydrogen
detachment by UV photons.  In dense clouds, grain surfaces catalyze
the formation of hydrogen-rich molecules (H$_2$, H$_2$O, CH$_4$, et
cetera), forming icy mantles. Here, grain growth is likely accelerated
by coagulation of the sticky, icy grains.  Subsequently, as the grains
are cycled back into the diffuse ISM or before they are incorporated
into comets and planets, their icy mantles might be processed by
energetic particles and radiation to form carbonaceous dust.
Exploration of questions related to ice formation and evolution forms
one of the central themes of SPHEREx science program; here we show
that SPHEREx will address other issues related to the properties of
interstellar grains. 

It is evident that as a result of all the physical and chemical
processes involved, dust properties should vary in space and time.
The SPHEREx all-sky spectroscopic survey (0.75-5.0 $\rm \mu m$) will
enable studies of specific dust property variations across
environments, for example:
\begin{itemize}
\item variations of the infrared continuum extinction curve 
\item variations in the infrared spectral characteristics of
  interstellar dust (due to, e.g., variations in the abundance of
  aliphatic carbon dust)
\end{itemize}
With these SPHEREx observations, a number of science questions can be
uniquely addressed:
\begin{itemize}
\item is there progressive grain growth from diffuse, to dense, to
  protostellar environments? SPHEREx is able to trace much higher
  extinctions ($A_{\rm V}>70$ mag) than is possible in the optical and
  UV.
\item what is the origin and evolution of hydrocarbon grains and what
  are the processes that drive this evolution? (this question is
  addressed in Sec.~\ref{ssec_hydro})
\end{itemize}

The SPHEREx all sky map of extinction due to interstellar dust can be
very usefully combined with the work of \cite{2015ApJ...810...25G},
who have produced a three-dimensional picture of the distribution of dust within $\sim$3~kpc of the Earth. Combining the two
data sets will allow determination, for example, of whether an anomaly
seen by SPHEREx in the grain properties along a particular line of
sight coincides with a particular feature of the galactic geography,
such as a young cluster or a supernova remnant. 

\subsubsection{Resolving Extinction Curves with SPHEREx}

\begin{figure}[t]     
\center
\includegraphics[width=0.6\textwidth,angle=+90]{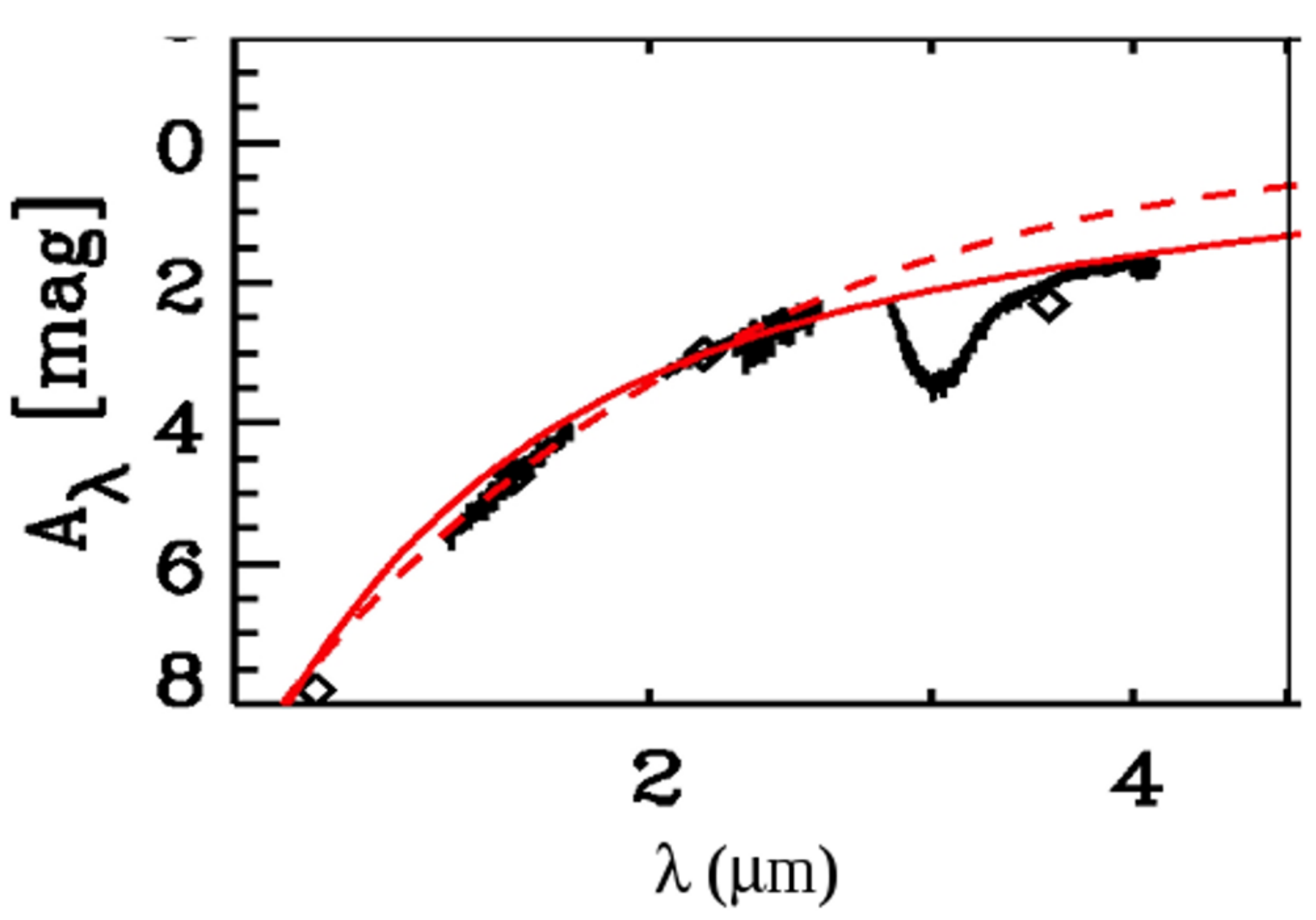}
 \caption{A spectrally resolved infrared extinction curve derived from
   ground-based observations of the dense core background star 2MASS
   J21240517+4959100 \citep[][in black]{2011ApJ...729...92B} compared to
   models of \cite{2001ApJ...548..296W} with $R {\rm _V}$=3.1 [dashed red],  and $R {\rm _V}$=5.5 [solid
     red]. The wavelength range covered
   by SPHEREx is shown.  This particular background star has a
   broad-band extinction $A_{\rm K}=3.1$ mag ($A_{\rm V}\sim 28$). The
   spectral type determined from the 2.25-2.6 $\mu$m CO overtone band
   head is M1 III. It has a $K$-band flux of 122 mJy. SPHEREx will be
   able to survey such objects at S/N$\geq$100 up to $A_{\rm K}\sim 8$
   mag ($A_{\rm V}\sim 72$), i.e., trace dust much deeper into the
   cloud core, without atmospheric gaps and uncertainties due to scaling
   spectra observed in different atmospheric bands.}\label{fig:extinction}
\end{figure}

The continuous coverage of the 0.75-5.0 $\rm \mu m$ wavelength range
will enable the determination of spectrally resolved extinction curves
(Fig.~\ref{fig:extinction}) for every star in the sky, limited by brightness and by the
ability to correct for the intrinsic stellar spectrum. 

\begin{figure}[t]     
\center
\includegraphics[width=0.5\textwidth]{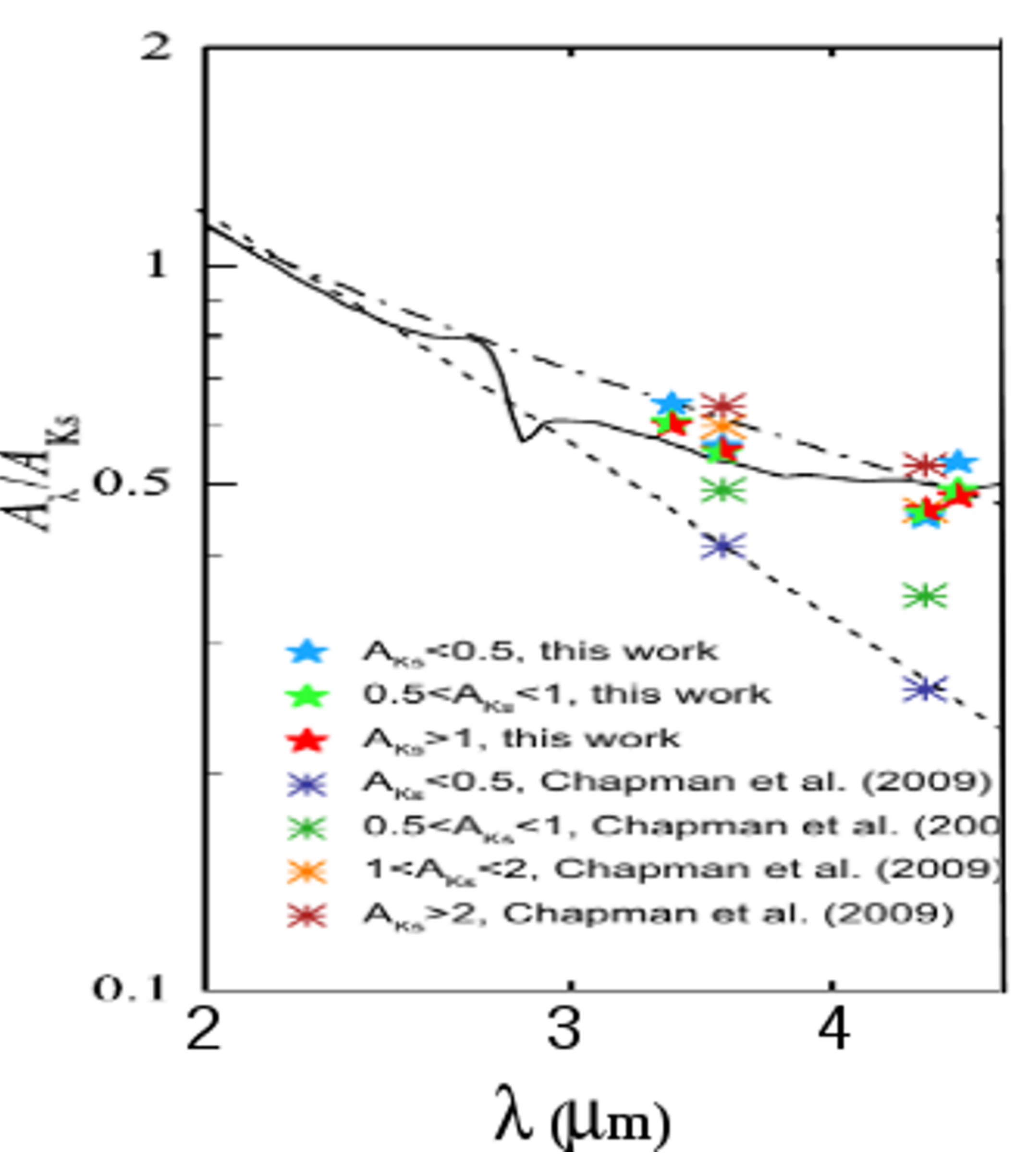}
 \caption{Broad band extinction curves from different studies and in
   different sight-lines (asterisks), showing much variation. The
   wavelength range starts at 2 $\mu$m, while SPHEREx will cover down
   to 0.75 $\mu$m as well. Note that SPHEREx covers the range where
   the different curves diverge. The solid line is a model including
   4 $\rm \mu m$-sized H$_2$O ice rich grains. The dotted and
   dash-dotted curves are for $R {\rm _V}$=3.1 and 5.5 models \citep{2001ApJ...548..296W}, respectively.  Adapted from \cite{2016arXiv160202928X}.}\label{fig:extinction2}
\end{figure}

These empirical extinction curves enable a separation of continuum and feature
extinction, which, in dense clouds, are in particular the ice
absorption bands at 3 (H$_2$O), 4.3 (CO$_2$), and 4.7 $\rm \mu m$
(CO), the major focus of SPHEREx' Ice investigation, and the aliphatic
features at 3.4 $\rm \mu$m in diffuse clouds. In broad band
measurements, feature and continuum extinction are combined. Also, for
broad band measurements, correction 
for the stellar photosphere often relies on typical spectral types in the
field, limiting the spatial resolution to arcminute scales, while with
SPHEREx it can be done for individual stars, whose spectral type is
determined by SPHEREx itself. This matters, for
example, across cloud boundaries where dust modification may well take
place.  Previous ground-based work on spectrally resolved extinction curve
measurements in this wavelength range rely on stitching together data
in different atmospheric bands. The gap around 2.8 $\rm \mu m$ limits
the H$_2$O ice band analysis. Ground-based observations above 4.1 $\rm
\mu m$ have poor sensitivity.  Broad band measurements show large
variations in derived extinction curves (Fig.~\ref{fig:extinction2}),
and it is unclear what the origin of these variations is, if real. It is important to
note that the divergence between the curves is around 3 $\rm \mu m$,
central in the SPHEREx wavelength coverage.

\subsubsection{SPHEREx Observations of Large Grain Signatures}

The flattening of the extinction curve above 3 $\rm \mu m$
(Figs.~\ref{fig:extinction} and \ref{fig:extinction2}) is a signature
of a population of large ($>$1 $\rm \mu m$) 
grains.  Dust in the diffuse ISM ($A{\rm _V} <$~1 mag) is well
characterized by $R{\rm _V}$=3.1 models \citep{2001ApJ...548..296W}
but that does not fit these data through dense cloud sight-lines
well. Grain models characterized by much larger sizes \citep[$R{\rm
  _V}$=5.5][]{2001ApJ...548..296W,2015MNRAS.454..569W} or grain
models including large ice grains \citep{2015ApJ...811...38W} do much
better. They would add 4-16\% to the dust mass. Large dust grains in
dense environments are also evident from the so-called coreshine
\citep[bright reflected light from dense cloud cores][]{2010Sci...329.1622P}. It has been long thought that grains grow extensively in
size in dense clouds through coagulation or through ice mantle
formation or both. SPHEREx can address such important questions as: how
does the grain growth depend on the cloud environment, including
column density, volume density, and depth of the star or the line of
sight within the cloud? Are the grain size and ice mantle volume
(traced by the 3 $\mu$m ice band) related? Are these large grains destroyed in the
environment of supernova remnants or inside HII regions?

Observationally, the presence of H$_2$O-rich $\mu$m-sized large grains
could be distinguished by a feature in the extinction curve near 2.7 $\mu$m, caused by
scattering (Fig.~\ref{fig:extinction2}). In addition, towards reflection nebulae
surrounding young stellar objects this feature appears in absorption as
a wing on the short-wavelength side of the 3.0 $\rm \mu m$ ice absorption feature. It is produced by grains
larger than 0.2~$\mu$m in radius and is thus diagnostic of grain growth.
So far this has been observed only in a 25 arcsec beam toward Orion
Molecular Cloud 2 using the Kuiper Airborne Observatory
\citep{1982ApJ...260..141K,1990ApJ...349..107P}. 
All-sky spectral maps by SPHEREx offer a unique tracer of such large
icy grains.  SPHEREx will allow us to probe the growth of dust and the
formation of ice mantles on a global scale in regions of star
formation. 

\subsubsection{Conclusions}

A SPHEREx all-sky 0.75-5.0 $\rm \mu m$ survey has the
unique capability to address a number of fundamental questions about
the origin and evolution of interstellar dust: is there progressive
grain growth from diffuse, to dense, to protostellar environments?
what is the distribution and origin of hydrocarbon grains? SPHEREx' full
coverage of the wavelength range, unhindered by the Earth's
atmosphere, its spectroscopic resolution and sensitivity, and in
particular its all-sky survey are unparalleled by past and other
future missions. 

\section{Stellar Science}
\label{sec:stellar}

\subsection{Normal Stars}

SPHEREx will enable exquisite characterization of fundamental stellar
parameters for the approximately two million Tycho stars with $V < 13$
mag. SPHEREx's VIS-near-IR spectrophotometry between $0.75-5.0 \mu$m will
cover the Rayleigh-Jeans tail of stellar spectral energy distributions
(SEDs) for all observed stars as well as the SED peak for stars with
$T_{eff}$ $\lesssim 4000$ K, which corresponds to mid-K and later
spectral types. These cool stars are the ones for which the models are most
uncertain and which are of particular interest as exoplanet host
stars (see Sec.~\ref{sec:TESS}). 

\begin{figure}[!th]
\center
\includegraphics[width=0.5\textwidth]{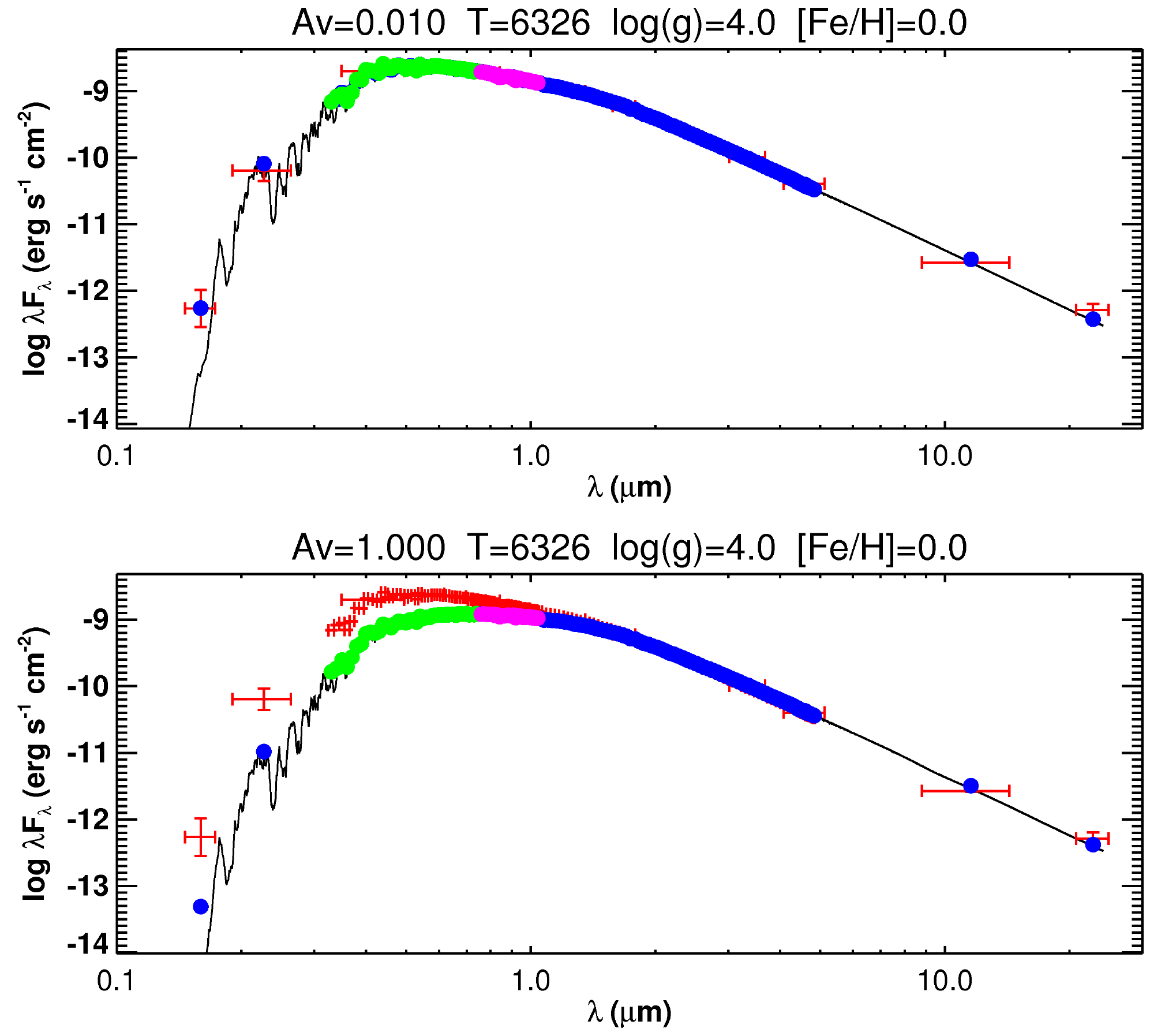}
\caption{\label{fig:av}Best-fit model spectral energy distribution of
  KELT-3, adapted from \cite{Pepper13}. The black line shows the
  model SED, while the red crosses show the literature broad-band flux
  measurements, the blue crosses show the estimated SPHEREx
  spectrophotometry, and the green crosses show the estimated
  \emph{Gaia} spectrophotometric flux coverage; the
  SPHEREx-\emph{Gaia} overlap is shown in magenta. The top panel
  assumes visual extinction of $A_{\it V} = 0.1$, while the bottom
  panel shows the same case in red and $A_{\it V} = 1.0$ in blue. The
  infrared SPHEREx measurements anchor the SED, enabling the
  extinction to be measured from the \emph{Gaia} coverage of the
  extincted SED peak.} 
\end{figure}

Adding other databases to SPHEREx fluxes, including \emph{Gaia's} overlapping
low-resolution spectrophotometry between 330nm and 1050nm, and broad-band
fluxes as blue as GALEX and as red as WISE, will yield
comprehensively-measured SEDs for hundreds of thousands of stars, with
bright, nearby stars benefitting from the most precise measurements.
The combined data base will capture the blackbody peak for stars with
$T_{eff}$ $\lesssim 8750$ K, corresponding to spectral type mid-A and
later spectral types. For bright (e.g. $V < 13$ mag) stars, which also
have broad-band flux measurements from the UV (e.g., GALEX) through the
near-IR (WISE), such blanket coverage will directly measure nearly all
of the flux: for the late F dwarf KELT-3 \cite{Pepper13}, literature
broad-band fluxes directly measure about 80\% of the fitted flux,
while adding SPHEREx and Gaia spectrophotometry increases the
measured flux to about 98\% (cf. Fig.~\ref{fig:av}). Thus, the
stellar SED is exceptionally well-constrained by the data, and the
best-fit effective temperatures, surface gravities, and [Fe/H] can be
determined to percent or sub-percent precision.  When Gaia astrometry
is included, the end result will be a large catalog of stars with
$V\leq$13 and well-determined distances, bolometric luminosities, proper motions, radii, effective
temperatures, metallicities, surface gravities, and extinctions. In particular, the combination of the
precise distance determination from Gaia and the other well-determined
parameters will yield stellar radii measured to around one percent.
This is approximately the precision achieved on interferometric radii
of bright, nearby stars \cite{Boyajian13}, but this
``spectrophotometric interferometry'' method works for stars hundreds
of parsecs away, versus tens of parsecs for interferometry. 

Additionally, while extinction negligibly affects the SED
normalization for nearby stars, moderate extinction $(A_{\it V} =
1.0)$ can reduce the peak flux for hot stars by an order of magnitude
(Fig.~\ref{fig:av}). Since the shape and slope of the Rayleigh-Jeans
tail is relatively insensitive to changes in interstellar extinction
-- particularly towards SPHEREx's $5\mu$m upper wavelength limit --
SPHEREx's spectrophotometric coverage of this wavelength range enables
one to measure directly the extinction in bluer bands and better
constrain the bolometric flux and effective temperature for stars with
$T_{eff}$ $\gtrsim 4600$~K; hotter stars can be studied by the
addition of Gaia.  A very important by-product of the improved
stellar parameters derived as described above is a consequence of the
fact that our knowledge of the masses, radii, and densities of
exoplanets discovered via the transit and radial velocity methods is
no better than our knowledge of the masses and radii of the host
stars.  This is addressed further in the discussion below concerning
synergy between Spitzer and TESS. 

Finally, the overlap in coverage between SPHEREx and \emph{Gaia}
spectrophotometry provides a handle on the systematic uncertainties
affecting each data set's flux measurements. With overlapping flux
measurements in the 750 --1050~nm range, systematic 
uncertainties in the SPHEREx and \emph{Gaia} data should introduce
relative offsets in the flux measurements across this range, enabling
us to quantify and correct for these systematics.  

\subsection{The Best and Brightest Metal-poor Stars with SPHEREx}

\textbf{The Brightest Metal-poor Stars:} Extremely metal-poor (EMP) stars
are very important for near-field cosmology and Galactic archaeology.
They constrain the high redshift initial mass function (e.g.,
\cite{Yoshida:2006}), the nucleosynthetic yields and explosive deaths
of Population III stars (e.g., \cite{Heger:2010}), the production of
lithium from Big Bang nucleosynthesis (e.g., \cite{Ryan:1999}), as well
as the formation and galactic chemical evolution of the Milky Way
(e.g., \cite{Cayrel:2004}). 

\begin{figure}[!th]
\center
\includegraphics[width=0.90\textwidth]{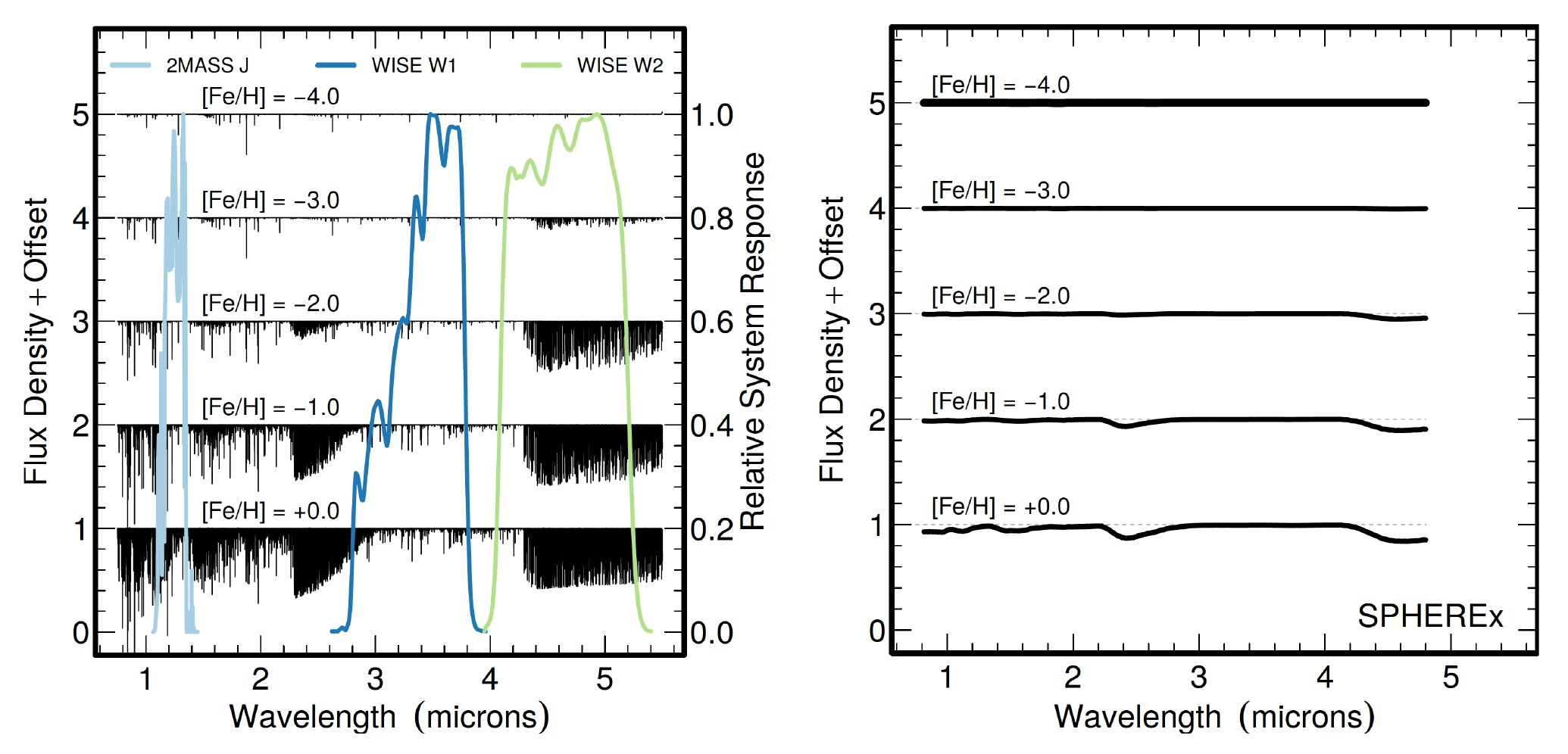}
\caption{\small Illustration of the effect of metallicity on
the spectra of giant stars.  Metal-rich stars have strong
molecular absorption at 2.3 and 4.6 microns, features that are
absent in the spectrum of a metal-poor star.  \textit{Left}:
the black curves are theoretical spectra with slope removed from the Gaia grid of
\cite{Brott:2005} for five K stars with $T_{\mathrm{eff}} = 4800$
K and [Fe/H] as indicated (the features are insensitive to $\log{g}$)
along with relative system response curves (RSRs) for the 2MASS $J$, WISE
$W1$, and WISE $W2$ bands.  \textit{Right}: the same spectra as before,
now smoothed to SPHEREx resolution.  While the infrared selection using
only 2MASS and WISE can only make use of the 4.6 micron feature, SPHEREx
would permit the use of the entire stellar SED -- including the 2.3
micron feature.  For that reason, SPHEREx data would enable a metal-poor
star selection that is nearly 100\% efficient.}
\label{fig:metal}
\end{figure}

However, current progress on the study of these ancient stars is
being limited by their faint apparent magnitudes.  High-resolution and
high signal-to-noise spectroscopy is necessary to measure the stellar
parameters and detailed abundances of EMP stars at visible wavelengths.  Because most known
EMP stars are faint, acquiring even a single spectrum can require
hours of exposure time even with 6--10 m telescopes.  These long
integration times make the construction of large samples of genuine
EMP stars prohibitively expensive, leaving their enormous scientific
potential tantalizingly close yet just out of reach.  Recently,
\cite{Schlaufman:2014} showed that infrared photometry can be used to efficiently
identify candidate metal-poor stars through their lack of strong
molecular absorption at 2.3 and 4.6~$\mu$m (Fig.~\ref{fig:metal}).  While the
current generation of this infrared selection using 2MASS and WISE data
is as efficient as other techniques used to identify EMP stars, more
than half of the candidates it identifies are not metal-poor stars.
It is also unable to identify the brightest and most easily studied
metal-poor stars, as WISE saturates for targets with $W2 \lesssim 8$.

SPHEREx will address both issues with the current generation of 
infrared selection of EMPs.  The current selection derives most of its power
from $W1-W2$ color, which measures only the strength of absorption near
4.6~$\mu$m.  The main false positives for the 2MASS- and WISE-based
selection are young stars with infrared excesses produced by disks of hot
dust, which can mask the molecular absorption present near 4.6 microns
in the atmospheres of metal-rich stars.  However, unlike the lack of
molecular absorption characteristic of metal-poor stars, hot dust would
also manifest itself as excess emission over a range of wavelength.
SPHEREx would easily differentiate between the two possibilities and
the SPHEREx spectrum would provide additional information (see Fig.~\ref{fig:metal}).
The saturation limit for SPHEREx is four magnitudes brighter than WISE,
so infrared data from SPHEREx could be used to find all of the brightest
undiscovered metal-poor stars in the solar neighborhood.

\textbf{The Most Ancient Stars in the Milky Way:}  The most ancient stars in
the Milky Way are thought to be the metal-poor stars in the inner Galaxy.
(e.g., \cite{Tumlinson:2010}). Already, two of the confirmed metal-poor stars from
\cite{Schlaufman:2014} with $\mathrm{[Fe/H]} \lesssim -2.7$ are within the
central 2 kpc of the bulge.  One $\mathrm{[Fe/H]} \lesssim -3.0$
star is approximately 4 kpc from the Galactic Center.  These three
stars are among the most metal-poor stars yet found in the bulge,
and there is a 70\% chance that at least one formed at $z \gtrsim 10$
\cite{Casey:2015}.  Because of the significant reddening towards the
bulge, infrared data is necessary to find these stars.  SPHEREx will
obtain spectra of hundreds of extremely metal-poor giants in the inner Galaxy, and
their detailed abundances will provide the strongest constraints yet
on the chemistry of the Milky Way and its progenitors at $z \gtrsim
10$. This is a good example of the power of the SPHEREx all sky data
base.

\textbf{The Mass of the Milky Way:}  The total mass of the Milky Way is
one of the most important parameters to models of Galaxy formation and
evolution.  It is also difficult to measure.  Accurate mass measurements
require distant tracers of the halo, like metal-poor giant stars.
Such tracers are rare and apparently faint, so blind searches are
extremely inefficient.  On the other hand, SPHEREx data could be used to
identify distant metal-poor giants at high Galactic latitude with nearly
100\% efficiency.  Once those stars are known, Gaia proper motions and
ground-based radial velocities would provide the necessary input for
the most precise mass measurement of the Milky Way that will be
possible for some time. 

\subsection{Nearby Cool Stars and Brown Dwarfs}
\label{sec:nearbycool}

SPHEREx will provide complete 0.75-5.0~$\mu$m spectra of many
thousands of low-mass stars and many hundreds of brown dwarfs. For known objects,
having such a complete spectral library will enable the computation of
accurate bolometric luminosities and effective temperatures, the
testing of model atmospheres over a wide range of wavelengths sampling
different atmospheric physics, and the identification of objects with
clear departures from solar C/O ratios. Although SPHEREx will clarify
our understanding of known low-mass stars and brown dwarfs, its
discovery aspects are equally exciting. Using spectroscopy alone,
SPHEREx will enable us to identify rarer objects -- low-metallicity L
and T dwarfs, low-gravity (young) brown dwarfs in nearby moving
groups, and perhaps even the occasional, overlooked cold Y dwarf --
while not having to resort to either the color or kinematic biases
present in all previous surveys for these objects. 

\begin{figure}
\includegraphics[width=\textwidth,angle=0]{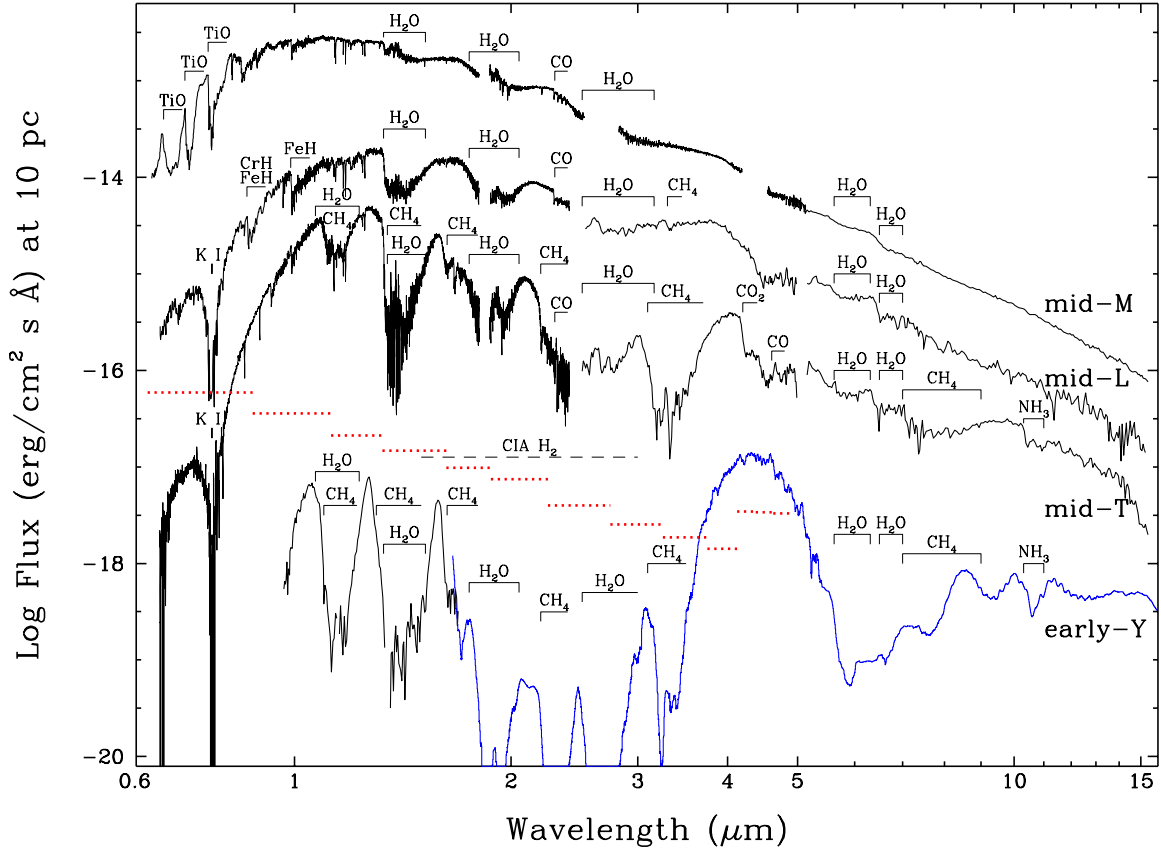}
\caption{Representative spectra of late-M through early-Y dwarfs from
  0.6 to 15 $\mu$m, adapted and updated from Figure 12 of
  \cite{Cushing:2006}. All spectra have been scaled to a distance of
  10   pc. Black spectra are actual observations taken from SDSS and
  Keck/LRIS for the optical, IRTF/SpeX and HST/WFC3 for the $JHK$
  bands, IRTF/SpeX and AKARI/IRC for the $LM$ bands, and {\it
    Spitzer}/IRS for wavelengths longer than 5 $\mu$m. Blue spectra
  are taken from Lyon models for wavelength ranges not yet observed
  for the Y dwarf. Major absorption features are labeled. Shown in red
  are the SPHEREx 5$\sigma$ detection limits at the $I$, $J$, $H$,
  $K$, W1, and W2 bands. 
\label{finder_chart}}
\end{figure}

Fig.~\ref{finder_chart} summarizes our current state of knowledge. Late-M dwarfs are
dominated by bands of TiO and H$_2$O; L dwarfs by hydrides, H$_2$O,
and CO; T dwarfs by CH$_4$, H$_2$O, and collision induced absorption
by H$_2$; and Y dwarfs by NH$_3$, CH$_4$, H$_2$O, and H$_2$. SPHEREx
makes available, largely for the first time, the window between 2.5
and 4.8 $\mu$m that is extremely difficult to observe from the
ground. In this range, fewer than two dozen late-M dwarfs have
observed spectra, and SPHEREx will provide $\sim$10,000 more. For L
dwarfs, SPHEREx will increase the number from $\sim$20 to $\sim$1,000,
and for T dwarfs the total will go from $\sim$10 to $\sim$200. Y
dwarfs will be difficult to detect, but SPHEREx should nonetheless
provide spectra of $\sim$9 of the 25 known examples, at least near
their 4.5-$\mu$m peak; others may be detected by coadding spectral
channels to increase the sensitivity. 

SPHEREx provides complete spectra of known stars with $T<$ 4500K over the
wavelength region where most of the flux emerges. Combined with {\it
  Gaia-}, {\it Spitzer-}, and ground-based parallaxes, these data
provide an unprecedented sample with which to accurately measure
absolute bolometric luminosities that, when paired with estimates of
(sub)stellar radii, can be used to measure effective temperatures,
providing an independent check of model atmosphere fits (e.g.,
\cite{Dupuy:2013}). These theoretical models can also be compared to 
SPHEREx spectra to test whether clouds best describe the emergent
spectra, as has been the standard paradigm, or whether fingering
convection models \cite{Tremblin:2016} do a more adequate
job. Also, these spectra can be used to check for the presence of
non-equilibrium chemistry -- such as the co-existence of CO and CH$_4$
-- which gives insights into vertical mixing \cite{Saumon:2003},
and to check at a fixed spectral type for object-to-object departures
from solar abundance ratios, as some of the data from AKARI have
suggested \cite{Sorahana:2013}.

SPHEREx is also capable of identifying new low-mass stars and brown
dwarfs of particular interest. Low-metallicity L and T dwarfs provide
excellent checks of exoplanet atmospheric theory because these
atmospheres are cold yet very simple, as they lack the complex
chemistry of more metal-rich objects
\cite{Kirkpatrick:2014,Kirkpatrick:2016}. These objects can be easily
selected based on their extremely blue colors across the $JHK$ bands; the blue colors are caused by the
increased relative importance of collision-induced absorption by
H$_2$, which is strongest near 2.2 $\mu$m. Young, low-gravity L dwarfs
also provide excellent checks of exoplanet models because these
objects have extremely low masses and solar composition. As with the
low-metallicity objects, these young objects also have unique spectral
signatures because of their extremely low gravity \cite{Faherty:2013}. Finally, there is also the possibility that SPHEREx will
identify new, very nearby Y dwarfs that WISE has imaged but not
selected due to a poorly measured W1$-$W2 color limit. As
Fig.~\ref{finder_chart} suggests, these objects will appear in SPHEREx data only at the
longest wavelengths and should be identifiable based on their unusual
spectral signature combined perhaps with their motions as measured by
SPHEREx or possibly from the ground.(see, e.g., the proper motion
discovery of the cold Y dwarf WISE 0855$-$0714 with {\it WISE} by
\cite{Luhman:2014}).

\subsection{Stars with circumstellar material: variability of extreme
  debris disks}
\begin{figure}
\center
\includegraphics[width=0.6\textwidth]{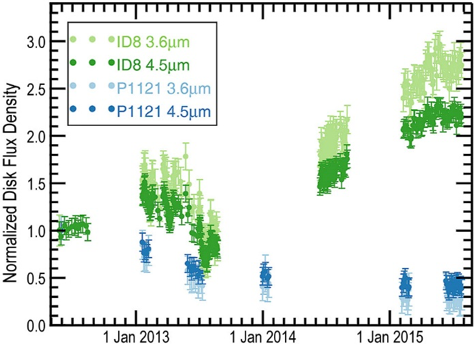}
\caption{Spitzer data identify other planetary systems that may be
  undergoing collisional events similar to our Late Heavy
  Bombardment. Plot shows IRAC flux versus time for two of the extreme
  debris disk systems monitored by \cite{Meng:2015}. Both stars are
  members of open clusters, with ages estimated as 35 and 80 Myr.} 
\label{spitzer_v1}
\end{figure}

The collisional process that creates planets leaves behind debris in
the form of planetesimals and dust (extrasolar analogs of asteroids
and comets on the one hand and the zodiacal and Kuiper Belt dust on the
other) as illustrated in Fig.~\ref{fig:nir_debris}. Debris disks around  main sequence stars were discovered by
IRAS, and were immediately recognized as indicators of at least the
first steps in the planet formation process, because the dust
particles they contain are much larger than interstellar grains.  In
addition, the particles in debris disks have only finite lifetimes and
must be replenished, presumably by collisions between rocky bodies and
evaporation of comets orbiting the stars.  Debris disks have been
extensively studied by infrared space facilities and, with the
prevalence of exoplanets demonstrated by Kepler, we study debris disks
not only because of their intriguing properties but also because they
hold clues to the nature of the exoplanetary systems which they occupy.  
Debris disk studies have concentrated on wavelengths longward of
$\simeq$ 10 $\mu$m because the disks generally appear to contain dust no warmer than
$\simeq$ 300K and because the contrast relative to the stellar photosphere
increases with wavelength.   However, recent Spitzer observations of extreme debris disks at 3.6 and 4.5 $\mu$m have returned striking
results which SPHEREx will build upon.  Extreme debris disks are systems
with $L_{dust}/L{\star} \simeq$ 0.01, about two orders of magnitude greater than
what had previously considered to be a bright debris disk.  The
brightness of the debris disk simply reflects the amount of dust
orbiting the star, so these systems are unusual in having both a large
amount of circumstellar dust and having dust warm enough to radiate
significantly shortward of 5~$\mu$m.  Several such disks have been
identified orbiting stars younger than $\simeq$ 100 Myr, and found to
have one other unusual property, which is that their infrared radiation shows
marked variability (Fig.~\ref{spitzer_v1}) on $\simeq$ 1 year time
scales.  At the same time, the optical radiation from the stars shows
little or no variability, so the infrared variability must result from
changes in the amount of dust orbiting the star, or in the visibility
of the dust. 

\begin{figure}
\center
\includegraphics[width=0.4\textwidth]{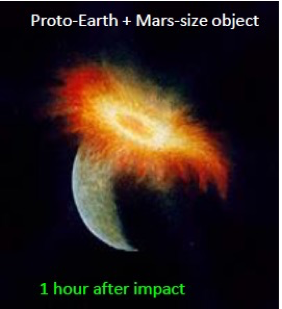}
\caption{Artists conception of a collision between the Proto-Earth
  and a Mars-size body.  It is thought that a collision like this when
  the Earth was only $\sim$50~Myr old may have produced the Moon.  Dust
  produced in collisions close to this in this scale appear to be
  required to account for the variability of the infrared emission
  from extreme debris disks.} 
\label{spitzer_v2}
\end{figure}

\begin{figure}
\includegraphics[scale=0.65,angle=270]{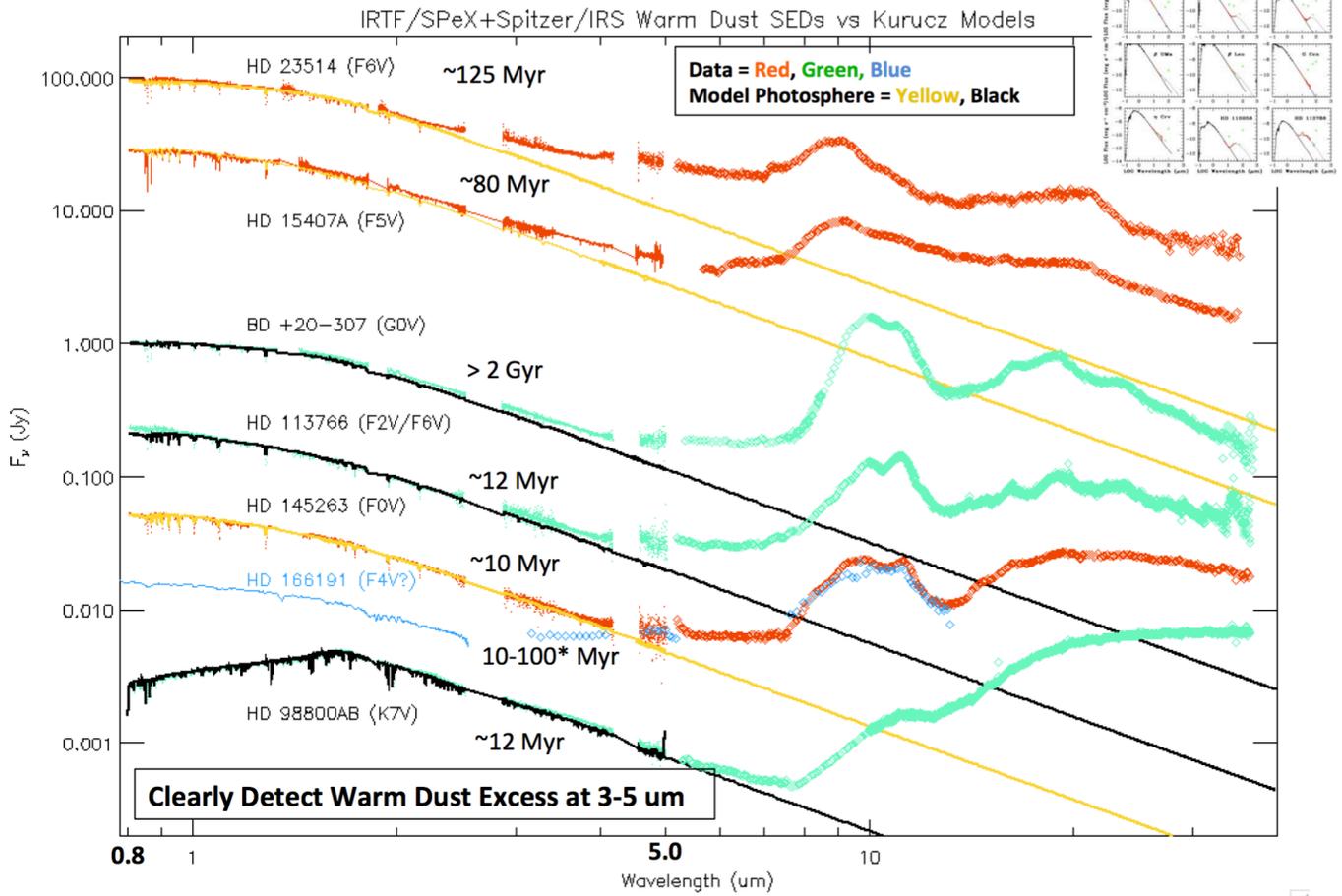}
\caption{SPHEREx has great potential for measuring 1-5 $\mu$m spectra
  of debris disks. Shown are NASA/IRTF R$\sim$1000 ground based
  spectra of well-known disks combined with Spitzer 5-35 $\mu$m space based spectra to make combined
  SEDs that span wavelength ranges dominated by stellar photospheric
  emission, scattered light from circumstellar material, and thermal
  emission from circumstellar dust. SPHEREx will be able to detect
  excess emission due to warm or hot (T $>$ 200 K) circumstellar dust,
  as seen around HD23514 and HD16191, for example, and produced by
  collisional grinding and giant impacts, as well as excesses due to CO (and
  possibly H$_2$O) gas found in abundance in very young  systems.}  
\label{fig:nir_debris}
\end{figure}

These and other extreme debris disks vary much more rapidly
than predicted for evolutionary models in which the dust in debris disks result from a gradual collisional cascade.  Instead, the
authors postulate that the variability results from major collisions
between two large bodies in which some of the colliding materials
might actually be vaporized and then condense rapidly into small
particles, degraded collisionally into the particles seen in the
infrared, which are blown out of the system but not replenished as
they would be in the case of a gradual collisional cascade (e.g., such
as the silica-rich but photometrically stable HD 172555 disk system,
\cite{Lisse:2009, Johnson:2012}).  The star 
ID8, in the young (35 MYr old) cluster NGC2547, hosts the most
dramatically variable debris disk (Fig.~\ref{spitzer_v1}).  Meng et al. estimate the mass of
the disk to be at least that of a $\simeq$ 180km diameter rocky asteroid (Fig.~\ref{spitzer_v2}).

As illustrated in Fig.~\ref{spitzer_v2}, truly titanic collisions may
be needed to produce the dust clouds required to account for the
variability of the extreme debris disks.  Several such collisions are
thought to have taken place in the first $\simeq$ 100 Myr of our Solar System,
and others may have been triggered in tandem with the Late Heavy
Bombardment (LHB) which caused comets and asteroids to rain down on the
terrestrial planets when the Solar System was 600 to 800 Myr old,
leaving the plethora of craters seen on the Moon today (see Sec.~\ref{sec:trojan}
for more detail of the LHB). At least one disk, that surrounding the
$\sim$1.4 Gyr old F2V star Eta Corvi, is thought to have been created by an
LHB  \cite{Wyatt:2005,Lisse:2012}), and the extreme Kepler
lightcurve system KIC 8462852 may be just starting one, as the best
explanation for the time variable spectrophotometric observations of the
system remains the close passage and breakup of a $>$ 100 km radius KBO
by the primary star \cite{Boyajian:2015,Marengo:2015,Lisse:2015}. 

It has been suggested  that such extremely dusty and perhaps variable stars may constitute up 1\% of young  stars in the
solar neighborhood \cite{Kennedy:2013}. SPHEREx has the sensitivity to
detect many of them and an appropriate time cadence to assess their
variability.  Of course, a main sequence star which shows excess
emission due to dust shortward of $\simeq$5 $\mu$m will be interesting
even it is not observed to vary, as the short lifetime of the dust
suggests an active exoplanetary system even if it is not variable over
one or two years \cite{Lisse:2013}. Particularly interesting candidates could be identified for JWST follow up while
SPHEREx refines our understanding of the statistics of their
occurrence.  The SPHEREx census will test the importance and frequency of
such violent events in the early stages of the planet formation epoch
and also determine whether the history of our own Solar System stands
out or fits in with that of many other stars. 

\section{Solar System Science}

\subsection{Executive Summary} 

SPHEREx has high potential for solar system science.  Numerous
solar system objects will be in the viewing zone during the
2 year lifetime of the survey. In fact, not only will these objects be
available for scientific analysis, but many, mainly asteroids, will
have to be removed from the data for SPHEREx to perform
the primary astrophysical science goals of the mission. By canvassing
the entire solar system for 2 years, SPHEREx has the
potential not only to achieve a relatively complete sensitivity
limited survey of the solar system's bodies, but also some capability to search for
variation in these bodies over time. SPHEREx will also map
in great temporal and spatial detail the zodiacal dust debris disk
cloud that these bodies produce, providing an unprecedented level of
information concerning the sources and sinks of this material. 

One way to look at SPHEREx's solar system science potential is to note
that its ultimate sensitivity per spectral channel is about the same
as the WISE prime mission's. I.e., SPHEREx will conduct the solar
system surveys ``\'a la'' WISE, but in 96 spectral channels, not 4
photometric bands. Based on the WISE results \cite{Bauer:2013,
  Bauer:2015,Mainzer:2011a,Mainzer:2012,Mainzer:2014,Masiero:2011,Masiero:2012,Nugent:2015},
we can expect a highly significant return for the small bodies of the
solar system in terms of tens of thousands of 0.8 - 5.0 $\mu$m, R
$\simeq$ 41. spectra of asteroids, of thousands of Trojan asteroids, of
hundreds of comets, and of several of KBOs, Centaurs and Scattered
Disk Objects (SDOs) 



Another way to look at SPHEREx's solar system science potential is to
realize that it will be mapping the zodiacal light over the whole sky
in a sun-synchronous fashion very similar to COBE/DIRBE, but with
updated detectors, a 6.2" x 6.2" pixel size (rather than 42' x 42'), and
with 96 spectral channels from 0.75 - 5.0 $\mu$m rather than 4 broad
photometric channels. The 2 year survey lifetime of SPHEREx will span
about the same length of time as the DIRBE cold + warm eras, but the
finer spatial and spectral resolution should produce much finer maps
of the cloud's structure in the asteroid belt and along comet trails
and near the planets, improving on the findings of
\cite{Franz:1996,Kelsall:1998,Reach:1997} and permitting searches for 
compositional signatures and variations within the zodiacal dust.   
 
\subsection{Major Solar System Science Programs} 

Below we list, in priority order, a number of key SPHEREx Solar System Science
programs. For each program we will discuss, below the main science
driver, the number of objects SPHEREx can expect to
detect, and any limitations the SPHEREx survey may encounter.   

\begin{enumerate}
\item Spectral survey of thousands of asteroids: Better reckoning of
  the origin and dynamical evolution of various types based on their
  reflectance spectra \& sizes/albedos from their thermal
  emission. Were they formed in layered zones of the protoplanetary
  disk and then scattered?  
\item Spectral classification of 100's of Trojan asteroids: Compare
  to Outer Main Belt. Were they formed with Jupiter, captured from the
  Outer Main Belt, or delivered in the LHB? 
\item Spectral determination of cometary chemical abundances, as
  tracers of the composition of the proto-planetary disk. Compare KB comets vs
  Oort Cloud Comets. Is CO$_2$ the dominant molecule in  carbon-ices in comets, and the fundamental carbon bearing molecule
  in the PPD? And by extension in all molecular clouds and hot cores?
  Finally, how do the comet sub-populations differ compositionally?
\item Spectral mapping of the zodiacal cloud in time and space. What
  produces this cloud, asteroids, comets, and/or KBOs? 
\item "Follow the Water" and other key species (CO, CO$_2$, Organics
  [CH$_4$, C$_2$H$_6$, HCN, CH$_3$OH, H$_2$CO, PAHs], NH$_3$)
  throughout the Solar System, as critical resources for life.   
\end{enumerate}

\subsection{SPHEREx Asteroid Spectral survey}  

\begin{figure}[!t]
\center
\includegraphics[width=0.7\textwidth,angle=90]{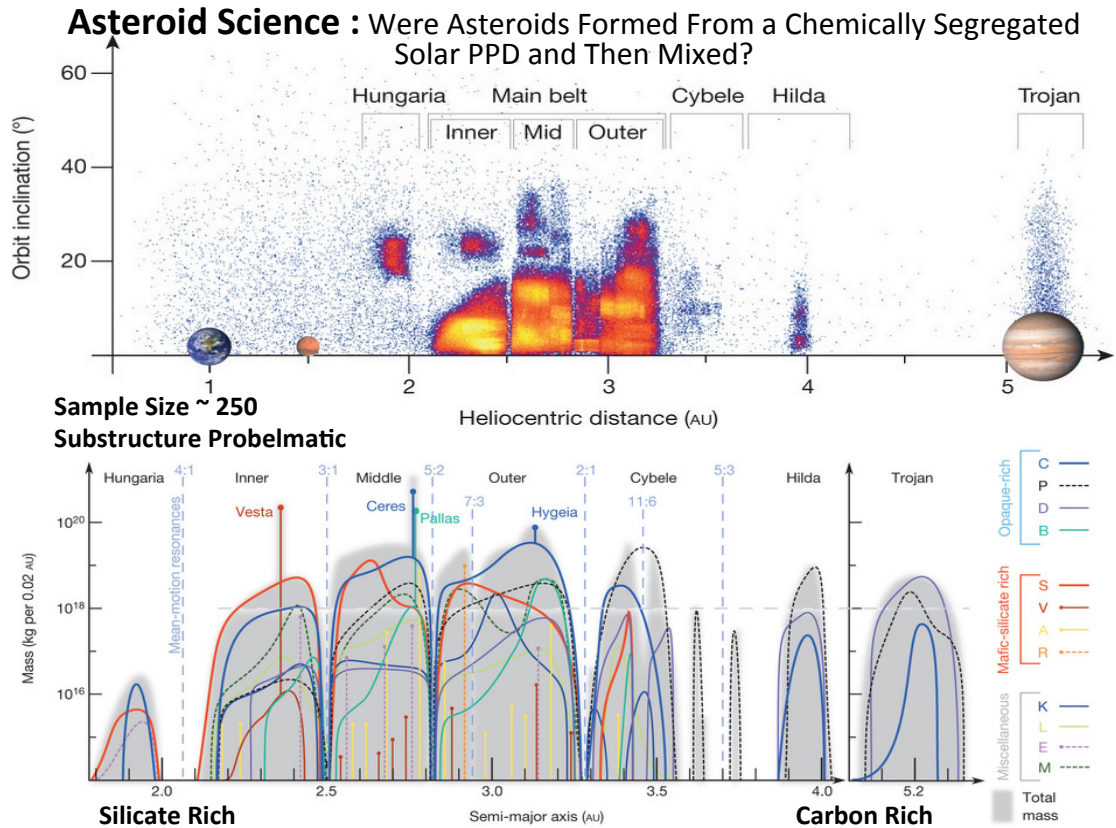}
\caption{Relative distribution of different asteroidal compositional
  classes with respect to distance from the Sun according to
  \cite{demeo:2014}. The composition of a asteroids has been
  determined using 0.8 - 2.5 $\mu$m spectroscopy obtained at the NASA/IRTF
  observatory on Mauna Kea. Overall trends for the largest asteroids
  show a concentration of the most refractory rocky material closest
  to the Sun, with more volatile carbonaceous materials becoming more
  abundant in the farthest reaches of the main asteroid belt.  There
  is an admixture of smaller bodies cutting across these general
  trends, though, presumably due to collisional grinding followed by
  impact redistribution of small fragments of the largest bodies.} 
\label{fig:lisse4}
\end{figure}

A major question in solar system formation science today is how and where the current
population of asteroids formed. In the last decade the formation of iron
meteorites, remnants of planetesimal cores, have been dated to within
1 to 3 Myr after  meteoritic CAI's ("Calcium Aluminum Inclusions",
mineral bits composed of the stablest and most refractory metal oxides
that form first out of a cooling solar abundance mixture) formed the oldest
known materials in the solar system. This has been interpreted as strong
evidence for a "top-down" formation of large Vesta-sized asteroidal
bodies in the innermost regions of the solar system, which then
underwent collisional disruption over the next 10-30 Myr to produce
collections of metallic (from the core), stony-iron (from the mantle),
and stony (from the crust) fragments that re-accreted into the
asteroid families we see today. Many of these bodies were absorbed in
the making of the terrestrial planets. Those that weren't underwent
further collisional grinding, to form the smaller bodies of the
asteroid families we know today. Recent spectral survey work of
$\simeq$ 300 asteroids at 0.8 - 2.5 $\mu$m by \cite{demeo:2014} has
shown that the asteroid belt also appears to be zoned by distance from the Sun, with
the largest bodies keeping the signatures of their formation location:
rockiest in the inner belt, mixed in the center regions, and most
carbon and water-rich in the outer regions. 

WISE performed an all-sky asteroidal survey from 2010 - 2015,
detecting $\sim$ 200,000 asteroids and determining their size, albedo,
and color frequency distributions. SPHEREx will be able to spectrally
characterize a large number of the WISE asteroids and determine their makeup
as a function of orbital location, bridging the gap between the WISE and the DeMeo
surveys \cite{demeo:2014}.  

SPHEREx will need to perform this asteroid survey for another reason -
given the NIR brightness frequency distribution of the asteroid
population and its extent across the sky, there is a few \% chance in
a SPHEREx observation of any given sky pixel that there will be a
significant asteroidal contribution to its measured flux. To make
robust measurements of extra-solar system objects, the SPHEREx team
will thus have to remove any foreground asteroid
contribution, naturally building up an asteroid spectral survey.
Fortunately, at SPHEREx sensitivity levels most of the asteroids
detected will be previously cataloged objects and asteroids would
appear as unusual rogue, variable sources in SPHEREx' redundant surveys. 

\begin{figure}[!t]
\center
\includegraphics[width=0.7\textwidth,angle=90]{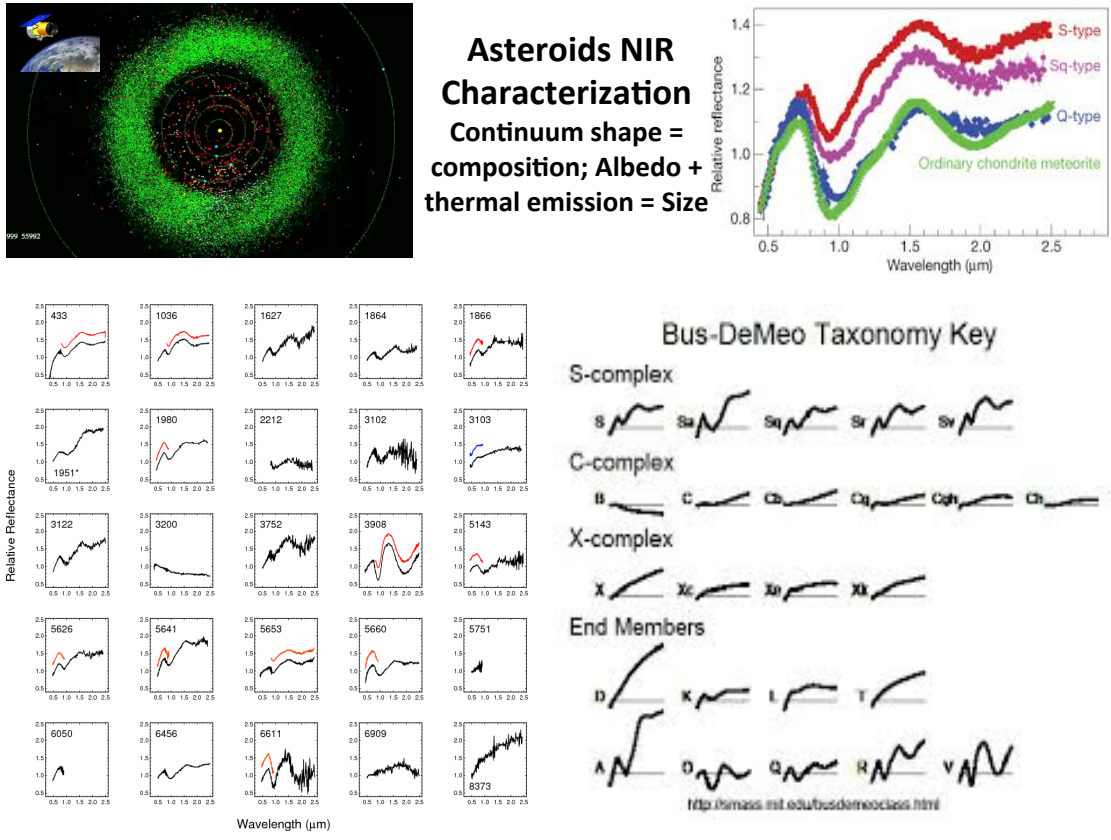}
\caption{Illustration of how asteroids are compositionally classified
 using near-infrared reflectance spectroscopy. Shown, starting in the
 upper left and moving counterclockwise are: the WISE spatial
 distribution of main belt asteroids (MBAs); 0.8 - 2.5 $\mu$m spectra for
 25 asteroids taken from the NASA/IRTF 3m on top of Mauna Kea;
 Bus-DeMeo \cite{Demeo:2009} spectral templates for different classes of asteroids,
 referenced to laboratory measurements of meteorite samples; and in
 the upper right hand corner an detailed comparison of 3 closely
 related rocky asteroids, showing the subtle differences in the
 olivine and pyroxene absorption features that sets them apart.  The
 dominant distinguishing spectral reflectance features observed are
 the 1.2 $\mu$m and 1.9 $\mu$m olivine absorptions, the 1.3 $\mu$m pyroxene
 absorption, and the broad shallow 0.8 - 2.5 $\mu$m continuum reddening
 due to carbonaceous material.} 
\label{fig:lisse5}
\end{figure}

Even though SPHEREx can easily achieve WISE-like sensitivity levels on
a fixed target, asteroids may pose challenges because of their motion
and rotation in combination with the piecewise way in which SPHEREx
will build up a spectrum.  Thus we have conservatively estimated that
SPHEREx will return useful data on tens of thousands, and not $\sim$2 $\times
10^5$ asteroids, as WISE has done. Even so, robust spectral
characterization from 0.75-to-$\sim$5$\mu$m of tens of thousands
of asteroids will be a major scientific advance over the $\sim$500
asteroids measured over the last 3 decades. 

\subsection{Trojan \& Greek Asteroid Survey} 
\label{sec:trojan}

\begin{figure}[!t]
\center
\includegraphics[width=0.7\textwidth,angle=90]{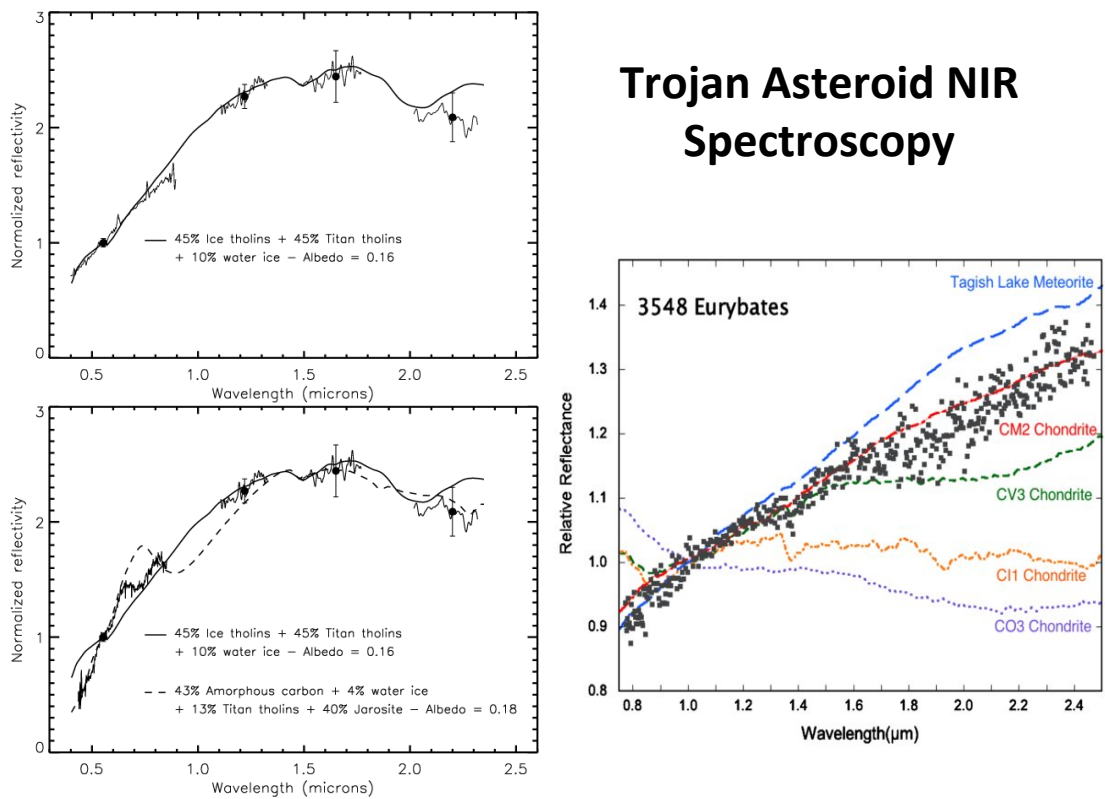}
\caption{\emph{Left}: Example of near-infrared spectral characterization of
Trojan and Greek asteroid composition. The spectra are in some ways similar
to the outer MBA spectra, as they are dominated by features due to
large amounts of very primitive carbonaceous material (here modeled by
laboratory grown "tholins"), but as they are located at $\sim$ 5.2 AU
outside the water ice line, they also show important features due to
water ice and highly hydrated minerals (e.g., Jarosite,
KFe$^{3+}_3$(OH)$_6$(SO$_4$)$_2$) not seen in the MBAs. \emph{Right:}
Comparison of one of  the larger Greek asteroids' reflectance
spectrum, 3548 Eurybates, to meteorite fall spectra. The best match is
to a CM2 spectrum, suggesting that like the Murchison meteorite, these
objects are very primitive and carbon and organics rich ($\sim$ 20\%
by weight) and have experienced extensive alteration by water-rich
fluids ($\sim$ 10\% by weight).}   
\label{fig:lisse2}
\end{figure}

In a similar fashion, SPHEREx will naturally observe and detect asteroidal bodies located in the
L4 and L5 resonances of Jupiter, the so-called Greek and Trojan
asteroids also known as "Jupiter Trojans". Characterizing these objects will be a natural offshoot of
the main belt asteroid survey. Even though their total number is
predicted to exceed the number of MBA's, only a fraction of these
objects have been studied spectroscopically in the
NIR. \cite{Grav:2010} studied 1742 Jupiter Trojan Asteroids using the
WISE spacecraft broad-band photometry, and found 3.4 $\mu$m albedo
differences between C \& P-spectral-type and D-types. SPHEREx will
provide a better understanding of the compositional causes of this
color correlation.  SPHEREx should be able to 
obtain good spectra of hundreds of Trojan and Greek asteroids; they move
across the sky more slowly and predictably than the main belt
asteroids and should present fewer observational problems.  

The importance of SPHEREx information about the composition of these bodies is bound up in our current understanding of
the migrational history of the solar system. The NICE model
\cite{Gomes:2005,Levison:2006} has predicted that the Kuiper Belt was disrupted
some 600 - 800 Myrs after the CAI and iron meteorite formation, when
Jupiter and Saturn moved into a 2:1 resonance which forced an inward
migration while Uranus and Neptune moved outward. This planetary
migration greatly disrupted the orbits of the Kuiper Belt
planetesimals, scattering some 99\% of them inward or outward in the
solar system and creating the system-wide Late Heavy Bombardment,
while sweeping up a small fraction ($\sim$100 are known today) into mean
motion resonances with Neptune (e.g., the Plutinos). It is thought
that while this was happening, some of the KBOs scattered inward would
be captured into the stable Lagrangian points around the giant
planets. Thus comparing the makeup of the Jovian Trojan population to
what is known about KBOs (and other purportedly captured KBOs, like
Saturn's moon Phoebe and Neptune's moon Triton) is an important test
of this model for solar system development. Other possibilities for
the Trojans' sourcing are capture of outer main belt asteroids, or
formation of the Trojans in situ from the PPD as Jupiter
formed. SPHEREx will be able to test each of these hypotheses by
comparing the Trojan spectral results to its MBA spectra
catalogue. SPHEREx would also be able to compare these Jupiter Trojan
spectra to spectra of more distant asteroidal bodies linked with the
KBOs, like the Centaur and SDO populations. SPHEREx is likely to
measure several of these bodies.  

The nature of the Trojan and Greek asteroids is thought to be such an
important issue that in 2015 NASA selected a proposed tour of 5 of
these bodies for a Phase A Discovery mission study. 

\begin{figure}[!t]
\center
\includegraphics[width=0.55\textwidth]{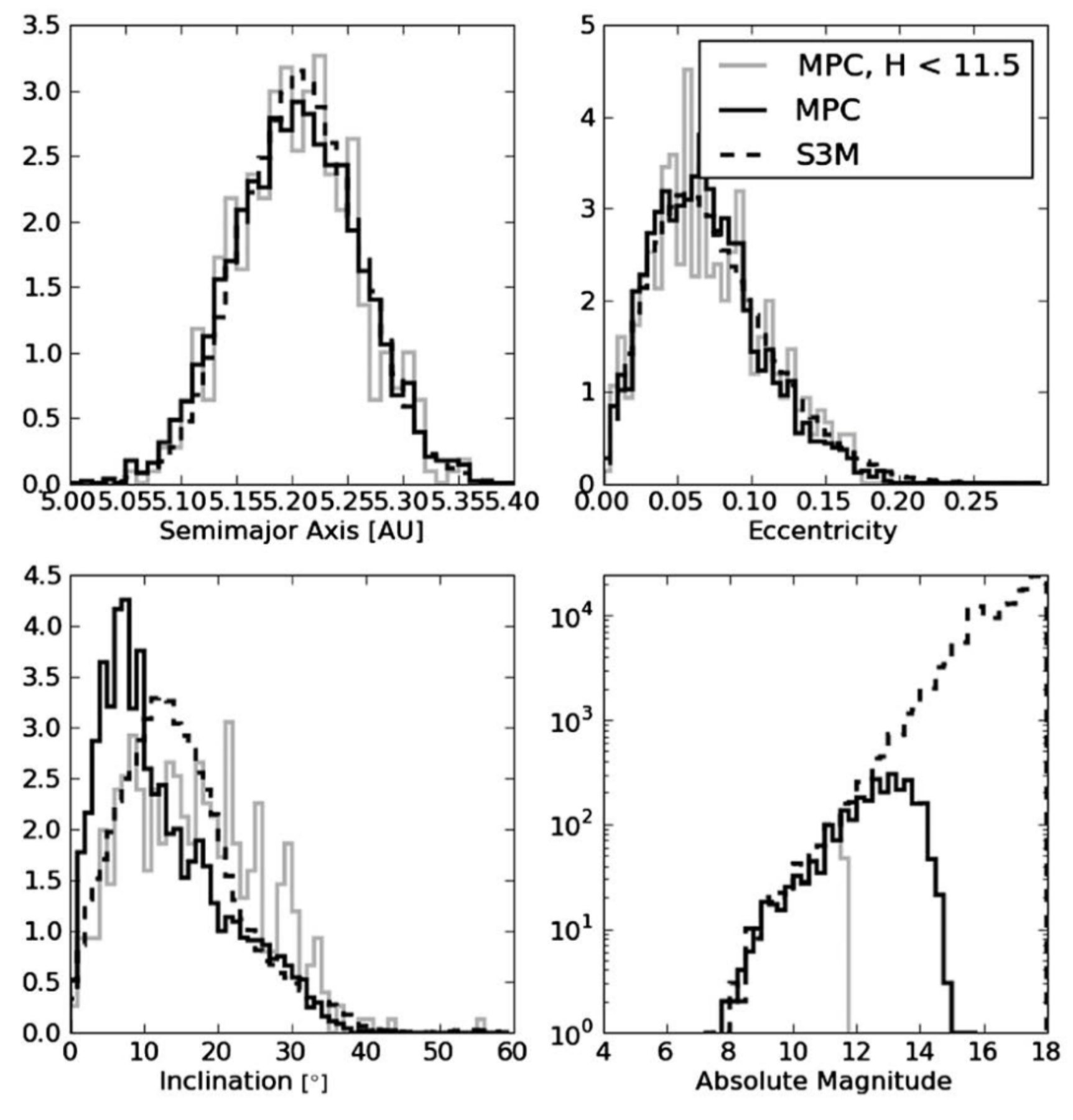}
\includegraphics[width=0.35\textwidth]{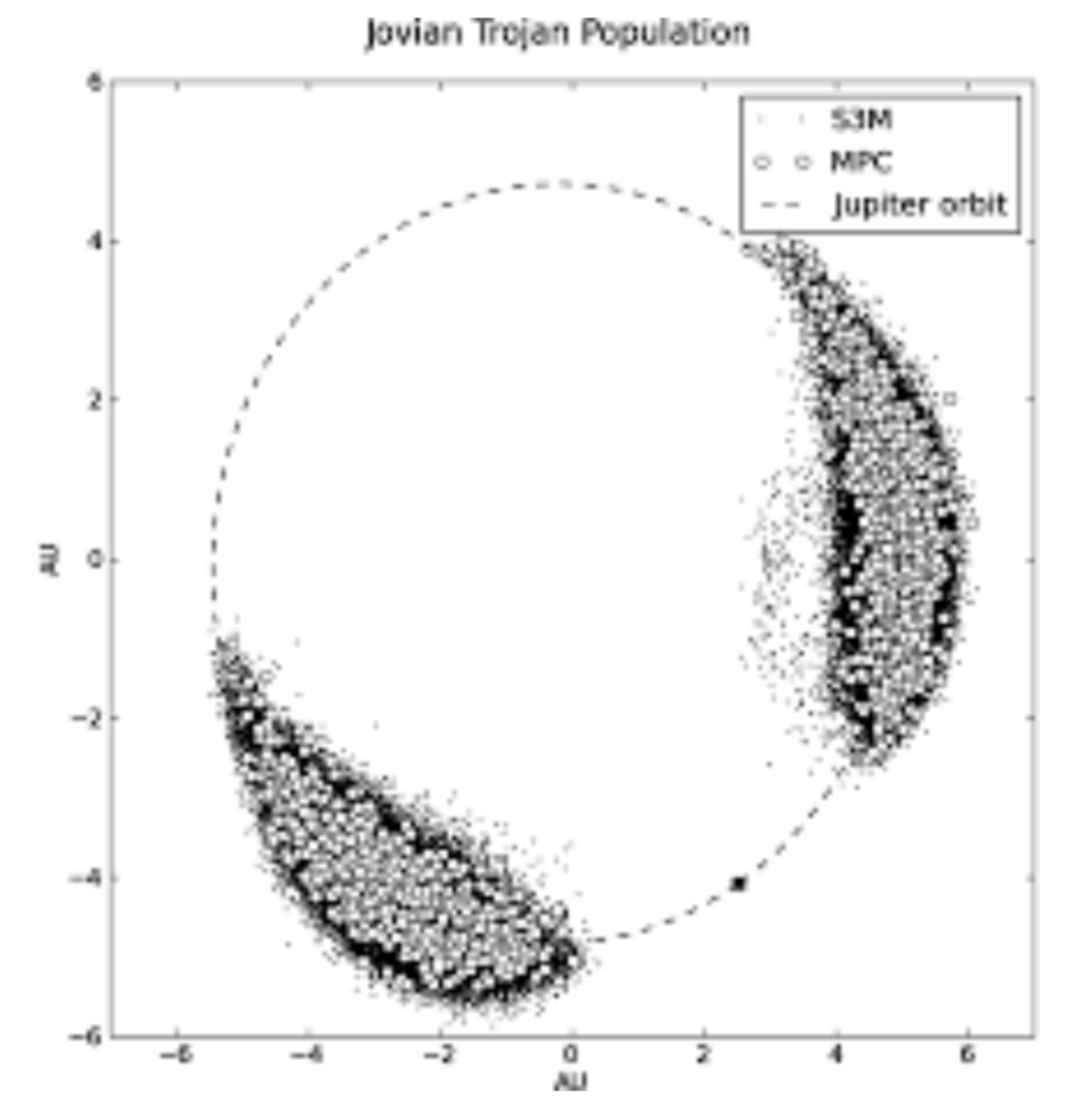}
\caption{\emph{Left:} Estimated brightness and orbital frequency
  distributions for the Trojan and Greek asteroids, after
  \cite{Grav:2010}. There are hundreds of known Trojan asteroids that 
  SPHEREx will be able to characterize spectrally for the first time
  during its 2-year mission. \emph{Right:} Spatial location of the Greek
  asteroids leading Jupiter in its orbit around the L4 resonance
  ($\sim$60 deg.  in the prograde direction) and of the Trojans trailing Jupiter in
  its orbit around the L5 resonance ($\sim$ 60 deg. in the retrograde
  direction). Jupiter is the small dot at (2.3, -4.2) AU. Note that
  the Trojans and Greeks are safely removed from Jupiter and each
  other, so that there is no likelihood of scattered Jovian light
  causing problems with the SPHEREx measurements, while at the same
  time it will require a dedicated survey like SPHEREx to sample
  the spatially extended swarms well.}  
\label{fig:lisse3}
\end{figure}

\subsection{Comet Chemical Abundance Survey} 

\begin{figure}[!t]
\center
\includegraphics[width=0.7\textwidth,angle=90]{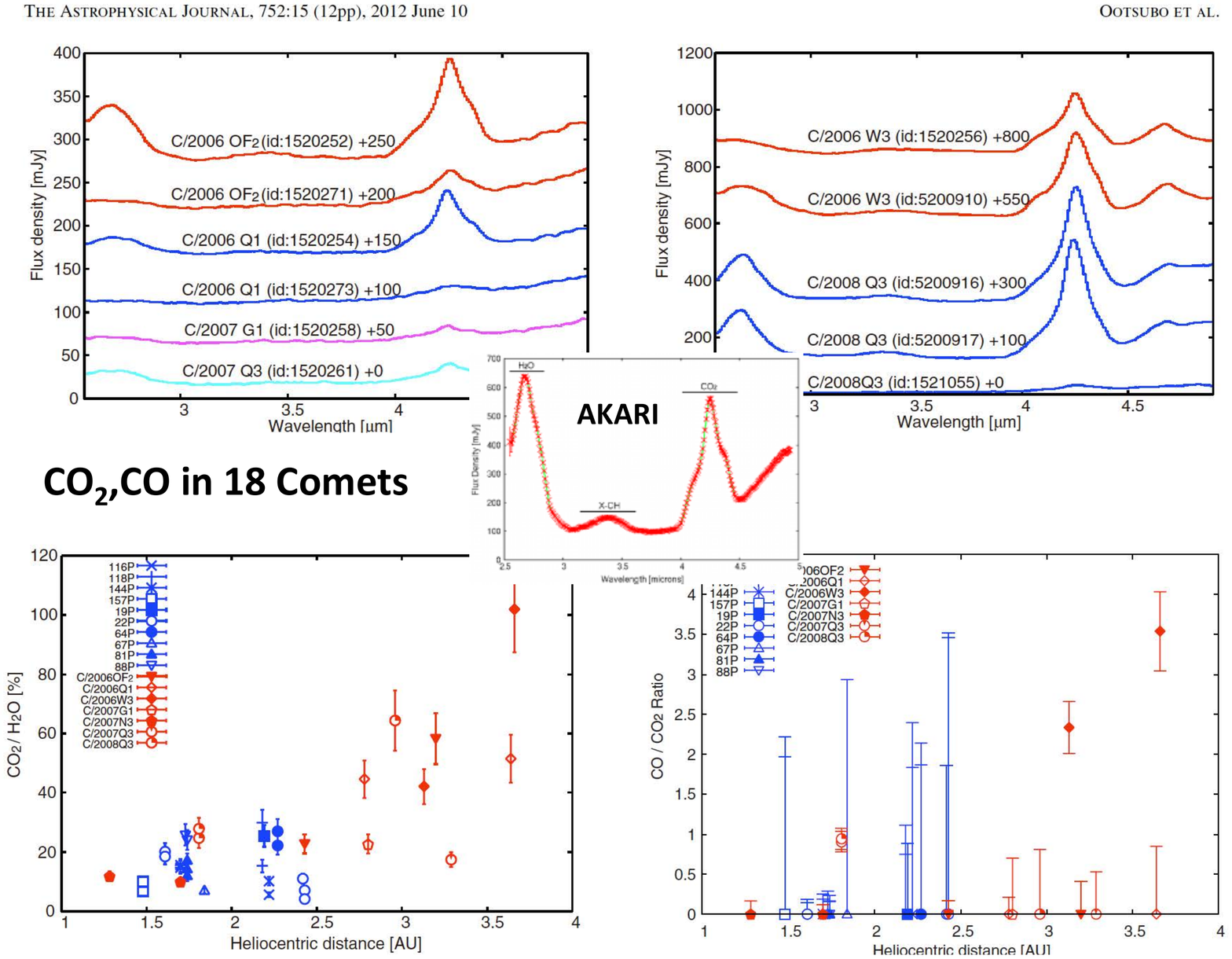}
\caption{(\emph{Top:}) Examples of the quality of cometary spectra
  from SPHEREx, after \emph{Ootsubo et al. (2012)}. Shown are AKARI
  1-5 $\mu$m R$\sim$50 spectra of 12 comets. The areas under each spectral
  feature are directly related to the total amount of the emitting
  species in the AKARI beam. The main features evinced are due to the
  O-H stretch in water and hydroxyl (2.4 - 2.8 $\mu$m), the aliphatic and
  aromatic C-H stretch in organics like CH$_4$, C$_2$H$_6$, H$_3$COH, H$_2$CO
  (methane, ethane methanol, and formaldehyde; 3.2 - 3.6 $\mu$m), the C=O
  stretch in CO$_2$ (carbon dioxide; the doublet from 4.0 - 4.5 $\mu$m
  centered around 4.25 $\mu$m), and the C=O stretch in CO at $\sim$4.7 $\mu$m
  (carbon monoxide). (\emph{Bottom:}) Trends in the estimated
  CO$_2$/H$_2$O and CO/CO$_2$ ratios. A clear trend of rising CO$_2$ vs H$_2$O
  production is seen as the observed comets move outside the water ice
  line at $\sim$ 2.5 AU. However, trends in the CO$_2$/CO ratio are hard to
  distinguish. SPHEREx will be able to greatly improve over AKARI in
  the 4-5 $\mu$m range while observing $\sim$ 100 comets because of its much
  colder optics bench, which will remove the confusion in previous 4.5
  um photometric surveys (e.g., WISE) between flux produced by CO$_2$ vs
  CO.}
\label{fig:lisse1}
\end{figure}

Comets, formed within the first Myr of the solar system's lifetime, are thought to be the
most primitive bodies left over from the Proto-Planetary Disk (PPD)
era. They are leftover relics from the era of 
planetary formation that have failed to aggregate into a planet (or
looked at another way, survived the era of violent early
aggregation). There are two main reservoirs of comets known today in the
solar system: the Kuiper Belt and the Oort Cloud. The Kuiper Belt
comets are small icy bodies that formed at the outer edges of the PPD,
where there was too little mass density to form another planet; looked
at another way, the PPD had to end somewhere, and in its regions of
lowest density planetesimals accretion was slow and truncated. These
objects are relatively well clustered around the plane of the ecliptic
with low inclinations (modulo the objects scattered by Neptune as it
migrated outward during the LHB), and exist in a radial zone extending
out to $\sim$50 AU from the Sun. We observe them when they become scattered
inward into the giant planet region by galactic tides and passing
stars, becoming first Centaurs and then finally short period
comets. The Oort Cloud comets, found in a roughly spherical
distribution from 10$^3$ to 10$^5$ AU, were counterintuitively formed inside
the Kuiper Belt in the giant planet region, as the feedstock for the
nascent giant planets; they represent the population of objects that
had near misses to the growing giant oligarchs in the first 1-10 Myr
of the solar system, and rather than accreting or being thrown into
the Sun or out of the system entirely, they were scattered into highly
elongated, barely bound, Myr orbits. One of the holy grails of comet
science over the last 2 decades has been to search for compositional
differences between the Kuiper Belt and Oort Cloud comets, as a
signature of radial chemical gradation in the PPD.  SPHEREx will
observe large enough numbers of comets to start to fill this data gap,
particularly if more than the expected number of Oort Cloud comets
appear during its 2 year mission. 

The search for PPD chemical signatures in comets has
produced another very interesting result. Cometary bodies are composed
of  $\sim$ 1/2 icy volatiles and 1/2 rocky refractory materials, with
the ices being dominated ($>$ 80\%) by H$_2$O ice
\cite{Crovisier:2004,Mumma:2011,Ootsubo:2012,DelloRusso:2014}. The
most important ice species after water are CO and CO$_2$
\cite{Bockel:2004}, with minor flavoring due to methane, 
ethane, formaldehyde, methanol, and ammonia. Until 2012, when Ootsubo
et al. used the Akari satellite to observe 18 comets from 1 - 5 $\mu$m at
R$\sim$50, it had long been thought that CO was the fundamental C-bearing
icy reservoir; but we now know that while CO$_2$ ranges from 5 to 25\%
vs. water in abundance in comets, CO can vary from 0.1 to 30\%
\cite{Ahearn:1992,Ahearn:2012}. Thus CO$_2$ may be the true leader of
the C-bearing family.

Much work has been done in the last 5 years to quantify the
amount of C-bearing gas in comets. As emission lines from these
species are best detected from space (and CO$_2$ is detectable
only from space) owing to the Earth's atmospheric absorption, space-based
platforms are the best-equipped facilities for characterizing their
production by comets. In addition to Akari,  WISE and Spitzer have
undertaken campaigns to identify CO and CO$_2$ emission in comets
(cf. \cite{Bauer:2015,Reach:2013}), and to characterize their
production as a function of distance from the Sun and of comet orbital
class. However, since both spacecraft detect the molecules' presence
through broad-band photometry, it is not possible to directly 
differentiate between CO and CO$_2$ production, as both lie in the
same ``4.5 $\mu$m'' filter. Furthermore, regarding
comparison to water production, because of the variable nature of
comets and the lack of the ability of these space platforms to
quantify water production simultaneously, placing the combined
production limits of CO and CO$_2$ in relation to the most plentiful
species produced in the inner solar system by comets is not
certain. Even though the water production may be characterized from
the ground, the ground-based observations are rarely simultaneous.
SPHEREx will be able to detect these C-bearing species separately, as
well as simultaneously with water, and other volatiles such as methane
or other organics. The imaging capabilities of the spacecraft will
play a special role in more detailed investigations of particular
species. For example, they will facilitate the accurate measure of the
dissociative scale length of CO and CO$_2$, a feature of the emission
which remains ambiguous (cf. \cite{Bauer:2015,Pittichova:2008}), and
so investigate the presently unaccounted physical processes which cause the shortened scale.  

By performing an unbiased spectral survey of $\sim$ 100 comets (based
on the WISE detection results, \cite{Bauer:2015}) SPHEREx will be able
to directly expand knowledge of the quantity of CO and CO$_2$ in comets
relative to H$_2$O, and so provide compositional constraints on the
PPD in which they formed. SPHEREx will further provide independent
estimates or constraints on the size, albedo, and activity level of
these comets. Finally, there is the potential for synergy with
contemporaneous planetary surveys, such as NEOCam, which have
bandpasses that encompass the 4 micron CO and CO$_2$ features, but
with much greater sensitivity. Such surveys will cover many of the
same comets several times, but at different times and while the comets
are at different heliocentric distances. This vast quantity of comet
detections will also yield size and albedo constraints, but SPHEREx
will provide the ``truth'' data set for the CO and CO$_2$
fractions seen in the thousands of comets the NEOCam survey
could likely detect, and will independently check the comet sizes and
albedos for the overlap sample. 

\subsection{Spectral Mapping of the Zodiacal Cloud}

SPHEREx can address the still  unsettled question of the origin of the zodiacal
dust cloud, which might have contributions from comets, asteroids, and
Kuiper belt objects.   Our solar system contains two known 
debris disk regions populated by relic planetesimals in collisional
equilibrium, the Main Asteroid Belt and the Kuiper Belt \cite{Farinella:1996,Stern:1996,Nesvorny:2010}. (A
third region, the Oort Cloud, while filled with relic comets from the
era of giant planet formation, has yet to be shown to be producing
dust due to comet collisions or activity.) Using infrared profiles of
the zodiacal cloud, various groups have argued for the relative amount of
dust produced by asteroid grinding and family formation in the main
belt vs. dust emitted from active comets and Centaurs
\cite{Dermott:1999,Durda:1997,Liou:1995,Rowan:2013,Nesvorny:2010}. The
recent passage of the New Horizons spacecraft carrying the Student Dust Counter Experiment  into
the Kuiper Belt has shown that dust is being created by KBO
collisional grinding and impact spallation.
While these debris disks in the solar system are relatively sparse and
low density compared to some of the well known IRAS, Spitzer, Herschel, and WISE
disks, likely due to the large number of surviving planets in our
system \cite{Greaves:2010}, they arise from the same phenomena.  

Studies of the zodiacal (``zodi'') cloud emission by IRAS in 1984 and COBE/DIRBE
spanning 1989 - 1992 showed that the brightness depends on the 
elongation, time of year, and wavelength of observation. We now
understand these effects as due to our vantage point moving with the Earth
along its slightly inclined and eccentric orbit through a zodiacal
dust cloud with its own eccentricity and tilt. Both surveys also detected
enhancements of the zodi, termed "bands", associated with asteroid
collisional families in the main belt (E.g. \cite{Espy:2008}). Trails
of heavy dust particles were also detected in association with
cometary activity \cite{Sykes:1996,Sykes:2002,Lisse:2008,Arendt:2014}. Cometary dust emission has been known for centuries, and
dynamical estimates of the dust input to the zodi suggest that the
largest amounts come from the highly gravitationally bound short
period comets \cite{Kresak:1987,Lisse:1998,Nesvorny:2010}. In the last few years, HST has
directly imaged asteroid-asteroid collisions
(e.g. \cite{Jewitt:2010}). ISOCAM MIR spectral measurements of the
zodi emission show silicate emission features similar to those of
cometary dust grains \cite{Reach:2006}, while NIR CIBER rocket flight
measurements show absorption features akin to stony asteroid surface
reflection spectra (\cite{Bock:2013}).    

The IRAS and COBE surveys were conducted using photometry at very
coarse spatial resolution, while the ISOCAM and CIBER observations
were taken for a few minutes over small patches of sky. SPHEREx will
provide detailed maps of the zodi emission in unprecedented spatial
and spectral detail over 2  years, allowing for in-depth searches of
the sources of the cloud amongst the asteroid families and active
comets. Like the asteroid survey described above, mapping the zodi is
also a SPHEREx science goal of high priority, as measurements of
diffuse emission from extra-solar system phenomena must contend with
the zodiacal light foreground.  

\subsection{Follow the Water \& Key Ices Throughout the Solar System}

The final two SPHEREx solar system science programs we discuss involve taking an
inventory of key astrobiological components (e.g. H$_2$O, CO, CO$_2$,
Organics [CH$_4$, C$_2$H$_6$, HCN, CH$_3$OH, H$_2$CO, PAHs], NH$_3$ ) throughout the
solar system. NASA has used the observation, that life as we know it on
Earth is only freely metabilizing if liquid water is present, to
motivate its "Follow the Water" methodology for searching for
extraterrestrial life in the solar system. Similarly, sources of
organic materials are required for life on Earth, as are sources of
nitrogen; NH$_3$ is also explicitly broken out in the list
above as it could possibly serve as a polar hydrogen-bonding
alternative solvent to H$_2$O in other environments. This program is broad, in that it will
attempt to find the relative abundances of these species on every body
in the solar system. But it also contains some focused sub-programs
where SPHEREx can greatly help as the most sensitive NIR spectrometer
above the Earth's atmosphere, by specifically searching for the H$_2$O (e.g., the 2.7 $\mu$m complex),
CO$_2$ (4.23 $\mu$m), CO (4.67 $\mu$m) and CH$_4$ (3.3 $\mu$m) signatures that are strongly blocked for ground based
telescopes.  

An example of the type of discovery which SPHEREx might make is
provided by the discovery of water emission from the dwarf planet
Ceres, reported by HST \cite{Ahearn:1992} and Herschel
\cite{Kuppers:2012}. Results from the DAWN mission argue strongly that
Ceres is a highly altered world, rich with water ice, ammonia, clays, and surface
evaporites, but water has been seen only in the Ivo crater region, and
it is not clear if Ceres is now a dead and frozen world.  Evidence for
the presence of water-bearing or water-altered species and organics
are also found in a variety of other asteroid spectra, especially in
the 2-4 $\mu$m region have been identified
(cf. \cite{Takir:2013,Rivkin:2015}). SPHEREx, obtaining the spectrum of every body which crosses its path, has the
potential to discover water and other astrobiologically interesting
molecules in numerous asteroids, while it is also studying these same
molecular species in comets.  A thorough inventory of solar system
ices and organics is of great scientific interest, and it also
connects directly to one of SPHEREx main scientific goals, which is
to study the evolution of icy materials from the interstellar medium
into protoplanetary and planetary systems. 

\section{Synergies with other NASA programs}

\subsection{JWST and SPHEREx Synergies}
\label{sec:jwst-intro}

The James Webb Space Telescope (JWST) is NASA's next major space observatory,
and scientific, technical, and schedule considerations conspire to
make SPHEREx an important synergistic partner.  JWST is the perfect
near--mid-IR sequel to HST and Spitzer \cite{Garner:2006}. SPHEREx
will be the perfect 0.75-5 $\mu$m\ {\it all-sky} object finder for
JWST, and play a critical role in the absolute flux calibration for
JWST at 0.7--4.8 $\mu$m. 
SPHEREx' wavelength range is totally encompassed within that accessible to JWST, so the
initial releases of the SPHEREx data (in 2021-22) and the all-sky
catalogs (in 2023) will provide an invaluable catalog of
scientifically interesting sources for full investigation with JWST.
JWST observations are slated to begin in the Spring of 2019 with a
planned five year lifetime; this already provides significant
schedule overlap with SPHEREx, which will increase considerably if, as seems very
probable, JWST operates beyond its current five year horizon. 


At the cost of a SMEX mission, SPHEREx's all-sky 0.7-5.0 $\mu$m spectral coverage
is a great bargain in preparation for JWST, and will result in truly maximizing
the enormous investment already made in JWST: Any (rare) classes of objects
that are discovered by SPHEREx which JWST must follow-up on during
its expected lifetime, will be observed by JWST only if SPHEREx flies well
before JWST expires. We give a couple of examples of such objects in this
document, and discuss broad areas of scientific synergy. 

\subsubsection{JWST--SPHEREx Synergy: Ground-Truth Calibrations}
\label{sec:cross-cal}

As of 2016, the JWST Project is planning to use its four instruments for
science parallels. This is currently being implemented for the most-used JWST
instrument combinations. JWST will therefore be able to provide the perfect
grism+imaging calibration data for SPHEREx (and also for WFIRST): JWST NIRCam
0.7--5 \mum\ images to AB\cle 29-31 mag and NIRISS 0.7--2 \mum\ grism spectra to
AB\cle 27-28 mag will critically help to disentangle the multi-band SPHEREx
database in crowded deep fields. Over very small FOV's at least (5--10\arcm 
after some mosaicking), JWST will therefore be able to provide the ground truth
for faint-object broad-band photometry and slitless spectra about 10 mag deeper
than what SPHEREx will routinely observe. 

Why is this so important? Because with its 6.2'' pixels, deep
multi-band SPHEREx images will be confusion limited, so that  an independent
assessment of the exact level of object confusion is absolutely critical. JWST
will provide that ground-truth in a number of very small-area but very deep and
well-studied survey fields. The good news is that SPHEREx' object confusion is
mathematically solvable. This is simply because the panchromatic normalized
differential galaxy counts --- measured over the range 10\cle AB\cle 30 mag and
0.2--2 \mum\ \cite{Windhorst:2011}, and recently also out to 500 \mum\
\cite{Driver:2016} --- converge with a sub-critical
magnitude slope of $<$0.4 dex/mag in the flux range 18\cle AB\cle 21 mag. That
is, at all wavelengths now measured with GALEX, HST, ground-based
facilities, Spitzer, WISE, and Herschel, the normalized differential galaxy
counts permanently reach a sub-Olbers slope for AB\cge 19 mag, or somewhat
fainter. Stated differently, the 0.75--5.0 \mum\ galaxy counts in the flux range
18\cle AB\cle 21 provide most of the measured integrated Extragalactic
Background Light (iEBL). Hence, most discrete 0.7--5.0 \mum\
object fluxes  fainter than AB$\sim$19--21 mag add relatively little additional flux to the total Extragalactic  Background Light. While
0.75--5.0 \mum\ objects fainter than 19--21 mag will occasionally confuse the
SPHEREx fluxes for brighter foreground objects, this will happen in a
manner that is statistically correctable. In summary,
\cite{Driver:2016} show that at all wavelengths 0.1--500 \mum, the EBL
integral clearly converges 
(\ie Olber's paradox is no longer relevant), and as a consequence, SPHEREx
object confusion for unresolved background objects at AB\cge 19 mag can be
statistically addressed and corrected for at 0.75--5.00 \mum. JWST will add the
essential ground-truth data to AB$\sim$29 mag from 0.75--5.0 \mum\ in selected
fields to test this in SPHEREx images of the same fields. The ultradeep JWST
images can then simply be used to measure the converging integral in each
SPHEREx beam beyond the SPHEREx detection limit, and verify that the
statistical confusion correction is as computed from \cite{Driver:2016} and
\cite{Windhorst:2008,Windhorst:2011}. 

\subsubsection{JWST and SPHEREx Synergies: Galactic Science}

\begin{figure}[!th]
\includegraphics[width=0.8\textwidth,angle=90]{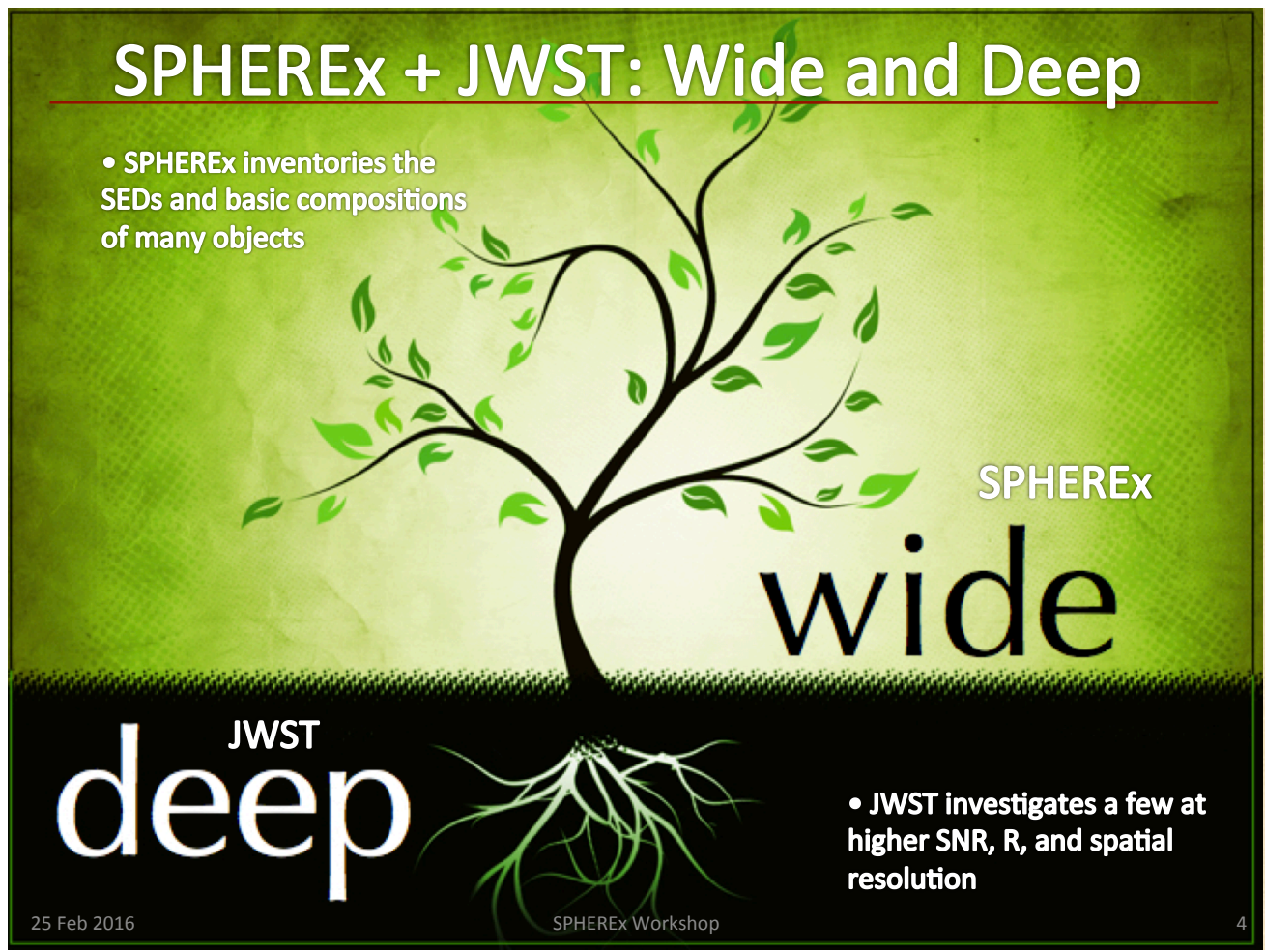}
\caption{Illustration of the synergies between SPHEREx and JWST.}
\label{fig:tree}
\end{figure}

SPHEREx and JWST are two very different missions, but they will both
play major roles in advancing our knowledge of galactic science,
including JWST's two major galactic science themes: {\it Birth of
  Stars and Planetary Systems} and {\it Planets and the Origins of
  Life} \citep[e.g., see][]{GMC06}. 

Their differences also complement each other very well. SPHEREx is a
moderate-aperture single-mode observatory that will provide R $\sim$
50 resolution spectra over $\lambda = 0.75 - 5.0$ $\mu$m for every
point in the sky JWST is a large-aperture observatory with 4 separate
high sensitivity and high spatial resolution science instruments having
specialized modes, each covering some part of JWST's 0.6 -- 28 $\mu$m
wavelength range. SPHEREx surveys the entire sky efficiently to
moderate depth, while JWST is optimized for very deep exposures; its
large overheads ($\sim$ 45 min for slew, guide star, and target
acquisitions) make JWST observations inefficient for all but the
faintest objects.

These characteristics produce a natural division: SPHEREx goes wide,
inventorying the SEDs and basic compositions of many objects in the
galaxy and beyond. JWST goes deep, studying many fewer objects but with much
higher spatial resolution, sensitivity, and spectral resolution as
well. SPHEREx data will be used to evaluate statistically the bulk
characteristics of whole classes of objects, while JWST data will give
detailed insights into the physical and chemical workings of
individual and modest groups of faint objects.  JWST can also follow
up on objects identified by SPHEREx as having unusual and intriguing
characteristics.  For example, JWST  can obtain higher spectral and
spatial resolution spectroscopy of objects with unusual ice spectra
which might be suggestive of stronger than expected isotopically
shifted species or atypical elemental abundance ratios. 

Much galactic science would benefit from a complete SPHEREx 0.75 --
5.0 $\mu$m  survey plus higher sensitivity and resolution JWST
observations of a smaller number of select objects at SPHEREx or
perhaps longer wavelengths. These include observations of dark cloud
extinctions, Class 0 and I protostars, circumstellar protoplanetary
and debris disks, interstellar radiation diagnostics (i.e., PAH
emission), planetary nebulae, T-class field brown dwarfs, stellar
populations, and potentially other objects as
well. Table~\ref{tbl-objects} illustrates the typical numbers and
sizes of some of these objects.  Although some sources seen by SPHEREx
might saturate some JWST observing modes, SPHEREx' dynamic range is
high enough, and JWST's instruments versatile enough, that millions of
SPHEREx detections will be readily observable by JWST. The expected
annual cadences of both JWST calls for proposals and SPHEREx
data releases should allow SPHEREx follow up observations to be
proposed during the first year of the SPHEREx mission and annually
thereafter.  

\begin{table}[h]
\centering
\caption{\normalsize Potential targets for SPHEREx + JWST Galactic Studies\label{tbl-objects}}
\vskip 1mm
\centering
\begin{tabular}{l r r l}
\hline
\hline
Objects & Number & Size('') & Comments \\
Dark Cloud Cores & $\sim$50 & $\sim$60 & Extinction studies for structure and ices \\
Class 0/I Protostars & 500 & 0.5 -- 2 & F$_\nu \sim 1$ mJy; extinctions and ices\\
Protoplanetary Disks & 1000 & $< 1$ & F$_\nu \sim 10+$ mJy; SEDs and ices \\
Debris disks, PNe, BDs... & Many & pt. -- 60 & Many object types\\
Solar System Giant Planets & 4 & 4 -- 50 & Full disk observations \\
Small Solar System Bodies & $>100$ & $<1$ & Surface compositions \\ 
\hline
\end{tabular}
\end{table}



\subsubsection{The near--IR Spectral Energy Distribution of Galaxies}

\begin{figure}[!t]
\center
\includegraphics[width=0.7\textwidth,angle=-0]{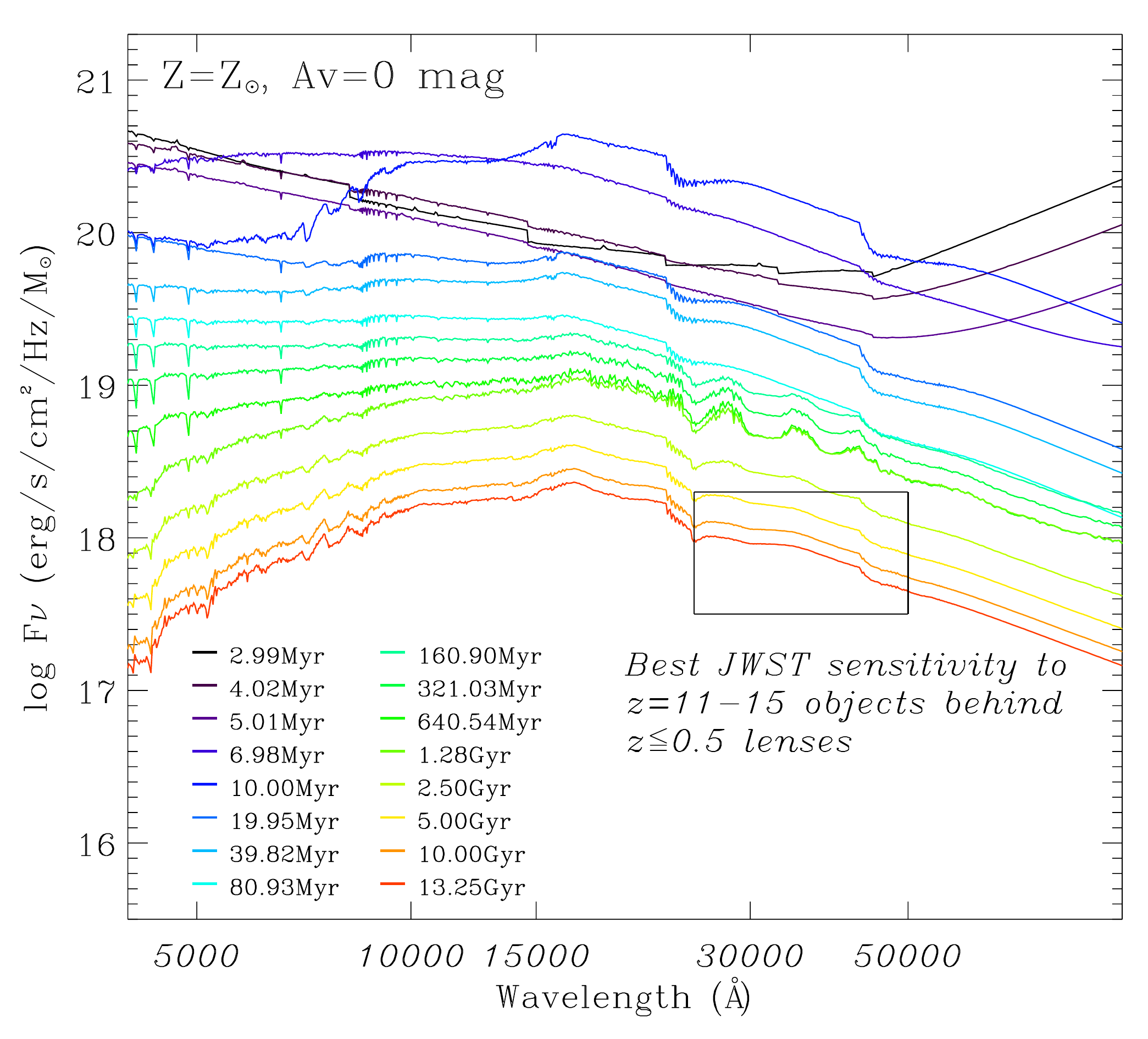}
\caption{Examples from an array of combined Simple Stellar Population
  (SSP) Spectral Energy Distributions (SEDs) from \cite{Kim:2016} for
  sixteen different ages, fixed solar metallicity ($Z_{\odot}$\,=\,0.02), and zero
extinction. SED ages are indicated by the colored lines. Other examples of
different metallicities and different extinction laws and $AV$-values are given
by \cite{Kim:2016}. SPHEREx is ideal for measuring SEDs of foreground 
gravitational lenses (z\cle 0.5) that JWST will use to find First Light
objects.}
\label{fig:seds} 
\end{figure}

Fig.~\ref{fig:seds} shows examples from an array of combined Simple Stellar Population
(SSP) Spectral Energy Distributions (SEDs) from \cite{Kim:2016} for 
different ages, solar metallicity, and zero extinction, as applicable for
early-type galaxies in groups or clusters at z\cle 0.5, which SPHEREx will
observe in very large numbers all across the sky (see \S\ 2). Note
that irrespective of age, nearly all SEDs older than 10 Myr show a
peak around $\lambda$$\sim$1.7 $\mu$m\ and 
a rapid decline at $\lambda$\cge 2.33 \mum. The Zodiacal
sky-brightness at L2 similarly declines rapidly at $\lambda$\cge 2.33 \mum\ for a 5 Gyr old star
(\ie\ the Sun). As a consequence, for any reasonable old lensing-galaxy
SED, the black box indicates that the sweet spot in JWST NIRCam sensitivity
to lensed First Light (z$\simeq$10--15) sources occurs at $\lambda$=2.5--4.5
\mum, where the Zodiacal foreground from L2 is the lowest. The best lensing
clusters to magnify First Light objects with JWST will be the most massive
lenses at z\cle 0.5, where the K-correction from the $\lambda$$\simeq$1.7 \mum\
peak in the foreground lensing SEDs is still modest. The gain in foreground
darkness of a z$\simeq$0.4 cluster not being corrupted by the redshifted ICL
beats the additional $(1+z)^4$ dimming that one would get for a z$\simeq$1
cluster. The latter generally also have by selection lower mass
($\sim$10$^{14}$\Mo), and are younger and dynamically less relaxed, making
them less suited for optimal lensing of First Light (z\cge 10--15)
targets to be seen by JWST compared to more massive clusters at z$\simeq$0.4. As discussed
in Sec.~\ref{sec:icl} and \ref{sec:lensingJWST}, many of the best lensing clusters used by JWST will need an
independent absolute measurement of the ICL, which may be complicated by
rogue-path straylight seen by JWST, but can be provided by SPHEREx. 

\subsubsection{Absolute Flux Calibration: IntraGroup \& IntraCluster
  Light Measurements}
\label{sec:icl}

Our Galaxy is a bright IR source at $\lambda$$\simeq$1--4 \mum. Because of 
JWST's unavoidably open architecture, rogue-path straylight will hit the JWST 
secondary mirror from certain directions of the sky via its sunshield. We will 
not know precisely how large the amplitude of this straylight and its gradients
are until JWST gets to its L2 orbit. Ray-tracing calculations suggest possible
stray-light amplitudes of $\sim$40--95\% of Zodi with gradients of 2--5\%
(typical--worst case). Since this straylight doesn't go through the optical
path of JWST, it doesn't carry the PSF of the telescope, and so any JWST
stray-light + gradients need to be disentangled from the astronomical
targets: Galactic nebulae, stars, galaxies, galaxy groups, and
galaxy clusters. If JWST rogue-path stray-light has a slight or complex
gradient, it may be hard to separate these from the real IntraCluster Light
(ICL) in gravitationally lensing clusters. As explained in
Sec.~\ref{sec:groups}, it will be critical for JWST to use the best gravitationally lensing clusters at z\cle 0.5
to see First Light objects (z\cge 10) in large numbers. 

Since SPHEREx has a closed architecture \cite{Bock:2013,Dore:2014cca}, it will
suffer very little stray-light, and so SPHEREx will provide an absolute
measurement of the IntraGroup Light (IGL) and the ICL for many tens of
thousands of groups and clusters across the sky, as shown in
Fig.~\ref{fig:groups}. Galaxy groups contain most of the mass in the
universe (rich, massive clusters are relatively rare), SPHEREx will be
critical to measure the stellar masses of galaxy groups at z\cle 0.5, including their diffuse  baryonic IGL
component. Similarly, since JWST will observe the best available 
gravitationally lensing clusters at z\cle 0.5 (see Fig.~\ref{fig:groups}) to detect the
maximum number of First Light objects, and SPHEREx will be essential to measure the
absolute value of the IGL at 0.75--5.0 \mum, which provides a critical
surface brightness-calibration for all JWST cluster work, including
any diffuse mass component  associated with their ICL. 

As discussed above, stellar masses of galaxies, galaxy groups, and galaxy
clusters will be directly estimated from JWST's 0.7--4.8 \mum\ photometric
measurements, but the absolute value of such estimates will be uncertain in
groups and clusters if they indeed have a substantial amount of {\it diffuse} 
IGL and ICL, \ie the diffuse light from unresolved stars and perhaps unresolved
globular clusters, which get dissociated from the individual galaxies in the
group or cluster due to their substantial velocity dispersions. Indeed, ROSAT
and Chandra X-ray observations have shown that both galaxy groups and galaxy
clusters contain a substantial reservoir of hot X-ray gas
\cite{Mulchaey:1996,Ebeling:1998} that was torn from individual
galaxies and escaped into the gravitational well of the group or the
cluster, together with stars and globular clusters. It is known that this baryonic mass fraction of hot gas is
substantial, but it is not known exactly how much mass is hidden in diffuse
unresolved stars and globular clusters that are bound to the group or cluster
as a whole, but {\it not bound to individual galaxies}. JWST will attempt to
measure the IGL and ICL, but it is quite possible that this will require {\it
accurate absolute} 0.7--4.8 \mum\ flux measurements of the diffuse component in
these groups and clusters. Spitzer will provide some anchor at 3.6 and 4.5 
\mum\ for the groups and clusters that it will have measured during its
life-time, but SPHEREx will provide this anchor at 0.75-5.0 \mum\ for {\it all}
groups and clusters in the universe that have individual members visible to
z\cle 0.5 (see Fig.~\ref{fig:groups}), thereby including {\it all} groups and
clusters that JWST will observe during its lifetime. 

To draw a comparison with radio astronomy: JWST will be the ``high-resolution
interferometer'' at 0.7-4.8 \mum\ that will map the First Galaxies in detail.
SPHEREx will provide the absolutely critical ``single-dish total-flux
measurements'' at 0.75--5.0 \mum\ of {\it all} foreground objects to AB\cle 19
mag and z\cle 0.5, especially for low-SB structures such as IGL and ICL, for
which JWST may not be able to accurately provide an absolute flux calibration. 

\subsubsection{Finding Dusty QSOs with SPHEREx' All-Sky Survey at  z$\simeq$2--7}

\begin{figure}
\center
\includegraphics[width=0.55\textwidth,angle=-0]{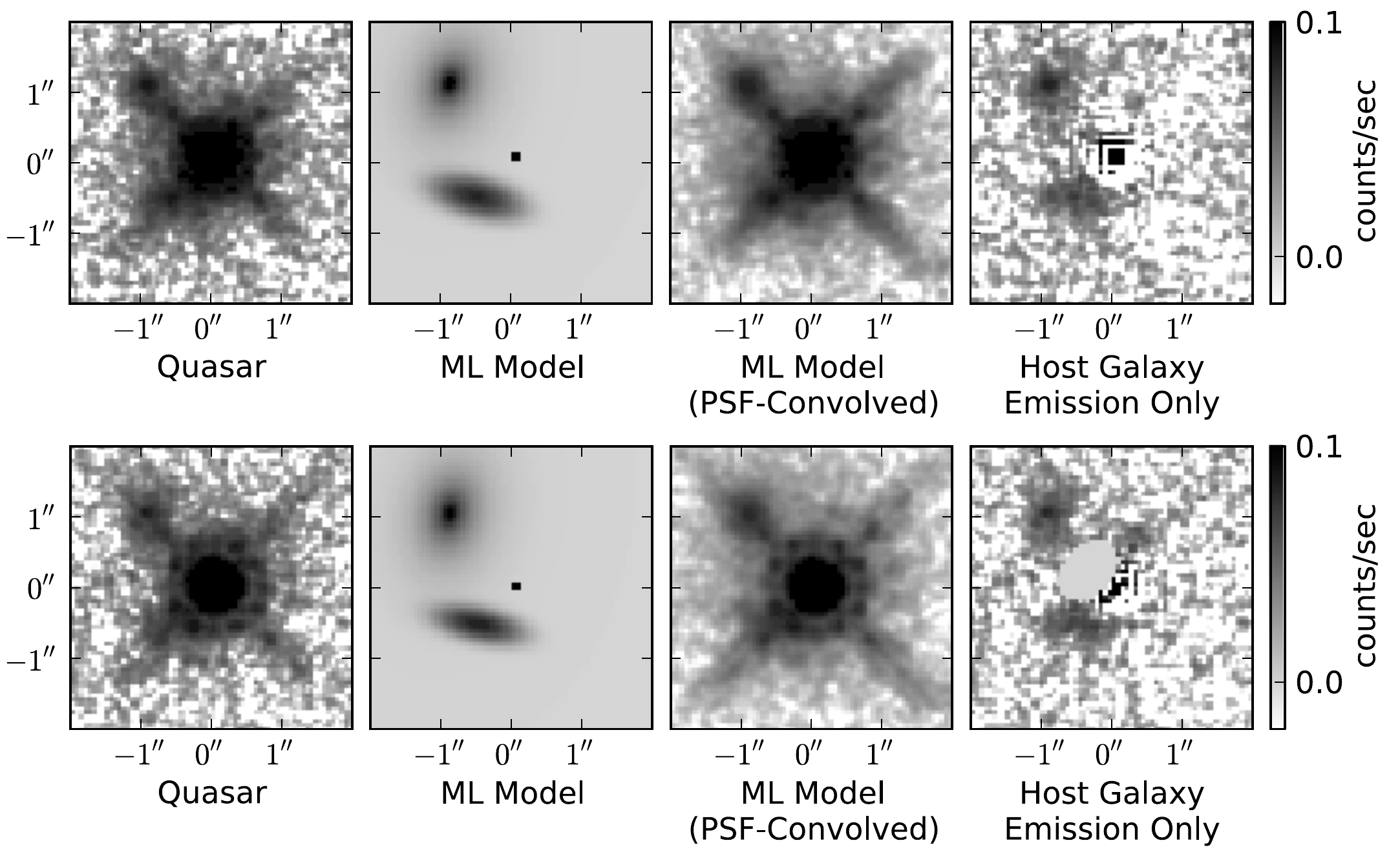}
\includegraphics[width=0.40\textwidth,height=0.35\textwidth]{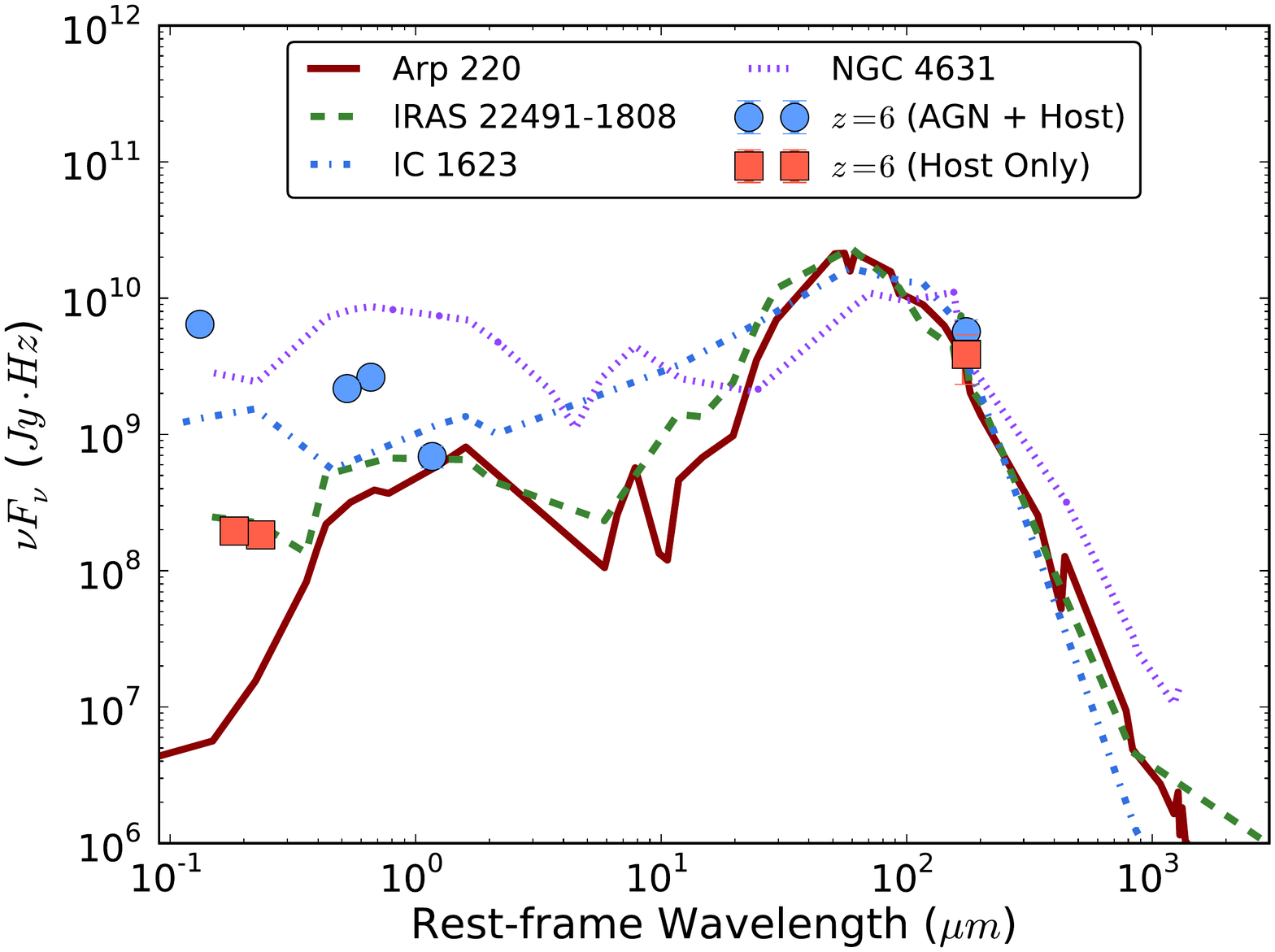}
\caption{\emph{Left:} HST WFC3 PSF+Sersic light-profile of the first QSO host
galaxy system detected at z$\simeq$6
\cite{Mechtley:2012,Mechtley:2016}. The host has a giant merger
morphology, plus very similar J (top) and H-band (bottom) 
structure.  \emph{Right:} Rest-frame far-UV---far-IR SED of its
z$\simeq$5.85 QSO host system. Blue points show the total SED of the
QSO+host galaxy. The red squares show the WFC3+submm SED of the
z$\simeq$6 host galaxy only. The plotted range of local (U)LIRG SED
models (Wang \etal\ 2011) show that the FIR/FUV flux ratio and
UV-slope are consistent with local {\it dusty} starburst system with
A$_{FUV}$$\simeq$1 mag and \MAB\cle --23.0 mag, or $\sim$2 mag
brighter than \Lstar at z$\simeq$6.}
\label{fig:QSO} 
\end{figure}

Fig.~\ref{fig:QSO} left shows a Monte Carlo Markov-Chain model of the HST WFC3 PSF-star
plus a host-galaxy Sersic light-profile for the first QSO host galaxy system
detected at z$\simeq$6 \cite{Mechtley:2012,Mechtley:2016}. Careful contemporaneous
orbital PSF-star subtraction removes most of the HST ``OTA spacecraft
breathing'' effects. For another QSO at z=6.42 with
AB$\simeq$18.5 mag, {\it no} host galaxy was detected 100$\times$fainter
nearly to the noise limit, or at AB\cge 23.5 mag \cite{Mechtley:2012}. 

For the QSO at z$\simeq$5.85 in Fig.~\ref{fig:QSO} left, however, a clear host system appears
with a merger morphology, plus perhaps some tidal features, and a rather
similar J- and H-band structure. Its UV-continuum slope is
$\beta_{\lambda}$$\simeq$--1.01, which constrains its dust content. This
rest-frame UV-color is significantly redder than LAEs and LBGs at z$\simeq$6,
indicating significant internal reddening due to
dust. Fig.~\ref{fig:QSO} right shows the rest-frame far-UV---far-IR SED of this z$\simeq$6 QSO host system. Plotted are
a range of fiducial local galaxy SEDs with starburst ages\cle 1 Gyr
\cite{Wang:2011}, normalized around $\sim$100 \mum. The WFC3 data on
the host galaxy rules out the unreddened spiral-like host galaxy SEDs (purple and blue
curves), but allows starbursting (U)LIRGs \& Arp 220-like SEDs
\cite{Mechtley:2012,Mechtley:2016}. The FIR/FUV flux ratio and
UV-slope are consistent with local {\it dusty} starburst systems with A$_{FUV}$$\simeq$1 mag and \MAB\cle
--23.0 mag, or $\sim$2 mag brighter than \Lstar at z$\simeq$6. Comparison to
the surface density of bright LBGs at z$\simeq$6 then suggests that the duty
cycle of this QSO may be \cle 10$^{-2}$ at z$\simeq$6 (\ie\ \cle 10 Myrs),
which would place it close to the Magorrian \etal\ relation \cite{Magorrian:1997}. 

Note how rapidly the z$\simeq$6 galaxy host SED gets brighter towards longer
near-IR wavelengths. It is possible that close to half of the energy in the EBL
comes from dust-obscured objects \cite{Driver:2016} and references therein),
and that as a consequence, many QSOs will only show up in near--mid-IR
selected surveys, which a combination of SPHEREx 
and WISE will provide at 0.7--24 \mum\ for 1000's of QSO's at z$\simeq$2 to
dozens of QSOs at z$\simeq$6--7. SPHEREx will thus provides all-sky
samples of z$\simeq$2--7 QSOs for JWST, including these very dusty QSOs not
selected in rest-frame UV--optical surveys. Careful JWST NIRCam+MIRI
PSF-subtraction and/or Coronagraphy after 2018 will image the host galaxies of
these z$\simeq$2--7 QSOs in the rest-frame optical--near-IR, constraining their
stellar masses, and the M$_{SMBH}$--M$_{bulge}$ relation of their host galaxies
at the highest redshifts. SPHEREx is thus the essential (dusty) object
finder for JWST.

\subsubsection{Galaxy Cluster Lensing with SPHEREx and JWST at  z\cle 0.5}
\label{sec:lensingJWST}

To see the largest number of First Light objects, JWST must cover the best 
lensing clusters. This is because the evolution of the Schechter UV luminosity
Function (LF) is very rapid \cite{Bouwens:2015,Oesch:2014}. While the
faint-end LF-slope does not get much steeper than $\alpha$$\simeq$--2 for z\cge
6 \cite{Yan:2004,Morgan:2015}, the characteristic density \Phistar and characteristic luminosity \Mstar potentially decline very rapidly
with redshift. At z\cge 8, we expect for JWST \Phistar \cle 10$^{-3.5}$
(Mpc$^{-3}$) and that \Mstar may drop below --18 mag at z\cge 10. If either
one of these were not the case, then a much larger number of
z$\simeq$9--11 candidates would have been observed in the HUDF. This
dramatic drop in \Phistar or \Mstar 
has significant consequences for the JWST survey strategy. Given the possible 
range in Schechter LF-parameter evolution with redshift for z\cge 8--10, one can
outline the optimal survey area and sensitivity covered per amount of JWST time
to detect the maximum number of First Light (z\cge 10--15) objects with JWST. 
These considerations imply that we will need to do a wedding-cake layered survey
with JWST to see First Light, \eg\ \cge 10 Webb Medium-Deep Fields to
AB\cle 29 mag, several Webb Deep Fields to AB\cle 31 mag, and perhaps
one Webb Ultra Deep Field to AB\cle 32 mag. In either case, the number
of objects anticipated at z\cge 12 will be very small, unless a
significant number of Webb Deep Fields is pointed at the best
gravitational lensing clusters. 

For this reason, the community has started to gather a significant amount of
data on the best lensing targets for JWST to detect z$\simeq$10--15 objects.
These clusters come from the ROSAT X-ray survey, the Planck and SPT SZ surveys,
and various other cluster surveys such as CLASH, MaDCoWS, RedMapper and the
Hubble Frontier Fields. As discussed in Sec.~\ref{sec:icl}, the sweet spot for JWST lensing of First Light
objects is for clusters at 0.3\cle z\cle 0.5 (see
Fig.~\ref{fig:seds}). To detect the largest number of lensed sources
at z\cge 10, these also need to have large masses of
10$^{15-15.6}$\Mo, and high concentrations of 4.5\cle C\cle 8.5. The
GAMA, SDSS, and WIGGLEz samples (see Fig.~\ref{fig:groups}) are an
excellent database to define such samples. The many available
spectroscopic redshifts provide accurate dynamical masses
\cite{Robotham:2011,Yang:2007}, and help remove chance projections. As
discussed in Sec.~\ref{sec:icl}, SPHEREx is needed to characterize the
total stellar light for these best lensing z\cle 0.5 clusters that
will be observed with JWST. This includes a careful absolute
measurement of the diffuse light (ICL) in these clusters that JWST may
not be able to disentangle from its anticipated rogue-path stray-light.

\subsection{WFIRST and Euclid}

NASA's Wide Field Infrared Survey Telescope (WFIRST;
\cite{Spergel:2015sza}) was the top ranked large space mission
recommendation in the 2010 Decadal Survey.  One of WFIRST's primary science goals is to
determine the nature of the dark energy that is driving the current
accelerated expansion of the Universe. WFIRST entered into Phase A in
February 2016 in preparation for launch in $\sim$ 2024. An existing
2.4 meter telescope will provide Hubble Space Telescope quality
imaging, but over a field of view that is $\sim$100 times that of Hubble.
WFIRST will have a 6 year primary mission at L2, but there are no
consumables that preclude a mission of 10 or more years.  The Wide
Field Instrument (WFI) on WFIRST will consist of 18 4k by 4k H4RG Near
infrared detectors that will provide grism spectroscopy (1.35-1.89$\mu$m;
R=461) and imaging (0.76-2$\mu$m) over 0.28 square degrees at a pixel
scale of 0.11 arcsec.  An Integral Field Unit spectroscopic channel
designed for supernova follow up will have a single 2k by 2k detector
and provide spectroscopy at R=80 to 120 at 0.6-2 $\mu$m.   
WFIRST will perform a $\simeq$ 2300 square degree High Latitude Survey (HLS)
in order to obtain weak lensing and galaxy clustering (BAO and RSD)
measurements.  A deeper survey of tens of square degrees will be used
to find and follow up $\simeq$ 2700 supernovae. Thus, WFIRST measure both the
expansion history of the Universe and the growth of structure using
multiple techniques with a survey designed for very tight systematics
control. 

The European Space Agency (ESA) Euclid mission
\cite{Laureijs:2011gra} will launch in 2020 for a 6 year mission at L2 to study dark energy
using weak lensing and galaxy clustering.  Euclid will have a 1.2 m
primary mirror and two instruments.  The visible instrument (VIS) will
have 36 CCDs (4k by 4k) that will perform photometry for galaxy shape
measurements using a single wide (Riz) filter.  The Near Infrared
Spectrometer and Photometer (NISP) instrument will use 16 2k by 2k
near infrared (NIR) detectors (provided by NASA) to perform imaging
and grism spectroscopy in the 1-2 $\mu$m range. Both instruments will
survey the same  0.5 square degree piece of the sky simultaneously via
a dichroic. Euclid will cover the lowest sky background 15,000 square 
degrees in Riz (single wide filer), Y, J and H filters and employ a simple grism
spectrometer to perform a survey with an unprecedented combination of area and resolution. 

Both WFIRST and Euclid are designed to do those
portions of the dark energy experiment that can be done only from
space.  There are three such measurements:
\begin{enumerate}
\item Very accurate galaxy shape measurements for weak gravitational
  lensing enabled by the small, stable PSF of a space-based telescope.  
\item Near infrared photometry for accurate photometric redshifts when
  combined with ground-based optical photometry.
\item Near infrared spectroscopy for accurate galaxy clustering
  measurements at redshifts $z>$1.
\end{enumerate}

There are a number of ways that the SPHEREx all sky infrared
measurements will be synergistic with Euclid and WFIRST:   
\begin{enumerate}
\item Filling in the low redshift (z$<$1) galaxy clustering field for
  baryon acoustic oscillations (BAO) and redshift-space distortions
  (RSD).  Since both WFIRST and Euclid concentrate on z$>$1 for 
  this cosmological probe, the combination with SPHEREx will allow for
  a full measurement of galaxy clustering over the full range at which
  the effects of dark energy are expected to begin to dominate the
  expansion history of the Universe.   
\item SPHEREx will acquire a huge number of low redshift spectra to
  calibrate photometric redshifts.   Every galaxy in the WFIRST and
  Euclid weak lensing surveys will require a distance measure and the
  vast majority of these will be calculated via photometric redshifts.
  Recent work has shown that up to 10$^5$ spectra may be needed to
  calibrate these photometric redshifts, and these spectra need to be
  a complete sample down to the magnitude used for weak lensing, 24.5
  for Euclid and $\simeq$ 27 for WFIRST and LSST (see,
  e.g. \cite{Newman:2014sra}).  While many of these spectra will need
  to be acquired with  ten and thirty meter class telescopes, a great number of them will
  be readily available from SPHEREx, especially at z$<$1.  Thus,
  SPHEREx will be a powerful resource in the required calibration of the
  photometric redshifts for WFIRST and Euclid.  
\item Intrinsic galaxy alignments (see
  \cite{Kirk:2015nma,Kiessling:2015sma,Joachimi:2015mma} for recent
reviews) represent the largest  astrophysical systematic for weak lensing measurements. Unmitigated,
  these intrinsic alignments can seriously bias cosmological
  inferences.   However, Krause et al (2016) \cite{Krause:2015jqa}, have shown that if
  intrinsic alignments are controlled at $z<$1, then the cosmological
  results for an LSST-like or Euclid-like survey will be virtually
  unbiased.  Intrinsic alignment mitigation is best achieved via very
  accurate redshifts for the foreground galaxies; thus, the immense
  z$<$1 SPHEREx spectroscopic redshift sample will prove to be a boon
  for intrinsic alignment mitigation for future lensing experiments.  
\item The same SPHEREx $z<$1 spectroscopic redshift sample that will
  provide intrinsic alignment mitigation will provide very accurate
  distance measures for foreground galaxies used for galaxy-galaxy
  lensing measurements.  Combining this with precision shape
  measurements of background galaxies via WFIRST and Euclid will allow
  for unprecedented galaxy-galaxy lensing measurements that will
  enable the study of the galaxy-halo connection as a function of
  luminosity, type and environment. This will also allow the study of
  baryonic feedback processes as a function of the above variables and
  enables higher order statistics such as galaxy-galaxy-galaxy
  lensing. This is detailed further below in Sec.~\ref{sec:ggl}
\item Both Euclid and WFIRST will have deep fields over several tens
  of square degrees that reach two or more magnitudes deeper than the
  wide cosmology surveys described above. Those fields will produce
  additional cosmology (for instance the SN survey for WFIRST), a huge
  amount of ancillary science, especially in galaxy evolution, and
  calibration.  Subject to only the orbital constraints of WFIRST and Euclid at
  L2, these deep fields can be coordinated with the deepest areas of
  the SPHEREx survey, creating a lasting legacy data that will be
  exploited for multiple science goals for years to come. In fact the
  SPHEREx deep survey field at the SEP coincides with one of
  Euclid's deep fields.   
\end{enumerate}

\subsubsection{Euclid and WFIRST Synergies with SPHEREx Example: Halo
  Masses for SPHEREx-selected galaxy populations}
\label{sec:ggl}

As an example of the powerful synergies between SPHEREx and other
surveys such as Euclid and WFIRST, we detail below how we could
measure better halo-masses using galaxy-galaxy lensing.

{\bf Motivation.} SPHEREx will discover interesting galaxy populations
as described in Sec.~\ref{sec:gal0.5} and
Sec.~\ref{sec:gal1.0}. Measuring the host halo mass scale for SPHEREx
selected galaxy populations will facilitate the connection with
broader galaxy evolution models, and will constrain galaxy evolution as
a function of environment. Here we forecast the accuracy with which
halo masses can be inferred via stacked galaxy-galaxy lensing, using
the SPHEREx selected galaxies as the lens sample, and Euclid weak
lensing source galaxies as the background sample. 

{\bf Forecast Details.} 
We consider a SPHEREx selected galaxy population with comoving density
$n_\mathrm{g} = 10^{-5} (h/\mathrm{Mpc})^3$, i.e., a rather rare
population, and assume a halo-mass -- observable relation with
characteristic mass scale $M_0$ and log-normal scatter
$\sigma_{M|\mathrm{obs}} = 0.75$.  
The accuracy of the halo mass measurement also depends on the redshift
of the SPHEREx selected galaxies, the redshift distribution of
background galaxies, and the survey area for which imaging and galaxy shape
measurements are available; we assume a Euclid-like survey for the
source galaxy population. 

\begin{figure}[!th]
\centering
\includegraphics[width=0.8\textwidth]{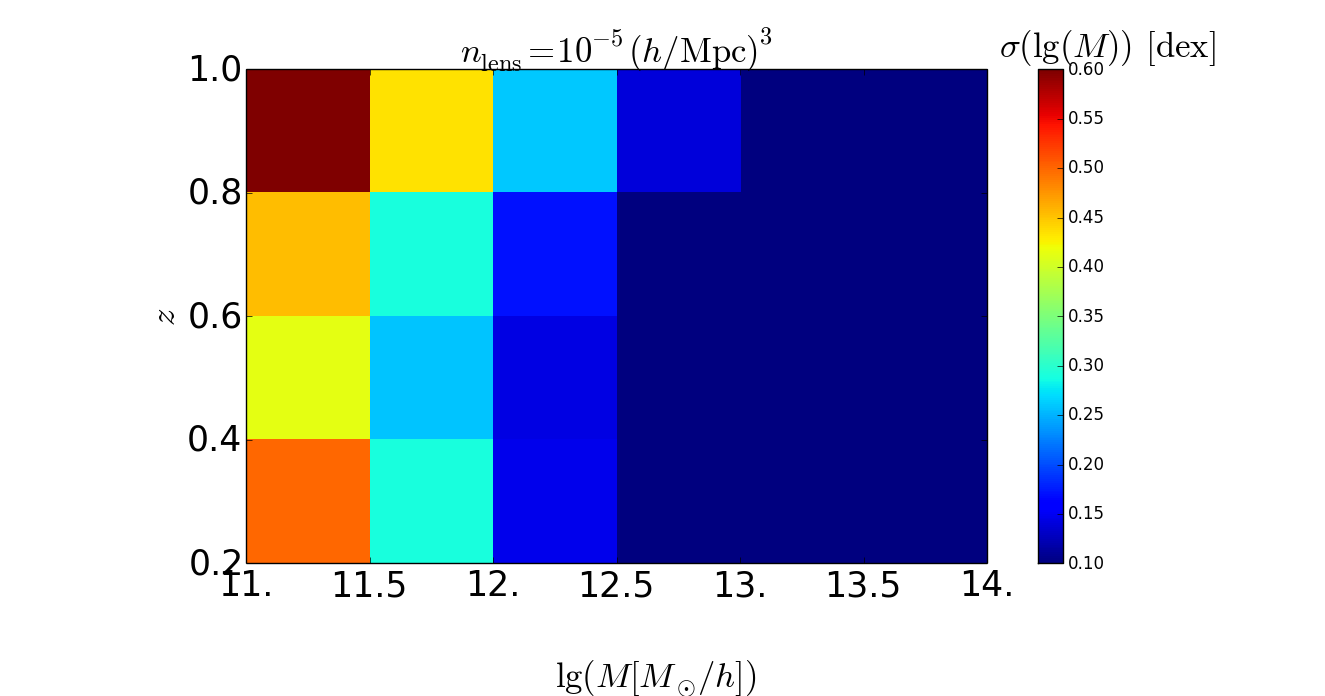}
\caption{\label{fig:sigma_M} Statistical uncertainty of weak-lensing
  based halo mass estimates for a rare, SPHEREx-selected galaxy
  population as a function of halo mass and redshift. The color bar is
  capped at $\sigma(lg(M)) = 0.1$.} 
\end{figure}

Fig.~\ref{fig:sigma_M} shows the uncertainty in the halo mass
determination as a function of host halo mass (x axis; stronger
lensing signal from more massive halos), and the lens redshift bin. At
fixed comoving density, the signal-to-noise ratio of galaxy-galaxy lensing
increases with halo mass, and the halo mass measurement uncertainty
decreases with increasing halo mass. The dependence of mass
uncertainty on lens redshift in this plot is a competition of two
effects: for the choice of redshift bins in this plot, the volume per
redshift bin (and hence the number of lens galaxies) increases with
redshift, while the number of background galaxies decreases. 
These uncertainties are marginalized over cosmology, source redshift 
uncertainties, shear calibration and scatter of the mass--observable
relation. For rare galaxy populations (i.e., in the shot-noise
dominated regime), the mass uncertainty scales approximately as 
\begin{equation}
\sigma\left(\lg(M)\right)\propto\sqrt{\frac{\left(10^{-5} (h/\mathrm{Mpc})^3\right)}{n_\mathrm{g}}}.
\end{equation}
Stacked weak lensing using SPHEREx selected lensing galaxies and
shapes of background galaxies from Euclid or WFIRST imaging will enable
interesting constraints on host halo masses for a range of halo masses
and lens redshifts. These halo mass measurements are complementary to
halo masses inferred from galaxy clustering, and with the assumption
of a halo mass -- bias relation can be used to validate galaxy bias
measurements. Note that this technique can be extended to higher lens
redshift using deeper imaging surveys, such as LSST (see Sec.~\ref{sec:LSST} below).

\subsection{SPHEREx and the Large Synoptic Survey Telescope}
\label{sec:LSST}

 As described earlier in Sec.~\ref{sec:intro}, many of the major new
astronomical facilities and capabilities that will come on-line in the
2020's will be dedicated to wide-field surveys of the sky.  This is
very much reflected in the priorities of the 2010 Decadal Survey: the
highest-ranked space-based mission is the Wide-Field Infrared Survey
Telescope (WFIRST), and the highest-ranked large ground-based
telescope is the Large Synoptic Survey Telescope (LSST).  Much of the
interest and excitement about such surveys is the tremendous breadth
of the science they can carry out.  This has been the lesson of the
Sloan Digital Sky Survey (SDSS), which uses a modest 2.5m telescope
with a large field of view to image and do follow-up spectroscopy of
1/4--1/3 of the celestial sphere.  It was originally designed to make
a three-dimensional map of the distribution of galaxies, but the data
required to make that happen has led to breakthroughs in gravitational
lensing, the structure of the Milky Way, the evolution of quasars, the
nature of brown dwarfs, and many other fields unanticipated at the
time the SDSS was designed.  The success of the initial SDSS, and its
utility to the community as a whole (at this writing, it has resulted
in more than 7000 refereed papers, the vast majority written
with publicly distributed data) has resulted in its operation being
extended multiple times, with continually evolving scientific goals.

 The LSST was in many ways inspired by the SDSS model: by gathering
 data designed for specific scientific goals, and analyzing those 
  data as well as the laws of physics and statistics allow, one
  enables a broad (and to some extent, unanticipated) set of science
  outcomes.  The LSST consists of a wide-field optical telescope with
  an effective diameter of 6.7 meters, and a 3.5 Gigapixel imaging
  camera with a field of view of 9.6 deg$^2$.  The combination of the
  large telescope aperture and the enormous field of view means that
  the LSST is able to simultaneously survey the sky ``Wide, Deep and
  Fast'', i.e., to cover a large solid angle to faint magnitudes, and
  do so repeatedly over its 10-year lifetime, to study variable and
  transient phenomena of all sorts.  The detailed characteristics of
  the survey may be found in the overview paper of \cite{Ivezic08},
  the LSST Science Book \cite{SciBook}, and the LSST website,
  \url{http://www.lsst.org}.

  The LSST main survey footprint will cover 18,000 deg$^2$ of
  high-latitude sky in the Southern hemisphere, in the declination
  range $-65^\circ < \delta < +2^\circ$.  Ancillary programs will push
  further North and South, and to lower Galactic latitudes.  Imaging
  will be done in six broad bands, $ugrizy$, carried out in 30-second
  ``visits'' (probably divided into two 15-second exposures).  The
  5-$\sigma$ point-source depth in a single visit will be $r \sim
  24.7$ in typical seeing (median 0.67'').  Each field will be visited
  over 800 times (summed over the six filters) over the 10-year survey
  period, yielding a full-survey depth of $r=27.5$ (and similarly in
  the other bands).  The key science goals include the study of dark
  matter and dark energy via gravitational lensing, supernovae, and
  large-scale structure; the mapping of the halo of the Milky Way; the
  characterization of the variable and transient sky; and the
  distribution of asteroids in the Solar System, from those which come
  close to Earth to the outer reaches of the Kuiper Belt.  However, as
  we stressed above, the LSST science impact will be much broader than
  that, and promises to influence essentially every area of
  observational astronomy.

  Analyzing the resulting tens of petabytes of data is a major
  component of the LSST project \cite{Juric15}.  The fully reduced and 
  calibrated object catalogs will be made available using a
  sophisticated user interface.  The data will have no proprietary
  period, but will be made public both on short timescales (all
  variable objects will be released worldwide as they are recognized
  by specialized software, following each exposure) and on long ones (yearly
  data releases of the data will be distributed to the US and Chilean
  communities).

  The construction of the  LSST system is supported by the US National
  Science Foundation and Department of Energy, as well as generous
  contributions from private donations.  The telescope is currently
  under construction, and will see first engineering light in late
  2019.  After an extensive commissioning period, it will begin its 
  10-year survey in 2022.  

  At first glance, the synergy between SPHEREx and LSST is not
  obvious, as LSST is focused on the faint universe, pushing to 27.5,
  while SPHEREx reaches its 5-sigma limit at an AB magnitude of 18.5
  for each 96 frequency elements.   (The saturation limit of LSST will depend on the detailed properties
  of the sensors, but will probably be between 16 and 17 in the $r$
  band, bright enough for easy detection by SPHEREx).  However, LSST is a purely imaging survey, and the
  spectroscopic capabilities of SPHEREx and the extension to the IR will be tremendously useful
  for characterizing the properties of the galaxies and stars it
  finds.  LSST plans to use photometric redshifts to determine the
  distances of its galaxies.  The combination of LSST and SPHEREx will
  give continuous wavelength coverage from the atmospheric cutoff at
  $\sim 3200$\AA\ to 5 $\mu$m, allowing far more than redshifts to be
  determined; with such detailed spectral energy density distribution
  information, one can also determine stellar masses, star formation
  rates, dust extinction, the presence of an AGN, and perhaps
  metallicities as well.  There is work ahead to determine
  specifically the impacts of the addition of LSST photometry to
  the SPHEREx spectroscopy will be, as a function of magnitude.  This
  can be carried out for every galaxy above the SPHEREx 5-sigma limit,
  and can be done to much fainter magnitudes in the SPHEREx deep field
  at the South Ecliptic Pole.  It will also be tremendously valuable
  to stack SPHEREx data at the positions of LSST-defined samples, to
  characterize their properties (see Padmanabhan et al. 2016, in preparation).

 The three-dimensional map of the galaxy distribution that SPHEREx
 will produce can be used to measure {\em clustering redshifts},
following the techniques of \cite{Newman08} and \cite{Rahman15}.  This
will allow exquisite calibration of   the redshift distribution of any subsample of galaxies in the LSST
  sample, at least over the redshift range that SPHEREx will probe.

  Another great synergy between SPHEREx and LSST will be in the search
  for and study of extremely red objects.  Such red objects fall into several categories:
\begin{itemize} 
\item  Objects at extremely high redshift ($z>6$), whose optical light is
  absorbed by neutral hydrogen in the intergalactic medium (the
  Gunn-Peterson trough).  The most luminous quasars at these redshifts
  are brighter than the $5\,\sigma$ limit of SPHEREx, and binning
  further in wavelength will allow identification of candidates much
  fainter than this.  
\item Brown dwarfs, whose surface temperatures will be determined to
  high accuracy with SPHEREx spectral coverage (see Sec.~\ref{sec:nearbycool}).
\item Highly reddened objects, whose spectra may show strong emission
  lines in the SPHEREx wavelength coverage. 
\end{itemize}
 Real work is needed in each of these cases to quantify what can be
 learned from the combination of LSST and SPHEREx data.  Some
 particularly interesting objects identified this way will require
 detailed follow-up, e.g., with the James Webb Space Telescope as
 outlined above (Sec.~\ref{sec:jwst-intro}), which wil be a powerful
 tool for characterizing unusual populations revealed from the LSST-SPHEREx synergy. 

  The fact that SPHEREx will survey the entire $4\,\pi$ steradians of
  the sky means that overlaps will occur with {\em all} the wide-field surveys of
  the 2020's.  This means that much of SPHEREx science will depend on
  the mutual comparison of different datasets, and building the data
  structures to allow these comparisons to be carried out efficiently
  and consistently will be key to ensuring the SPHEREx scientific
  legacy.  


\subsection{SPHEREx and TESS}
\label{sec:TESS} 

Direct and precise stellar masses and radii are essential for stellar
astrophysics, as they enable the calibration of correlations between
fundamental stellar  properties, and tests of stellar evolutionary
models. Additionally, the masses and radii of exoplanets discovered
via the transit and radial velocity methods  depend on the masses and
radii of the host stars. The best sample to date -- that of Torres
\cite{Torres} -- has 94 double-lined eclipsing binaries (and $\alpha$Cen)   with precise
($\leq 3\%$ fractional uncertainty), model-independent masses and
radii.  However, this sample contains only four stars with  masses
below half that of the Sun -- a region in which measured radii differ
significantly from radii predicted by models (typically a 10\%
discrepancy). Single M dwarfs also exhibit radius inflation, but the
masses of these stars cannot be measured directly and are typically
inferred from a mass-luminosity relation (cf.
\cite{Mann15}). Furthermore, very few of the Torres systems have
abundance measurements.   

A much larger sample of single-lined eclipsing systems discovered by
exoplanet transit surveys can -- with radial velocity follow-up,
well-constrained SEDs, and \emph{Gaia} parallaxes -- also yield
model-independent masses and radii of both the primary and secondary
stars.  Many of these systems will have low-mass stellar secondaries,
thereby filling in a relatively poorly populated region of parameter
space of stars with directly measured parameters, given precise
follow-up eclipse photometry and high-resolution radial velocity
measurements.  

\begin{figure}[!th]
\center
\includegraphics[width=0.5\textwidth]{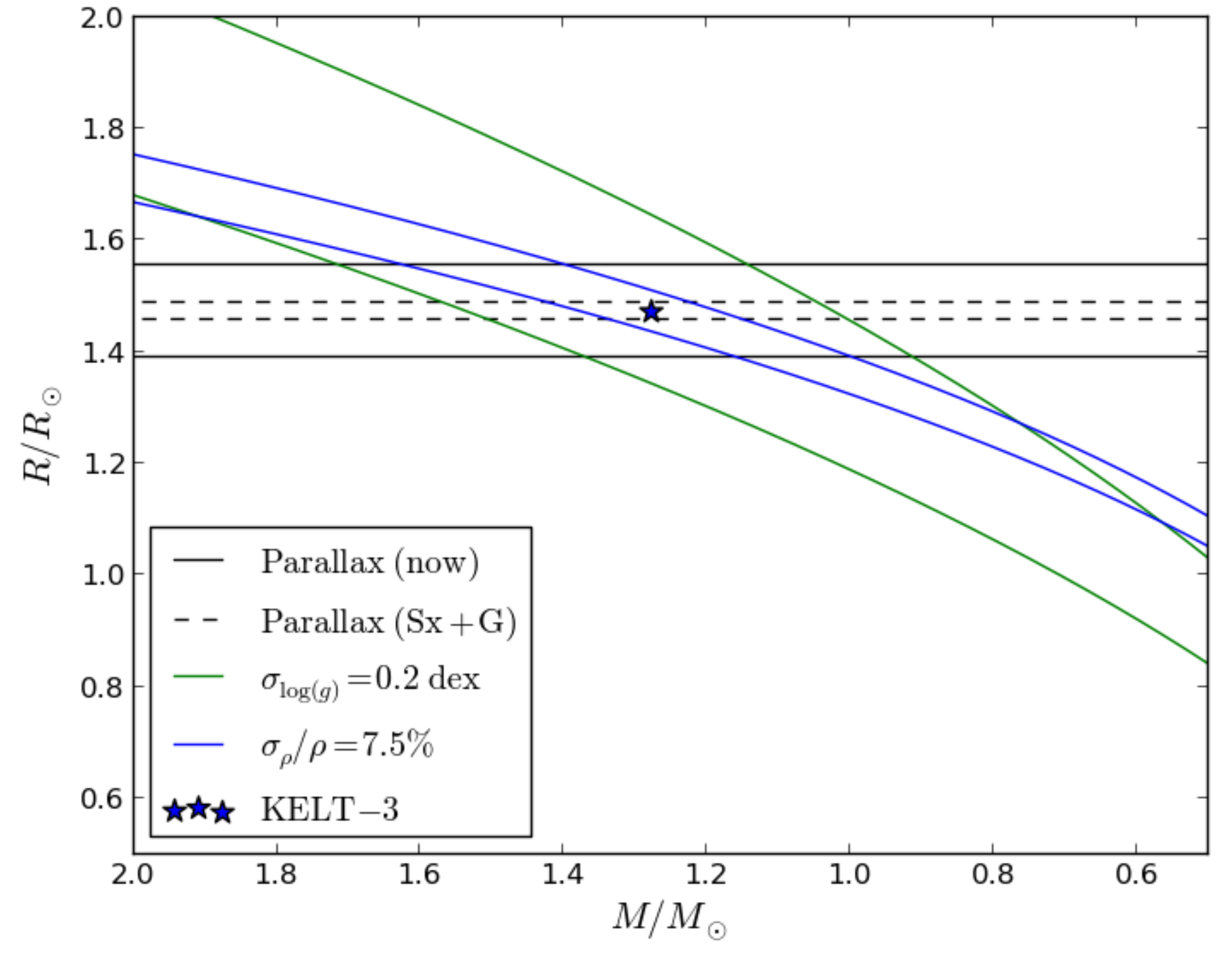}
\caption{\label{fig:mr} Constraints on the mass and radius of KELT-3
  (star) from the transit density (blue), spectroscopic surface
  gravity (green), parallax with only broad-band flux measurements
  (black solid), and parallax with broad-band fluxes as well as
  estimated SPHEREx and \emph{Gaia} spectrophotometry assuming
  Poisson-noise dominated errors (black dashed). Uncertainties are
  assumed typical values, and the KELT-3 parameters and broad-band
  flux measurements can be found in \cite{Pepper13}.} 
\end{figure}

The all-sky Transiting Exoplanet Survey Satellite (TESS) will target
200,000 bright ($I_C \leq 13$) stars with spectral types ranging from
early F to late M for transiting exoplanets, with expected yields of
1,700 planetary systems and 1,150 eclipsing binaries
\citep{Sullivan15}. Thus, TESS will observe many single-lined
eclipsing binaries (including transiting exoplanets) which will have
SPHEREx spectrophotometry as well as \emph{Gaia} parallaxes and
spectrophotometry and literature broad-band UV-NIR fluxes;
high-resolution RV and precise follow-up photometry will also be
possible. Fig.~\ref{fig:mr} illustrates the anticipated power of
spectrophotometric information in constraining the radius of a
typical TESS target. Reducing the uncertainty in the stars radius from
$\sim$ 5\% \cite{Pepper13} to $\sim$ 1\%, as shown in Fig.~\ref{fig:mr} for
KELT-3, will reduce the uncertainty in the radius of the exoplanet to $\sim$ 1\% as
well. Applied to many systems, this precision will greatly increase
our knowledge of exoplanet properties. 

In particular, a large number of low-mass stars will have precisely
determined masses and radii this way, both as eclipsing companions to
more massive stars and as transiting exoplanet hosts, so the low-mass
end of the mass-radius relation can be much more strongly anchored
with the inclusion of these systems. Specifically, due to SPHEREx coverage of the SED peak for cooler stars, the radii (and thus masses) 
of M dwarf planet hosts \emph{and} the planets around them will be
known extremely well. Additionally, TESS is sensitive enough to detect
$p-$mode oscillations for around 2,000 main-sequence and sub-giant
stars with $V \leq 7.5$. For the subset of eclipsing systems for
which TESS will obtain asteroseismic measurements of the primary
star, the asteroseismic density places an additional constraint
(albeit not an orthogonal one) on the primary's stellar parameters and
tightens the uncertainties on the host star's mass and radius and,
therefore, on the companion's mass and radius. 

In addition, measuring spectroscopic parameters for stars in
double-lined eclipsing systems requires accurate disentanglement of
each star's spectral features from the composite spectra. TESS,
however, can provide effective temperatures of the secondary directly
from the secondary eclipse depths, given the primary's effective
temperature as determined from spectrophotometric interferometry.   

\section{Conclusion}

The SPHEREx Community Workshop identified and discussed many
scientific investigations enabled by the unique 0.75-to-5 $\mu$m all
sky spectral database which will be the legacy of SPHEREx.  This
scientific return from this legacy would make SPHEREx a worthy
participant in the Decade of the Surveys, as the 2020s have been
called.  The Workshop also showed that SPHEREx is synergistic with
numerous other major astronomical facilities  - space and ground
including GAIA, JWST, LSST, Kepler, eROSITA, TESS, Euclid and WFIRST.  JWST and 
WFIRST users in particular could mine the SPHEREx data bases to select optimum targets for a range
of General Observer investigations.  For the case of JWST, this
synergy relies on the temporal overlap of the two missions.  Selection
of SPHEREx for flight on the current schedule would lead to initial
releases of SPHEREx data and catalogs in 2021-to-2022. This is well
synchronized with JWST, which is planned to operate at least through
2024.  While JWST is expected to operate well beyond 2024, delaying
the selection of SPHEREx to a subsequent opportunity two or three
years hence would undermine this valuable synergy.  The scientific
examples discussed above illuminate the power of an all-sky spectral
survey, which would be a new scientific tool for astronomers.  On the
one hand, it would permit comprehensive studies of known classes of
objects, providing spectra for thousands of brown dwarfs or hundreds
of millions of galaxies, as an example.  It would also identify and highlight for further study
objects such as low metallicity stars or high redshift QSOs, known to exist but
frequently difficult to identify on the basis of photometric surveys
alone.  Finally, of course, like other all sky surveys which open new
parameter space, SPHEREx has the potential to yield new discoveries
which change our view of the astronomical Universe. 

\acknowledgments

O.D and M.W. would like to warmly thank Michael Blanton, Davy
Kirkpatrick, Casey Lisse,  Gary Melnick, Massimo Robberto, Michael
Strauss, Stephen Unwin, Meg Urry and Roger Windorst. They served as 
  the Scientific Organizing Committee for this workshop  and their
  time and suggestions were invaluable.  We would like to also thank Kathy
  Deniston, Michele Judd  and the Keck Institute for Space Sciences
  staff for logistical support  during our workshop. Part of the
  research described in this paper was carried out at the Jet
  Propulsion Laboratory, California Institute of Technology, under a
  contract with the National Aeronautics and Space Administration. 

\bibliographystyle{JHEP}
\bibliography{merged_refs}

\end{document}